\documentclass[structabstract]{aa}
\usepackage{graphicx}
\usepackage{natbib} \bibpunct{(}{)}{;}{a}{}{,}
\usepackage{color}
\usepackage{rotating}
\usepackage{txfonts}
\usepackage{siunitx}
\usepackage{caption}
\usepackage{rotating}
\usepackage{mwe,tikz}\usepackage[percent]{overpic}
\usepackage[colorlinks=true,linkcolor=blue,citecolor=blue,urlcolor=blue]{hyperref}
\graphicspath{{graphics/}}

 \def\mso{\,\mathrm{M}_\odot}

 \def\llso{\log\, L/{\rm L}_\odot \,}
 \def\simle{\mathrel{\hbox{\rlap{\hbox{\lower4pt\hbox{$\sim$}}}\hbox{$<$}}}}
 \def\simgr{\mathrel{\hbox{\rlap{\hbox{\lower4pt\hbox{$\sim$}}}\hbox{$>$}}}}

\newcommand{\msunpyr}{\ifmmode{\,M_{\odot}\,\mbox{yr}^{-1}} \else{ M$_{\odot}$/yr}\fi}

\newcommand{\kms}{\ifmmode{\,\mbox{km}\,\mbox{s}^{-1}}\else{km/s}\fi}
\newcommand{\kpc}{\ifmmode {\,\mbox{kpc}} \else{kpc}\fi}
\newcommand{\msun}{\ifmmode M_{\odot} \else M$_{\odot}$\fi}
\newcommand{\rsun}{\ifmmode R_{\odot} \else R$_{\odot}$\fi}
\newcommand{\lsun}{\ifmmode L_{\odot} \else L$_{\odot}$\fi}
\newcommand{\zsun}{\ifmmode Z_{\odot} \else $Z_{\odot}$\fi}
\newcommand{\xsun}{\ifmmode X_{\odot} \else $X_{\odot}$\fi}
\newcommand{\velo}{\ifmmode\varv\else$\varv$\fi}
\newcommand{\vinf}{\ifmmode\velo_\infty\else$\velo_\infty$\fi}
\newcommand{\rgal}{\ifmmode \,R_{\mathrm{gal}} \else R$_{\mathrm{gal}}$\fi}

\begin{document} 
 
\title{A synthetic population of Wolf-Rayet stars in the LMC based on detailed single and
binary star evolution models}

\subtitle{}
\titlerunning{A synthetic WR star population in the LMC}
 
\author{D. Pauli$^{1,2}$, N. Langer$^{1,3}$, D. R. Aguilera-Dena $^{4}$, C. Wang$^{1,5}$, P. Marchant$^{6}$
}
 
\institute{Argelander-Institut f{\"u}r Astronomie der Universit{\"a}t Bonn, Auf dem H{\"u}gel 71, 53121 Bonn, Germany\label{inst1}
\and Institut f{\"u}r Physik und Astronomie, Universit{\"a}t Potsdam, Karl-Liebknecht-Str. 24/25, 14476 Potsdam, Germany\label{inst2}
\and Max-Planck-Institut für Radioastronomie, Auf dem H\"ugel 69, 53121 Bonn, Germany
\and Institute of Astrophysics, FORTH, Dept. of Physics, University of Crete, Voutes, University Campus, GR-71003, Heraklion, Greece 
\and Max Planck Institute for Astrophysics, Karl-Schwarzschild-Strasse 1, 85748 Garching, Germany
\and Institute of Astronomy, KU Leuven, Celestijnenlaan 200D, 3001 Leuven, Belgium
}
 
\date{Received ; Accepted}

\abstract{Without doubt, mass transfer in close binary systems contributes to the
 populations of Wolf-Rayet (WR) stars in the Milky Way and the Magellanic Clouds. 
 However, the binary formation channel is so far not well explored.}
{We want to remedy this by exploring large grids of detailed binary and single star
    evolution models computed with the publicly available MESA code, for a metallicity appropriate for the Large Magellanic Cloud (LMC).}
{The binary models are calculated through
    Roche-lobe overflow and mass transfer, until the initially more massive star exhausts helium in its core. 
    We distinguish models of WR and helium stars based on the estimated stellar wind optical depth.
    We use these models to build a synthetic WR population, assuming constant star formation.}
{Our models can reproduce the WR
    population of the LMC to significant detail,
    including the number and luminosity functions of the main WR subtypes. 
    We find that for binary fractions of 100\% (50\%), all LMC WR stars below $10^6\,L_{\odot}$
    ($10^{5.7}\,L_{\odot}$) are stripped binary mass donors.
    We also identify several insightful mismatches.
    With a single star fraction of 50\%, our models produce too many yellow supergiants,
    calling either for a larger initial binary fraction, or for enhanced mass-loss
    near the Humphreys-Davidson limit. 
    Our models predict more long-period WR binaries than observed, arguably due to an
    observational bias towards short periods. 
    Our models also underpredict the
    shortest-period WR binaries, which may have implications for understanding the progenitors of
    double black hole mergers.}
{
    The fraction of binary produced WR stars may be larger than often assumed, and outline the risk
    to mis-calibrate stellar physics when only single star models are used to reproduce the
    observed WR stars. 
}
\keywords{stars: Wolf-Rayet - stars: evolution - stars: massive - binaries: close - galaxies: Magellanic Clouds}
\maketitle

\section{Introduction} 
\label{sec:intro} 

    Wolf-Rayet (WR) stars are helium-rich massive stars with strong, optically thick winds 
    which lead to emission-line dominated spectra \citep{cro1:07}. 
    Their high surface temperatures place many of them to the left of the zero-age main sequence in the HR diagram \citep{abb1:87,sch1:89,koe1:95,hai1:14}, allowing them to emit copious amounts of ionizing radiation \citep{wal1:04}, which may strongly influence star forming regions in galaxies \citep{ram1:18,cro1:19}, and the whole appearance of star-forming so called ``Wolf-Rayet galaxies'' \citep{con1:91,bri2:08}. 
    
    The strong winds and ionizing photons of WR stars drive shocks through the interstellar medium
    and sweep up interstellar bubbles \citep{gar2:96,gar1:96,wea1:97}. Their ejection of hydrogen- or helium-burning products \citep{mae1:83,dra1:03} and dust \citep{che1:00} drives the chemical evolution of galaxies. Finally, WR stars may also give rise to hydrogen-free \citep{des1:11,gro3:13} and superluminous supernovae \citep{ins1:13,des1:20,agu1:20,agu1:22}, to long-duration gamma-ray bursts \citep{woo2:93,yoo1:05}, and black hole binaries \citep{rem1:06,lan1:20}.
    
    Despite their importance, the formation process of WR stars is still poorly constrained. 
    There are two main ways for massive stars to have their envelope stripped. In the single star scenario, the star develops strong winds that 
    remove its hydrogen-rich envelope on a timescale comparable to its life time \citep{loo1:78,mae2:87,lan1:87}. Sufficiently strong winds are only achieved by the most massive stars. Due to the metallicity dependence of hot star winds \citep{mok2:07}, it is unknown whether self-stripping
    of single stars is possible at low metallicity. Alternatively, stars of any mass and metallicity may lose most of their hydrogen envelope
    due to interaction with a close binary companion \citep{pac1:67,van1:80,wel1:99}. 
    A third, but not dominant channel of forming WR stars would be  through internal mixing, which requires quasi-chemically homogeneous evolution \citep{mae1:87,yoo1:06, woo1:06,mar1:16,agu1:18,has1:20}. It
    is suggested only to be important at low metallicity, high mass, and extreme rotation. 
    
    It is yet unclear which of the two envelope stripping channels, the single star and the binary channel, is more important for forming WR stars.  The main reason for this is that the mass loss rate of cool massive stars are not yet well understood.
    In particular, the role of the Luminous Blue Variables stage \citep{wei1:20} in removing the H-rich envelope in stars near the Eddington-limit \citep{lan1:94,smi1:06,gra1:21} and the metallicity dependence 
    of LBV mass loss \citep{kal1:18}, are uncertain.
    Similarly, the mass loss rates of red supergiants are weakly constrained \citep{mau1:11}
    as well as, again, the impact of metallicity \citep{kee1:21}.     
    
    The single star path for forming WR stars has been explored extensively \citep[and references therein]{mae1:12,lan1:12}. \citet{mey1:03,mey1:05} found that single star models which include
    strong rotationally induced internal mixing can explain the major properties of the WR populations in the Galaxy and the Magellanic Clouds.
    However, the magnitude of rotational mixing has been questioned since then, based on the spectroscopic analysis of large samples of O and early B stars \citep{hun1:08,gri1:17,mar1:18}, and the realization that the majority of stars massive enough to potentially form WR stars rotate relatively slowly \citep{bes1:14,sab1:17,ram1:17}.
    
    Over the last decade several works have argued that binary interactions are crucial for our understanding of stellar evolution and that their importance cannot be neglected, as its stellar products --- compared to single star models --- contribute to different stellar populations \citep[e.g.][]{dem2:13,sch1:15,wan1:20}. Nonetheless, it is still unclear to which extend binaries contribute to these populations \citep[e.g.][]{She1:20,mas1:21}. Recent observations of massive stars show that their evolution is dominated by binary interaction \citep{san2:12,san1:14,moe1:17}, meaning that at least half of all massive stars are born with a close companion
    such that mass transfer is unavoidable. The immediate implication is that binarity plays a crucial role for WR star formation.
    Fortunately, the uncertainties in the process of the formation of WR stars through binary systems are much smaller than those for single stars. The result of binary induced mass stripping
    is well defined; the donor in mass transferring binary systems contract and end the mass stripping once most of
    their hydrogen envelope is gone. Even though binary evolution in general
    needs to consider a larger number of initial parameters, and contains additional physics uncertainties \citep{eld1:08}, the predictions of the properties of WR star models formed through binary interaction
    may be more credible than comparable single star predictions, since the uncertain wind mass loss rates for cool stars, or their metallicity dependence,
    hardly matters.

    Here, we explore WR star formation in single stars and close binaries in order to bring clarity to the question on how the individual channels contribute to an entire WR population \citep[e.g.][]{goe1:18,She1:20}. For this purpose, 
    we use a dense grid of detailed binary evolution models. The computation of a complementary grid
    of single star models using the same input physics allows us to address the relative importance of both WR star formation channels.
    The ideal testing ground for our models turns out to be the Large Magellanic Cloud, because with more than 100 WR stars, it provides a statistically significant 
    WR sample \citep{bre1:99,bar1:01,hai1:14}. At the same time, the LMC WR sample is thought to be essentially complete \citep{neu1:18},
    and likely originates from a more chemically homogeneous O star population than the Galactic WR stars. 
    
    Our paper is organized as follows. In the next section, we describe the set-up and choice of physics parameters for our single star and binary evolution models,
    and provide examples of evolutionary tracks in the HR diagram for both. In Sect.\,\ref{sec:the_models}, we outline our procedure to obtain a synthetic LMC WR star population
    from our stellar evolution models. Sect.\,\ref{sec:results} provides a discussion of the obtained synthetic populations of WR stars, separating between WN stars with and without
    hydrogen, as well as WC stars, and compares the result with the observed WR populations. In Sect.\,\ref{sec:orbtial_periods_and_mass_ratios}, we discuss the properties of our our computed
    WR binary systems, and compare them with the properties of observed WR binaries. Sect.\,\ref{sec:disc} gives a discussion of the uncertainties involved
    in our analysis, before we provide our final conclusions in Sect.\,\ref{sec:conclusions}.

\section{Methods}
    In this section, we describe the methods employed to produce and analyze a grid of detailed stellar models. In Sect.\,\ref{sec:binary_model_grids} we provide a brief overview of the chosen input parameters of our single and binary evolutionary models. In Sect.\,\ref{sec:optical_depth_of_WR_winds} we introduce a criterion based on the optical depth to differentiate stripped-envelope (He-stars) and WR stars. Finally, in Sect.\,\ref{sec:pop_sys} we describe the assumptions made to create a synthetic WR stellar population.

\subsection{Input parameters for our stellar evolution models}
\label{sec:binary_model_grids}
    We use version 10398 of the Modules for Experiments in Stellar Astrophysics (MESA) code \citep[e.g.][]{pax1:11,pax1:13,pax1:15,pax1:18,pax1:19} to calculate a grid of detailed binary stellar evolution models at LMC metallicity. The initial chemical composition of our models by mass fraction is $X_\mathrm{H} = 0.7383$, $X_\mathrm{He} = 0.2569$ and $Z=0.0048$, where $Z$ includes all elements heavier than helium. 
    Solar-scaled abundances for the heavy elements in metal poor dwarf galaxies, like the LMC, are commonly used in stellar evolutionary models \citep[i.e.][]{egg1:21,eld1:17}. However, dwarf galaxies have different metallicity distributions than galaxies like the Milky Way \citep[i.e.][]{mar1:19}. Therefore, we employ for our models a tailored initial chemical compositions, similar as in the work \citet{bro1:11}. For the initial C, N and O mixtures we use the abundances as they are observed from \ion{H}{II} regions \citep{kur1:98}, namely ${\log(\mathrm{C}/\mathrm{H})+12=7.75}$, ${\log(\mathrm{N}/\mathrm{H})+12=6.90}$, and ${\log(\mathrm{O}/\mathrm{H})+12=7.35}$. For Mg and Si we use ${\log(\mathrm{Mg}/\mathrm{H})+12=7.05}$, and, ${\log(\mathrm{Si}/\mathrm{H})+12=7.20}$ based on observations from B-type stars in the LMC \citep{tru1:07,hun1:07,hun1:09}. As initial iron abundance we use ${\log(\mathrm{Fe}/\mathrm{H})+12=7.05}$ \citep[see][]{bro1:11} which is in agreement with the observed abundances in the LMC \citep[e.g.][]{fer1:06}.  For the remaining metals we adopt the initial chemical abundances of \citet{asp1:09} reduced by 0.4\,dex, accounting for the lower metallicity in the LMC.
    
    The resulting model grid covers initial primary masses in the range of ${M_{1,\mathrm{i}}\simeq\SIrange{28}{89}{M_\odot}}$ in steps of ${\Delta \log(M_{1,\mathrm{i}}/\msun)=0.05}$. Stellar models with lower initial mass are not expected to contribute to the WR population (cf. Fig.\,\ref{fig:Phase_Binary}) and thus are not included in this work. The initial orbital periods cover the range of $P_\mathrm{\,i}\simeq \SIrange{1.4}{10000}{d}$ in steps of ${\Delta \log(P_\mathrm{\,i}/\si{d})=0.05}$, and the initial mass ratios cover the range of ${q_\mathrm{\,i}=\numrange{0.25}{0.95}}$ in steps of ${\Delta q_\mathrm{\,i}=0.05}$. Our grid contains a total of $10780$ models. We assume that our binary models are initially tidally synchronized at the zero-age main-sequence (ZAMS). As the models with the highest initial orbital periods (i.e., ${P_\mathrm{\,i} \simeq \SI{10000}{d}}$) never interact during their evolution, they serve as our grid of quasi non-rotating single star models. While initial tidal synchronization is not realistic for wide binary systems, it ensures that both components evolve
    essentially as our single star models before the onset of mass-transfer via Roche-lobe overflow (RLOF).
    
    Stellar wind mass-loss is included as in \citet{bro1:11} with slight modifications, described below. The mass-loss rates of \citet{vin1:01} are used for hydrogen-rich main-sequence stars. For temperatures below the bi-stability jump \citep[][their equation 14 and 15]{vin1:01}, the maximum value of either \citet{vin1:01} or \citet{nie1:90}  wind prescriptions is adopted.
    To model mass-loss during the WR phase, we use the mass-loss rates of \citet{nug1:00} for the hydrogen-rich WN phase where the surface hydrogen abundance is below $X_\mathrm{H} < 0.4$, and for the hydrogen-free WN and WC phase we use the mass-loss recipe of \citet{yoo1:17} with a clumping factor of $D=4$, meaning that this leads to a variation of the original mass-loss rate as ${\dot{M}\propto D^{-1/2}}$. We note, that some of the mass-loss recipes are originally calibrated for $D=10$ and thus when using a clumping factor of $D=4$ the mass-loss rates are increased by a factor of $1.58$. For surface hydrogen abundances between $0.7$ and $0.4$, the wind mass-loss rate is linearly interpolated between those of \citet{vin1:01} and \citet{nug1:00}. For all used mass-loss rates a solar baseline of  $\zsun=0.017$ is used.

    Rotational mixing is modeled as a diffusive process including the effects of dynamical and secular shear instabilities, the Goldreich-Schubert-Fricke instability, and Eddington-Sweet circulations \citep{heg1:00}. The efficiency of rotational mixing is calibrated following \citet{bro1:11}, with an efficiency factors $f_c=1/30$ and $f_\mu=0.1$. In addition, internal angular momentum transport via magnetic fields is included according to \citet{spr1:02}.

    Convection is modeled using the Ledoux criterion and the standard mixing length theory \citep{boe1:58} with a mixing length parameter of ${\alpha_\mathrm{mlt}=l/H_\mathrm{P}=1.5}$. As such, the
    effect of envelope inflation near the Eddington limit \citep{san1:15}
    is not suppressed. However, to avoid convergence failure during the late evolutionary stages of the primaries and secondaries
    in our binary models, the option MLT++ \citep[see section 7.2 of][]{pax1:13} is turned on for stellar models at core helium ignition if their helium core mass exceeds $>14\mathrm{\msun}$.
    For core hydrogen burning models, step overshooting is used so that the convective core is extended by ${0.335\,H_\mathrm{P}}$ \citep{bro1:11} where ${H_\mathrm{P}}$ is the pressure scale height at the boundary of the convective core. Thermohaline mixing is included with an efficiency parameter of ${\alpha_\mathrm{th}=1}$ \citep{kip1:80}, and semiconvective mixing is included with an efficiency parameter of ${\alpha_\mathrm{sc} = 1}$ \citep{lan1:83}.
    
    Mass transfer is modeled by using the 'contact' scheme from MESA \citep{mar1:16}. This is an implicit method in which the mass transfer rate is adjusted in such a way that the radius of the donor star stays inside the Roche lobe. Moreover, as the name already suggests, it is capable of modeling a contact phase where 
    both stars overfill their Roche volumes, as long as mass outflow via the
    2nd Lagrange point is avoided \citep{mar1:16,men1:21}.
    
    The mass donors in our binary models are calculated until core helium depletion, after which we assume that they form a compact object. We assume that the system becomes disrupted and thus set the orbital separation to infinity, implying that from then on the mass gainer is modeled without further binary interaction, until core carbon depletion.
    
\subsection{The optical depth of WR winds}
\label{sec:optical_depth_of_WR_winds}

    Classical WR stars are hydrogen-poor stars with optically thick winds that show strong emission lines. \citet{She1:20} found that WR stars have a minimum luminosity that depends on metallicity. Stripped-envelope stars below this luminosity are believed to have optically thin atmospheres and do not posses the emission features that characterizes WR-stars \citep[i.e.][]{goe1:18}. Stars below the minimum luminosity are also believed to have mass-loss rates that are lower than those that would be inferred from extrapolating empirical mass-loss rate prescriptions to lower luminosities \citep{vin1:17,san1:20}. 
    
    Therefore, we employ a criterion to select WR stars from our models, and discard transparent-wind stripped-envelope stars.
    Following \citet{agu1:21} we estimate the surface optical depth of our stellar models, and only include those with large enough optical depths in our synthetic WR population. This criterion assumes that the velocity structure in a stellar wind can be described by a $\beta$-law with $\beta = 1$, and that the opacity $\kappa$ can be approximated by the electron scattering opacity $\kappa_\mathrm{es}=0.20\,(1+X_\mathrm{H})\,\si{cm^2\,g^{-1}}$. We then compute the optical depth  according to \citep{lan1:89} as
    \begin{equation}
        \tau(R) = \dfrac{-\kappa \dot{M}}{4\pi R (\varv_\infty-\varv_0) } \ln\left(\dfrac{\varv_\infty}{\varv_0}\right).
        \label{eq:optical_depth}
    \end{equation}
    Here, $\dot{M}$ is the wind mass-loss rate, $R$ the radius of the star, $\varv_\infty$ is the terminal wind velocity, and $\varv_0$ the expansion velocity at the base of the wind, which is on the order of the sound speed $\varv_0 = \SI{20}{km\,s^{-1}}$.

    Following \citet{gra1:17}, the terminal wind velocity $\varv_\infty$ of H-free WR stars is related to their escape velocity $\varv_\mathrm{esc}$ and can be expressed as
    \begin{equation}
        \varv_\infty=1.3\, \varv_\mathrm{esc}=1.3\sqrt{\frac{2GM}{R}(1-\Gamma})\,,
        \label{eq:terminal_wind_velocity}
    \end{equation}
    where $\Gamma$ is the Eddington factor. For WC stars a steeper relation of ${\varv_\infty=1.6\, \varv_\mathrm{esc}}$ is used \citep{gra1:13}.
    For hot OB stars (${\SI{27500}{K} \leq T_\mathrm{eff} \leq \SI{50000}{K}}$) we use a steep relation of ${\varv_\infty=2.6\, v_\mathrm{esc}}$, and a more shallow relation for cool OB stars (${\SI{12500}{K} \leq T_\mathrm{eff} \leq \SI{22500}{K}}$), given by ${\varv_\infty=1.3\, v_\mathrm{esc}}$ \citep{vin1:01}.  Regarding H-rich and H-poor WN stars, no such relation has been derived yet. We assume that ${\varv_\infty=1.3\, v_\mathrm{esc}}$ is a valid choice for this phase, as it is also used for the later H-free WN phase. For the supergiant stage the maximum value of the mass-loss rates of \citet{vin1:01} or \citet{nie1:90} are used. As in \citet{nie1:90} no such correlation is given, the correlation of ${\varv_\infty=0.7\,\varv_\mathrm{esc}}$ for temperatures below ${T_\mathrm{eff}\leq \SI{10000}{K}}$ is used \citep{vin1:01}.
    
    \citet{agu1:21} calibrated this method for hydrogen-free WR stars using the minimum luminosities inferred from the SMC, LMC and Milky Way by \citet{She1:20}, and find that a wind optical depth of $\tau=1.45$ describes the borderline between WR-stars and helium stars with transparent winds best (see their figure 1). For simplicity in this work $\tau_\mathrm{thresh}=1.5$ is used. 
    For WR stars with hydrogen, no such calibration is available. However, using the above scheme to estimate the optical depth of observed LMC WR stars with hydrogen indicates optical depth
    values between 0.3 and 1. Therefore, this criterion is employed in our models to distinguish between classical WRs, hydrogen burning WRs and transparent-wind stripped-envelope stars (see also Appendix\,\ref{app:optical_depth}).
 
\subsection{Population synthesis}
\label{sec:pop_sys}
    
    To create a synthetic WR population we employ the intrinsic mass, period and mass-ratio distributions obtained by \citet{san1:13} for stars in the LMC. To summarize, we assume that the initial masses of single and donor stars are given by an initial mass function of the form ${f(M/\msun) \propto (M/\msun)^{-\alpha}}$ with an exponent of ${\alpha = 2.3}$ \citep{sal1:55}, that the intrinsic orbital period distribution is given by ${f(\log(P/\si{d})) \propto \log(P/\si{d})^\pi}$ with an exponent of ${\pi = -0.45}$ \citep{san1:13} , and that the intrinsic mass-ratio distribution is given by ${f(q) \propto q^{\,\kappa}}$ with an exponent of ${\kappa=-1.0}$ \citep{san1:13}. The intrinsic binary fraction derived by \citet{san1:13} is ${f_\mathrm{bin}=0.51\pm0.04}$, but for simplicity we use ${f_\mathrm{bin}=0.5}$ in the following analysis.

    To minimize statistical fluctuations, we draw ${15\,000\,000}$ binary and ${15\,000\,000}$ single star models according to the initial distributions. To link the drawn initial parameters to a model in our grid, a nearest neighbor approach is used.
    
    It is assumed that there was a constant star formation rate in the LMC for the last $\SI{10e6}{yr}$ thus the age of all stars follows a uniform distribution. The time resolution of our population model is limited by the time resolution of the stellar evolution models. With typical time steps of $\lesssim\SI{1000}{yr}$ in our models, it is ensured that also short lived evolutionary phases are covered. A model only contributes to the synthetic WR population if its calculated optical depth is $\tau\geq1.5$ (see Sect.\,\ref{sec:optical_depth_of_WR_winds}).
    
    To make the synthetic WR population comparable to observations it is divided into four subgroups: H-rich WN, H-poor WN, H-free WN and WC. The synthetic H-free WN population is used to obtain a normalization factor such that the number of synthetic H-free WN stars matches the observed number of H-free WN stars in the LMC. The resulting normalization factor is then applied to the other WR subgroups. \citet{har1:09} found that within the last $\SI{50e6}{yr}$ the average star formation rate in the LMC was ${\mathrm{SFR}_\mathrm{obs}\approx\SI{0.4}{M_\odot\,yr^{-1}}}$. With the normalization required to reproduce the WR population we obtain a similar result of ${\mathrm{SFR}_\mathrm{req}\approx\SI{0.4}{M_\odot\,yr^{-1}}}$.
    
    We note, that our synthetic population is not complete, in a sense, as not all possible evolutionary scenarios can be modeled with the MESA code yet. The supernova explosion of the primary and its effect on the binary orbit is not included. Afterwards the secondary is evolved as a single star. Common envelope evolution and its descendant stellar products are not modeled and thus not included in the final population. We note, that common envelope evolution, especially for high mass stars, is not well understood, and the final contribution of systems that have undergone a common envelope evolution to the WR population remains highly uncertain \citep{iva1:20}. We discuss possible impacts of post common envelope evolution systems on our synthetic WR population in more detail in Sect.\,\ref{sec:unconsidered_binary_outcomes}.

\section{Single star and binary evolution models}
\label{sec:the_models}

    In the following section, we discuss the most important aspects of the evolutionary tracks of our single star and binary models, as function of the initial stellar masses, and the initial orbital periods. For a comparison with other stellar evolutionary models and their predictions on the WR population, we refer to Appendix\,\ref{sec:comparision_with_previous_works}.

    \subsection{Single star models}

\begin{figure}[t]
        \centering
        \begin{tikzpicture}
            \node [anchor=south west] at (0,0)
            {\includegraphics[trim= 0.7cm 0.7cm 0.7cm 0.7cm ,clip ,width=0.47\textwidth]{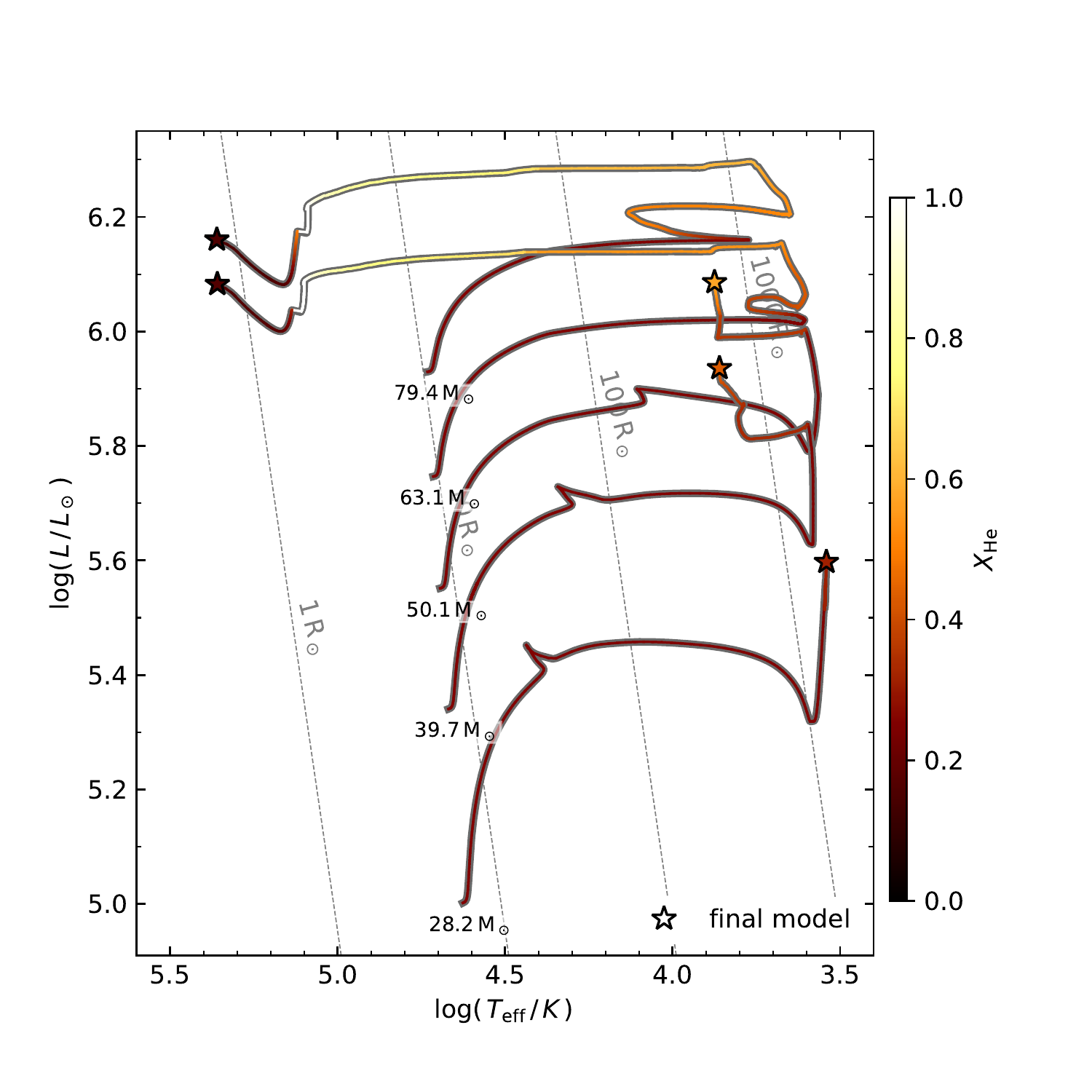}};
        \end{tikzpicture}
        \caption{Evolutionary tracks of selected single star models in the Hertzsprung-Russell diagram. The tracks are labeled by the initial stellar mass, 
        and color-coded by the surface helium abundance $X_\mathrm{He}$
        (see color bar to the right). A star symbol (also color-coded), marks the point of 
        core helium depletion. Lines of constant radius are drawn as thin straight lines.}
        \label{fig:HRD_single}
    \end{figure}

    We present five selected evolutionary tracks of our single star models in Fig.\,\ref{fig:HRD_single} that are representative for the possible outcomes of our single star grid. All models start their life on the the zero-age main-sequence, and they expand on the nuclear timescale during core hydrogen burning. 
    Core hydrogen exhaustion can be identified by a hook in the evolutionary track, which corresponds to a short contraction phase before hydrogen shell ignition. For the two tracks with the highest initial masses this contraction occurs at effective temperatures below 10\,kK
    \citep[see also][]{bro1:11,koe1:15}. 
    
    All stellar models evolve into cool supergiants, with large envelope
    masses and effective temperatures below 10\,kK. 
    Here, models below $\lesssim 30\mso$ become classical red supergiants, while the more massive models develop He-enriched surfaces and 
    become yellow supergiants. With the mass loss rates adopted as described above, only  
    models with masses above $M_\mathrm{\,i} \gtrsim 50\msun$ 
    manage to remove most of their hydrogen-rich envelope and
    develop into a WR star. Since the mass loss rates
    of cool supergiants are uncertain, so is the mass limit above
    which our single star models can become WR stars. We discuss the
    consequence of this for our synthetic WR population in detail in Sect.\,\ref{sec:reliability}.
    
    In order for our models to reach a WR phase, they need to achieve
    a surface helium mass fraction of $Y_{\rm s}\simeq 0.4$ or more \citep[see][for a model-independent view on this]{sch1:18}. Only then, their surface temperatures can exceed 
    30\,kK. 
    During their evolution as WR stars, the models go first though a 
    phase during which the surface still contains hydrogen (H-poor WN phase). In case they lose their hydrogen envelope completely,
    as the two most massive models shown in Fig.\,\ref{fig:HRD_single},
    they settle on the helium main sequence (we do not correct the
    surface temperature of our WR models for optical depth effects of their winds) as H-free WN stars. In case the models also lose their
    helium-envelope, they may appear as WC\,stars. We do not consider
    a possible WO phase here \citep[see][]{agu1:21}.
     
    \subsection{Binary star models}
        
    \begin{sidewaysfigure*}
        \centering
        \includegraphics[trim= 3.0cm 2.5cm 3cm 2.5cm ,clip ,width=0.9\textwidth]{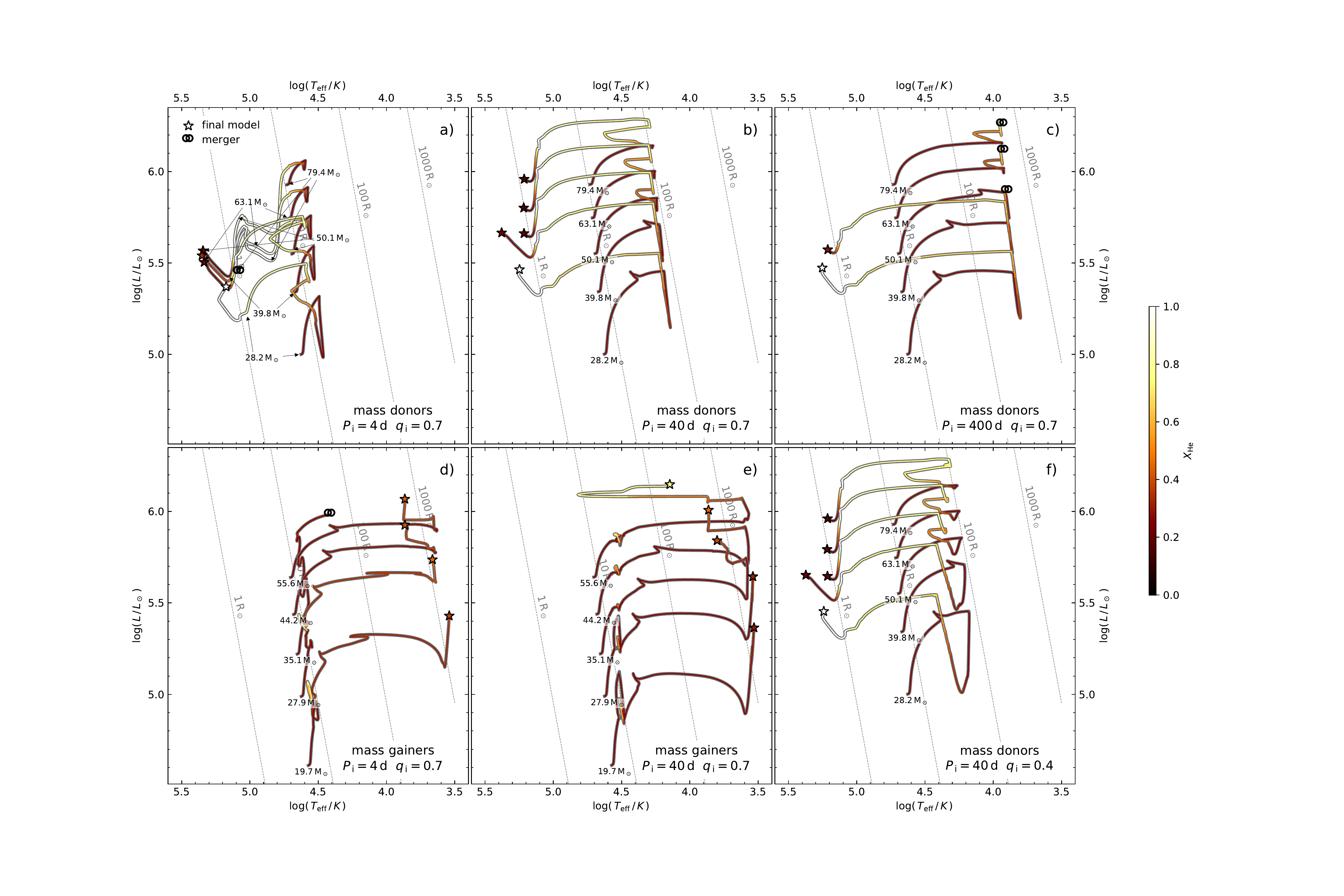}
        \caption{Evolutionary tracks of mass donors (panels a)-c) and f)), and of mass gainers
        (panels d) and e)), for selected binary models in the Hertzsprung-Russell diagram. 
        The tracks are labeled with the selected initial stellar masses. 
        The mass gainers shown in panels d) and e) are the counterparts to the mass donors
        shown in panels a) and b), respectively. The top row depicts the donor evolution for
        three different initial orbital periods, for binaries with a fixed initial mass ratio of $q_{\rm i}=0.7$. Panels b) and f) show donors for the same orbital period ($P_{\rm i}=40\,$d) but for two different initial mass ratios.
        In the panel a) diagram, arrows identify different parts of the same tracks. The tracks and star symbols are color-coded as in Fig.\,\ref{fig:HRD_single}. 
        Double rings marks a predicted merger situation; our calculations end there. 
        }
        \label{fig:HRD_binary}
    \end{sidewaysfigure*}

    The evolution of a star in a binary system differs drastically from the single star scenario as it is possible that both components interact during their lifetime. In a binary system mass transfer is initiated when one of the stellar components grows and exceeds the so-called Roche-lobe radius $R_\mathrm{RL}$ as the overflowing material is gravitational attracted by the other component \citep{Tau:06}. 
    In Fig.\,\ref{fig:HRD_binary} the evolutionary tracks of the mass-donors and mass-gainers for different initial orbital periods and mass ratios are shown.
    
    Panel a) of Fig.\,\ref{fig:HRD_binary} shows the evolutionary tracks of the mass donors
    for five initial donor masses of systems with an
    initial orbital period of $P_\mathrm{\,i}=\SI{4}{d}$ and an initial mass ratio $q_\mathrm{\,i}=0.7$. All five donors start mass-transfer while they are still undergoing core hydrogen burning. From the indicated helium surface abundances it becomes evident that most of the hydrogen rich envelope is removed from the donor star model
    during the mass transfer. While the donor models with initial masses $M_\mathrm{\,i}= 28.2\msun$, $39.2\msun$, $50.1\msun$ and $63.1\msun$ are calculated up to core helium depletion, the binary model with an initial donor mass of $79.4\msun$ enters a common envelope phase which leads to a merger of both components.
    
    Panel d) shows the evolution of the mass-gainers corresponding to the mass-donors of panel a). These models accrete a large amount of the transferred matter, as indicated by the large
    luminosity increase compared to single star models.
    Their surface helium abundance is enriched during the mass transfer, but later reduced by
    thermohaline mixing. 
    
    The evolutionary tracks of mass-donors of systems with initial orbital period of $P_\mathrm{\,i}=\SI{40}{d}$ are shown in panel b) of Fig.\,\ref{fig:HRD_binary}. Here, the two lowest mass models start mass transfer after core hydrogen exhaustion.
    Models above $\gtrsim50.1\msun$ inflate during their main-sequence evolution and initiate mass transfer during core hydrogen burning. However, the effect of the mass-transfer phase is similar for all models, as almost all of the hydrogen-rich material is removed from the donor star model, followed by a blueward evolution. A comparison to the models shown in panel a)
    shows that the resulting helium star models cover a larger luminosity range. 
    The more limited luminosity increase of the corresponding mass gainers (panel e)) 
    compared to the case of the $P_\mathrm{\,i}=\SI{4}{d}$ 
    indicates that the mass transfer in the wider binaries is less efficient and most of the mass is lost from the systems.
    
    The tracks of the mass-donors in binaries with an initial orbital period of $P_\mathrm{\,i}=\SI{400}{d}$ are depicted in panel c) of Fig.\,\ref{fig:HRD_binary}. As expected, the mass-transfer phase is initiated at a later evolutionary phase. The models have a rapid mass-transfer phase near or shortly after the end of their main-sequence. Due to their convective envelopes, the three more massive models
    are expected merge during the mass transfer.
    In contrast to the short period binary models which are shown in panel a) the intermediate and long period models can produce more luminous WR stars if they are able to avoid a common envelope phase.
    
    Lastly, panel f) shows the evolutionary tracks of mass-donors with an initial orbital period of $P_\mathrm{\,i}=\SI{40}{d}$, similar as panel b), but for an initial mass ratio of $q_\mathrm{\,i}=0.4$. By comparing the stellar evolution tracks shown in panel b) and f) it becomes evident that the mass-transfer phase starts approximately at the same evolutionary stage. Even though the
    tracks differ slightly during the mass transfer phase, the resulting helium star models are very similar. This shows that the mass ratio is of secondary importance, as long as
    a binary merger can be avoided, which is inevitable for too small initial mass ratios (see Appendix\,\ref{app:phase_diagramm}).
    
    In summary, from the comparison between panel a) and b) as well as d and e we have learned two things: First, the initial orbital period indeed has a direct impact on the final products of the mass-donors as well as mass-gainers and second, the accretion efficiency decreases with increasing initial orbital period. Panel b) and f) show that the initial mass ratio is expected to have only a small effect on the outcomes of a binary evolution.
      
\section{The observed WR population in the LMC}
\label{sec:WR_pop_LMC}
    Based on the LMC WR cataloge of \citet{bre1:99}, the WN population in the LMC as analyzed by \citet{hai1:14} contains 107 stars. Among those are 17 confirmed and 22 suspected  binary system.  \citet{hai1:14} analyzed the WN spectra with the Potsdam Wolf-Rayet (PoWR) atmospheric models and determined their stellar properties. The most important parameters for our analysis are listed in Table\,\ref{tab:WN_LMC}. 
    
    Even though the so called ``stellar'' temperatures of the WN stars are estimated by \citet{hai1:14} we do not include them in our analysis, for two reasons. First, the temperature estimation of WR stars is a non trivial process as they are covered by optically thick material thus the star cannot be observed directly. The differences between the observed stellar temperatures and the ones predicted by stellar evolutionary models is still debated in the literature \citep[e.g.][]{gra1:12}. Secondly we do not have temperature estimates for all WR stars, especially for the WC stars. 
    
    The luminosities are more reliable than the temperatures and are therefore used for comparison with our models. The uncertainties on luminosity are not discussed in \citet{hai1:14}. Therefore, a typical uncertainties of ${\Delta \log( L/\lsun) = 0.2}$ is adopted which is used in similar works like the studies of galactic WN stars  \citep{ham1:19}.
    
    Binary parameters, namely the orbital period and mass ratio, of known WN binaries are adopted from \citet{She1:19} and are listed in Table\,\ref{tab:WN_LMC} as well. The uncertainties of the orbital period are below 1\% and the typical uncertainties in the mass ratio is in the order of 10\%.
    
    In addition to the WN stars, the LMC contains 24 known WC type stars. Only 8 WC have been analyzed with stellar atmosphere codes by \citet{cro1:02}, \citet{ram1:18} and \citet{hil1:21}. \citet{bar1:01} determined the absolute visual magnitudes and when applying a typical bolometric correction of ${M_\mathrm{bol} = 4.5}$ \citep{smi1:94} it is possible to obtain the luminosity of all 24 WC stars. The calculated luminosities, as well as the luminosities obtained by stellar atmosphere codes which are used for comparison, are listed in Table\,\ref{tab:WC_LMC}. The uncertainties of the calculated luminosities are about ${\Delta \log( L/\lsun) = 0.3}$, which is similar to the typical uncertainties from stellar atmospheric models \citep{san1:12}. One can see that the discrepancy for the single stars is rather small and within the given uncertainties. Only the value of the WC binary BAT99 53 differs from the calculated luminosity by ${\Delta \log(L\,/\lsun) = 0.5}$.
    
    The orbital periods of known WC binaries are adopted from \citet{bar2:01} and are listed in Table\,\ref{tab:WC_LMC}. The uncertainties of the orbital period are below 1\%.
    
    Complementary to the WN and WC stars, the LMC hosts 3 known WO type stars, namely LH41-1042, LMC195-1 and Sanduleak 2. 
    Typically one assumes that these stars have enriched surface oxygen abundances, but in the recent work of \citet{aad1:22} it is shown that WO type stars only have a higher C content than WC type stars while their surface O abundance is comparable to those of WC type stars. It is still believed that these stars reflect the final stages of core He- and C-burning and hence cannot be not fully covered by our model grids \citet{tra1:15,san1:19,agu1:21}. Nonetheless, we want to give a brief comparison of the fundamental stellar parameters of these WO-type stars and the predictions of our model grid.
    
    LH41-1042 was studied by \citet{tra1:15} and has a luminosity of  ${\log\,L/\lsun=5.26^{+0.12}_{-0.14}}$ and a surface carbon abundance of ${X_\mathrm{C}=0.60}$. \citet{aad1:22} recently reanalyzed LMC195-1 and Sanduleak 2, finding that both have similar luminosities of ${\log\,L/\lsun=5.41}$ and surface carbon abundances of $X_\mathrm{C}=0.62$. Our models predict a maximum surface carbon abundance of $X_\mathrm{C}=0.53$. \citet{aad1:22} have shown in their work that also other stellar evolutionary models have problems in predicting this high carbon surface abundances. They find that it is linked to a too efficient ${^{12}\mathrm{C}+^4\mathrm{He}\rightarrow^{16}\mathrm{O}}$ nuclear reaction rate which already becomes important during core He-burning. We believe that this might also be the case in our models. However, a more in-depth study of this problem would be beyond the scope of our paper. Despite the disagreement of the carbon surface abundance, we find that our models with $X_\mathrm{C}\gtrsim0.50$ have luminosities in the range of ${\log\,L/\lsun=\numrange{5.3}{5.5}}$ which is close to those of the observed WO stars.
    
    Our models only cover a luminosity range until $\log\,L/\lsun \leq 6.35$, thus observed stars with higher luminosity are excluded from the sample. Furthermore, we exclude stars with spectral types Of as they are on the verge of becoming WR stars but in most of the cases they are quite cool and have a high amounts of H in their envelopes, which is not predicted by our stellar models when applying the optical depth criterion.

\section{The synthetic WR population}
\label{sec:results}

\subsection{Luminosity distributions}

    In this section, we evaluate the number of WR stars of different types, obtained from our models via the single and the binary channels, 
    and derive the luminosity distributions of the different WR types from our synthetic WR populations. We then compare these findings with the observed WR population in the LMC. In Sect.\,\ref{sec:orbtial_periods_and_mass_ratios}, we show the main properties of our WR binary models,
    and compare them with the --- more limited --- observed properties of WR binaries in the LMC.

    \begin{figure*}[t]
        \centering
        \begin{tikzpicture}
        \node [anchor=south west] at (0,0) {\includegraphics[trim= 2.cm 0cm 2cm 0cm ,clip ,width=1.0\textwidth]{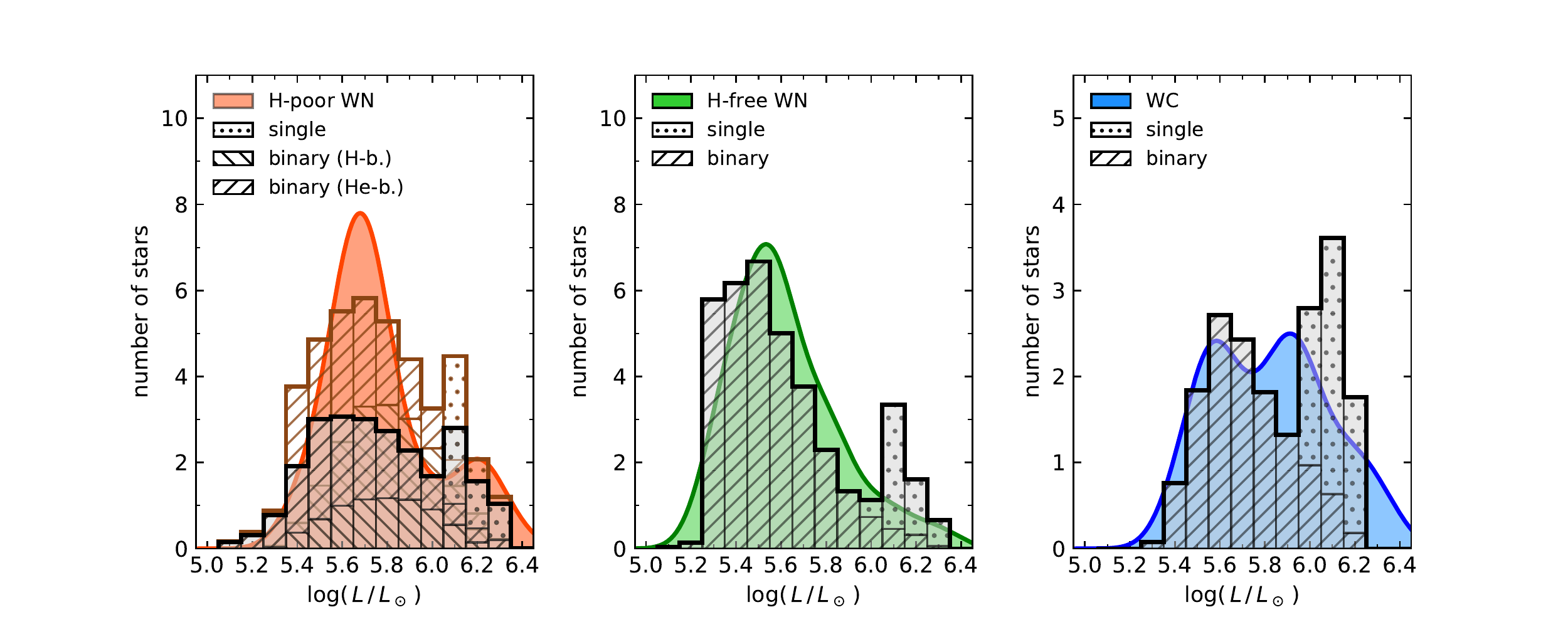}};
        % WC
        \draw[blue] (17.34,7.45) node {\tiny \textbf{19.0}};
        \draw[black!80!gray] (17.41,7.05) node {\tiny {6.5}};
        \draw[black!80!gray] (17.35,6.65) node {\tiny {12.8}};
        \draw[black!80!gray] (17.04,6.45) -- (17.74,6.45);
        \draw[black!80!gray] (17.34,6.25) node {\tiny \textbf{19.3}};
        % H-free WN
        \draw[green!40!black] (11.22,7.45) node {\tiny \textbf{38.0}};
        \draw[black!80!gray] (11.29,7.05) node {\tiny {5.7}};
        \draw[black!80!gray] (11.23,6.65) node {\tiny {32.3}};
        \draw[black!80!gray] (10.88,6.45) -- (11.58,6.45);
        \draw[black!80!gray] (11.22,6.25) node {\tiny \textbf{38.0}};
        % H-rich WN
        \draw[orange!40!red] (5.75-1.2,7.45) node {\tiny \textbf{37.0}};
        \draw[black!80!gray] (5.82-1.2,7.05) node {\tiny {4.2}};
        \draw[black!80!gray] (5.82-1.2,6.65) node {\tiny {6.0}};
        \draw[black!80!gray] (5.76-1.2,6.25) node {\tiny {14.1}};
        \draw[black!80!gray] (5.4-1.2,6.05) -- (6.1-1.2,6.05);
        \draw[black!80!gray] (5.75-1.2,5.85) node {\tiny \textbf{24.3}};
        \draw[orange!40!red] (5.35,7.45) node {\tiny \textbf{37.0}};
        \draw[black!40!red] (5.42,7.05) node {\tiny {4.4}};
        \draw[black!40!red] (5.36,6.65) node {\tiny {18.4}};
        \draw[black!40!red] (5.36,6.25) node {\tiny {19.1}};
        \draw[black!80!gray] (5.0,6.05) -- (5.7,6.05);
        \draw[black!40!red] (5.35,5.85) node {\tiny \textbf{41.9}};
        \end{tikzpicture}
        \caption{Luminosity distribution of observed H-poor WN (orange), H-free WN (green), and WC (blue) star populations (colored shading)
        with our synthetic WR population (black histogram, ${\tau\geq1.5}$). The synthetic WR population distinguishes WR models obtained from single star (dotted~bins) and  binary models (dashed~bins). The observed H-poor WN population excludes stars with spectral types Of and Of/WN. 
        In the upper right corner of each plot, we give the number of observed stars, the numbers of predicted  single stars and binary produced WR stars 
        of the respective WR subtype. For the synthetic H-poor WN populations an additional histogram (brown) for lower optical depths (${\tau\geq0.5}$) is shown.}
        \label{fig:WR_both_0.5}
    \end{figure*}
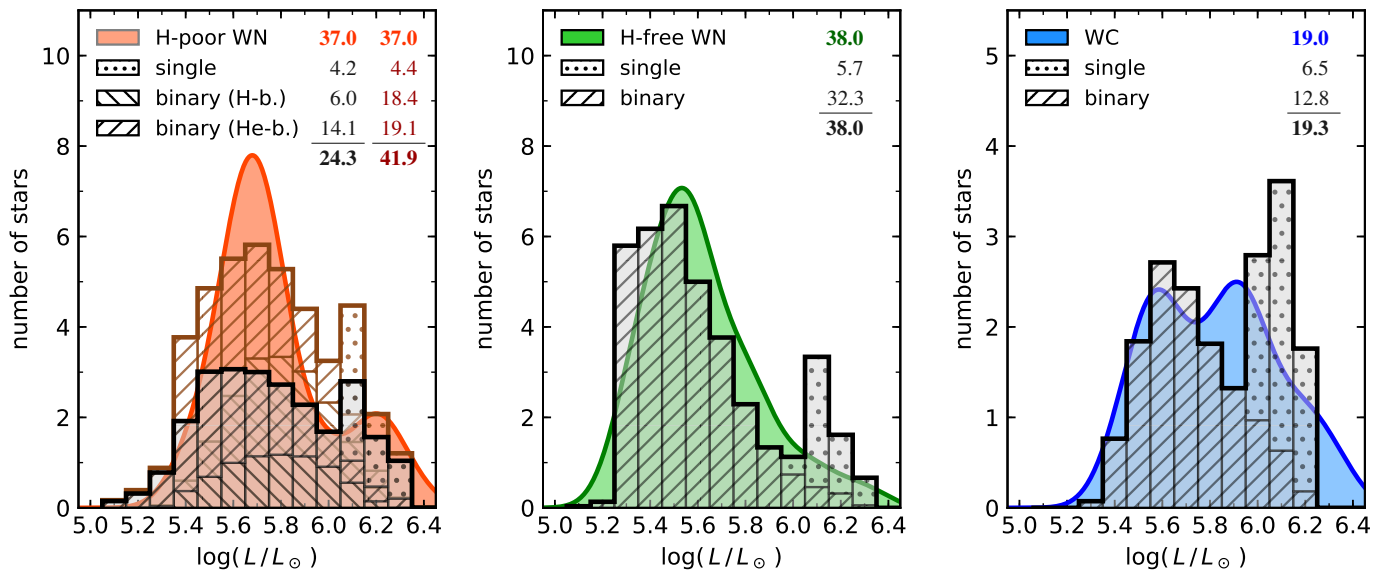
    
    As discussed above, we restrict ourselves to compare the luminosity distributions of the synthetic and observed WR population in the LMC,
    temperatures and radii are not always available and are subject to larger uncertainties. In the work of \citet{hai1:14} the WN population of the LMC was studied spectroscopically by using a model atmosphere grid with surface hydrogen abundances of $X_\mathrm{H} = 0.0$, $0.2$ and $0.4$. 
    Here, we follow their differentiation and distinguish ``H-free WN stars'' as having  $X_\mathrm{H} < 0.05$ (undetectable) and ``H-poor WN stars'' with $X_\mathrm{H}\lesssim0.4$, and  ``H-rich WN stars'' with $X_\mathrm{H}>0.4$.  
    Notably, the observed WN population in the LMC shows a strong drop in
    the number of objects with hydrogen abundances above 0.4 (see below). 
     
    According to  \citet{hai1:14}, four genuine WN stars show
    hydrogen abundances above 0.4, with a hydrogen mass fraction of either 0.6 or 0.7. 
    Two of them have luminosities of $\llso=6.7$ and $\llso=6.8$, well above the luminosity limit of our model grid ($\llso=6.35$) and thus not considered here. The other two
    have a luminosity close to this limit (BAT99-49 with $\llso=6.34$, and BAT99-111 with $\llso=6.25$). BAT99-49 is member of an eccentric
    ($e=0.35$) binary system \citep{foe1:03}(Foellmi+2003, MNRAS 338, 1025), which excludes a previous Roche-lobe overflow phase. And BAT99-111 is a bright X-ray source and as such likely
    a colliding wind binary.
    In agreement with \citet{hai1:14} and others, we conclude that these four stars are not stripped envelope stars, but rather 
    core H-burning stars with extreme mass and luminosity such that their stellar wind becomes optically thick. 
    
    Of the stars with luminosities below $\llso=6.35$ in the list of \citet{hai1:14}, 37 have a WN spectral type and a H-abundance in the range 
    0.1 to 0.4, 38 have a WN spectral type and no hydrogen detected, and 19 stars have a WC spectral type.
    The resulting luminosity distributions of the observed and predicted WR population are shown in Fig.\,\ref{fig:WR_both_0.5}. The predicted WR population contains 24.3 H-poor WN star models, 38 H-free WN star models and 19.3 WC star models, where the matching number of H-free WN stars is due to our applied normalization.
    
    The synthetic H-free WN and the (also H-free) WC star populations show an overall agreement with the observed luminosity distribution. All of these models are predicted to be core helium burning which is in agreement with the overall picture of the evolution of WR stars. Our synthetic populations predict that the majority of the less luminous stars ($\log\,L/\lsun \lesssim6.0$) are expected to be in binary systems and that the more luminous WR stars originate from stars evolved in isolation. 
    
    The synthetic luminosity distribution of the H-poor WN stars with an optical depth above $\tau \geq 1.5$ contains only 24.3 stars and clearly underestimates the observed population. Our synthetic population shows again, that the majority of these stars is expected to be in binary systems and that the single star models dominate at luminosities above $\log\,L/\lsun \gtrsim6.0$. The observed distribution of H-poor WN stars also shows a double peaked distribution which might be linked to the binary nature of the systems. However, the observed distribution contains many more stars at luminosities around $\log\,L/\lsun \sim5.7$ which are not reproduced by the synthetic population. 
    
    The threshold of $\tau\geq1.5$ was determined by using theoretical prescription on the stellar properties and mass-loss rates of H-free WN stars. However, the values used as input in Eq.\,\ref{eq:optical_depth} may differ for H-poor WN stars, especially the core hydrogen burning ones. Therefore, the wind optical depths for such stars may be different (i.e. lower) as well as their threshold value. We explore this possibility by modifying the optical depth threshold for the H-poor WN distribution. This is roughly equivalent to performing a new calibration for the optical depth threshold in this population, which is otherwise impossible to perform due to the uncertainties in their luminosity distribution. The best agreement between the number of H-poor WNs and their shape can be found when using $\tau\geq0.5$. The resulting synthetic population predicts that almost 50\% of the H-poor WN stars are currently core hydrogen burning.
    
    In all predicted luminosity functions, the single stars produce a peak above
    $10^6\,$L$_{\odot}$. Except perhaps for the WN stars with a bit of hydrogen, this peak
    does not find a correspondence in the observed luminosity distributions. 
    An initial binary fraction of the most massive stars above 50\% could remedy this situation (cf., Sect.\,\ref{sec:ss}). 
    
    A further discrepancy between the observed and predicted distributions is the following. Our synthetic WR population predicts that about 80\% of all WR stars are in binary system, while only $\sim40\%$ of the WR stars in the LMC are known to be in a binary system \citep{hai1:14}. According to \citet{hai1:14}, it is possible that a significant fraction of the apparently single WR stars are in fact in a binary system with a yet undetected companion. We return to this point in Sect.\ref{sec:wr-bin}.

\subsection{Surface hydrogen abundances}
\label{sec:H}    
    
\begin{figure}[t]
        \centering
        \resizebox{\hsize}{!}{\includegraphics{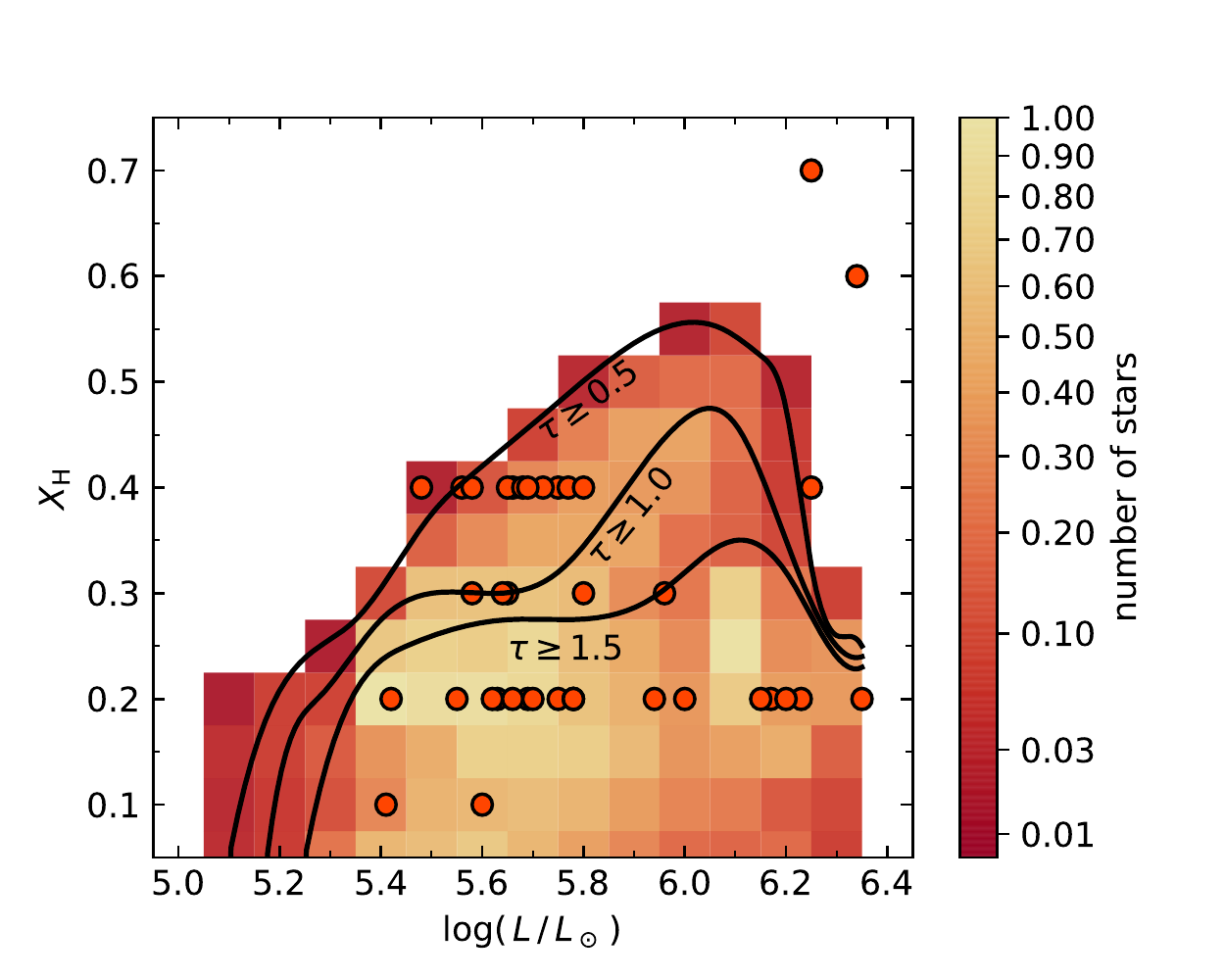}}
        \caption{Surface hydrogen abundances as a function of luminosity of the observed H-poor WN stars (orange dots), and of our synthetic population (histogram). Black lines indicate borderlines of the theoretical distributions obtained for different optical depth thresholds ($\tau\geq0.5$, $1.0$, and $1.5$).}
        \label{fig:H}
    \end{figure}
    
Fig.\,\ref{fig:H} shows the predicted distribution of surface hydrogen abundances and luminosities of our WN models with hydrogen. We find that the most likely hydrogen abundances are in the range
$X_{\rm H}=\numrange{0.15}{0.25}$ throughout most of the considered luminosity range. 
Furthermore, our models predict the possibility of higher hydrogen abundances
for more luminous stars. Focussing on the distribution obtained for a limiting optical depth of $\tau = 0.5$, we predict luminous WN stars with hydrogen abundances
of up tp  $X_{\rm H}=0.6$, of which, according to Fig.\,\ref{fig:WR_both_0.5}  the majority still undergoes core hydrogen burning. We note that for luminosities below $10^6\,{\rm L}_\odot$, where our single star models do not contribute, 
H-rich WR stars may be produced during the slow Case\,A mass transfer phase,
as detailed in \citet{sen1:22}.

As for the lumnosity distributions, when comparing to the observed distribution, the agreement is fair when $\tau = 0.5$ is used to defined the WR star models.
Except for two very luminous ones, all observed stars fall within the predicted
range. From the trend for the borderline of the predicted distributions as function of
threshold optical depth, one can anticipate that these two objects would also be recovered if a smaller optical depth threshold would be adopted (cf., Sect.\,\ref{sec:optical_depth_of_WR_winds}).

The area where we expect the largest number of objects in Fig.\,\ref{fig:H} is well populated with observed WR stars. However, we see perhaps more stars at $X_{\rm H}=0.4$ than expected, and we predict more stars with low hydrogen abundances $X_{\rm H}\simeq 0.1$ than observed. On the other hand, we note that the observed hydrogen abundances are derived by \cite{hai1:14} based on three grids of atmosphere models, using $X_{\rm H} = 0$, $0.2$, and $0.4$, which are the values preferentially found in the observed stars
($X_{\rm H} = 0$ is not shown in Fig.\,\ref{fig:H}). We conclude that the 
hydrogen abundances in the models of our synthetic WR population agrees rather well
with the observations.

\subsection{WR binary properties}
\label{sec:orbtial_periods_and_mass_ratios}

    In the previous sections we focused on the stellar properties of the WR stars in the LMC. In this Section, we investigate their binary properties of our WR population.
    We derive the predicted distributions of orbital periods and mass ratios for
    WR binaries with different WR sub-types, and compare the with the available
    observations. In contrast to luminosity and hydrogen abundance, where the
    available information is rather complete, orbital periods are known for only
    18 LMC WR binaries, and mass ratios only for 8 of them.

    \begin{figure*}[thbp]
        \centering
        \begin{tikzpicture}
            \node [anchor=south west] at (0,0)
            {\includegraphics[trim= 1.5cm 1.5cm 2.0cm 2.44cm ,clip ,width=0.73\textwidth]{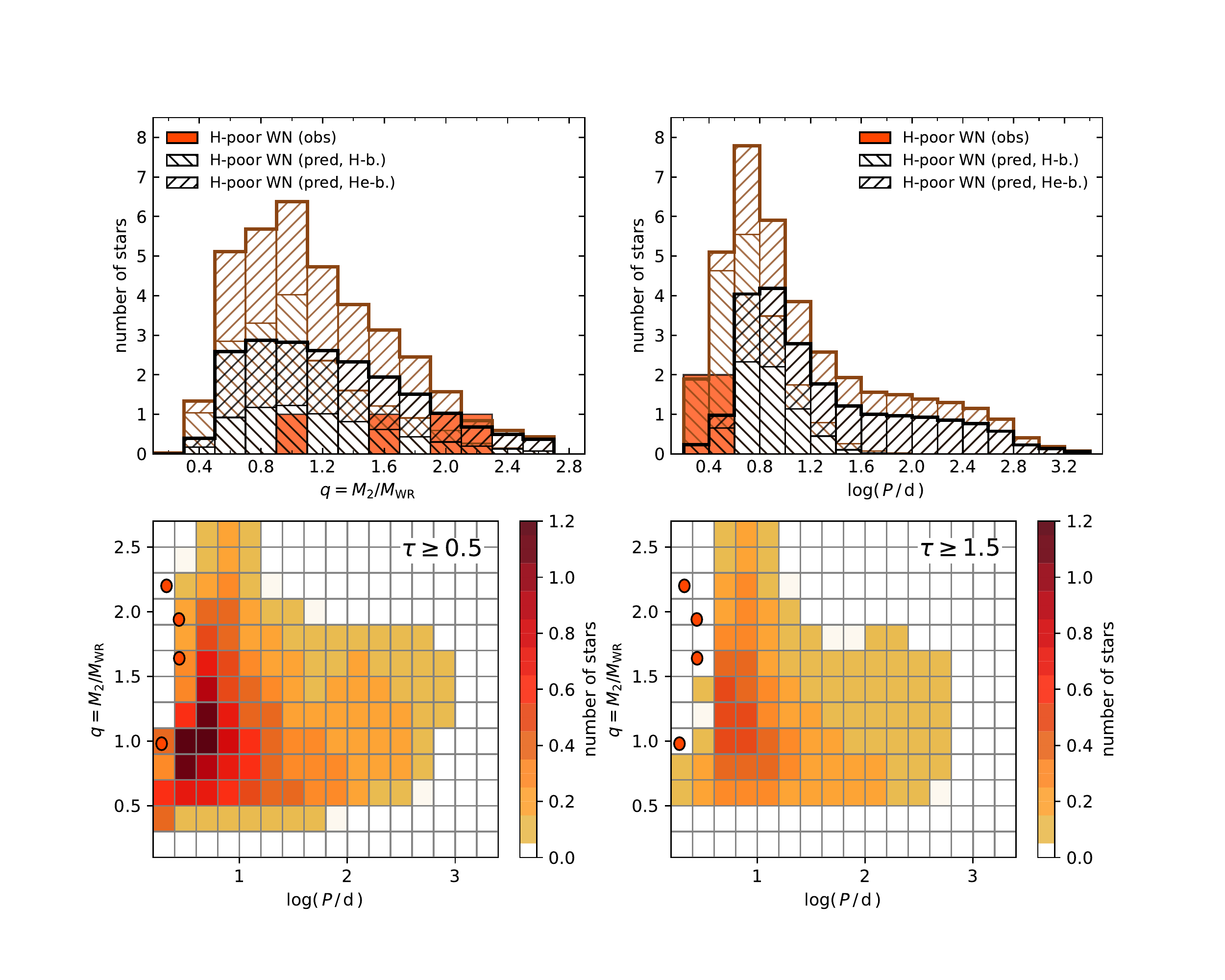}};
            \draw[orange!40!red] (5.76+0.5,10.11-0.2) node {\tiny 4.0};
            \draw[black!80!gray] (5.76+0.5-0.6,9.83-0.2) node {\tiny 6.0};
            \draw[black!80!gray] (5.71+0.5-0.6,9.57-0.2) node {\tiny 14.1};
            \draw[black!40!red] (5.71+0.5,9.83-0.2) node {\tiny 18.4};
            \draw[black!40!red] (5.71+0.5,9.57-0.2) node {\tiny 19.1};
            \draw[orange!40!red] (7.90+0.6+1.3,10.11-0.2) node {\tiny 4.0};
            \draw[black!80!gray] (7.90+1.3,9.83-0.2) node {\tiny 6.0};
            \draw[black!80!gray] (7.85+1.3,9.57-0.2) node {\tiny 14.1};
            \draw[black!40!red] (7.85+0.6+1.3,9.83-0.2) node {\tiny 18.4};
            \draw[black!40!red] (7.85+0.6+1.3,9.57-0.2) node {\tiny 19.1};
        \end{tikzpicture}
        \caption{\textit{Upper left}: Predicted mass ratio distribution of the H-poor WN stars in our synthetic population with optical depths $\tau\geq0.5$ (brown dashed histogram) and $\tau\geq1.5$ (black dashed histogram), where we distinguish WR stars during core hydrogen and core helium burning. 
        Overlayed is the histogram of the observed mass ratios in LMC WN stars with hydrogen (orange). \textit{Upper right}: Same as in the upper left panel, but for the
        orbital periods. \textit{Lower left}: 2D histogram of mass ratios and orbital periods of our synthetic population with optical depths $\tau\geq0.5$, represented by colored squares, where the color indicates the expected number of objects (see color bar to the right). Individual observed binaries are shown as orange dots. \textit{Lower right}: Same as the lower left panel but now for optical depths $\tau\geq1.5$. The observational data is take from \citet{She1:19}.}
        \label{fig:H-rich-WN-PQ}
        
        \begin{tikzpicture}
            \node [anchor=south west] at (0,0)
            {\includegraphics[trim= 2.3cm 1.5cm 2.5cm 2.2cm ,clip ,width=0.7\textwidth]{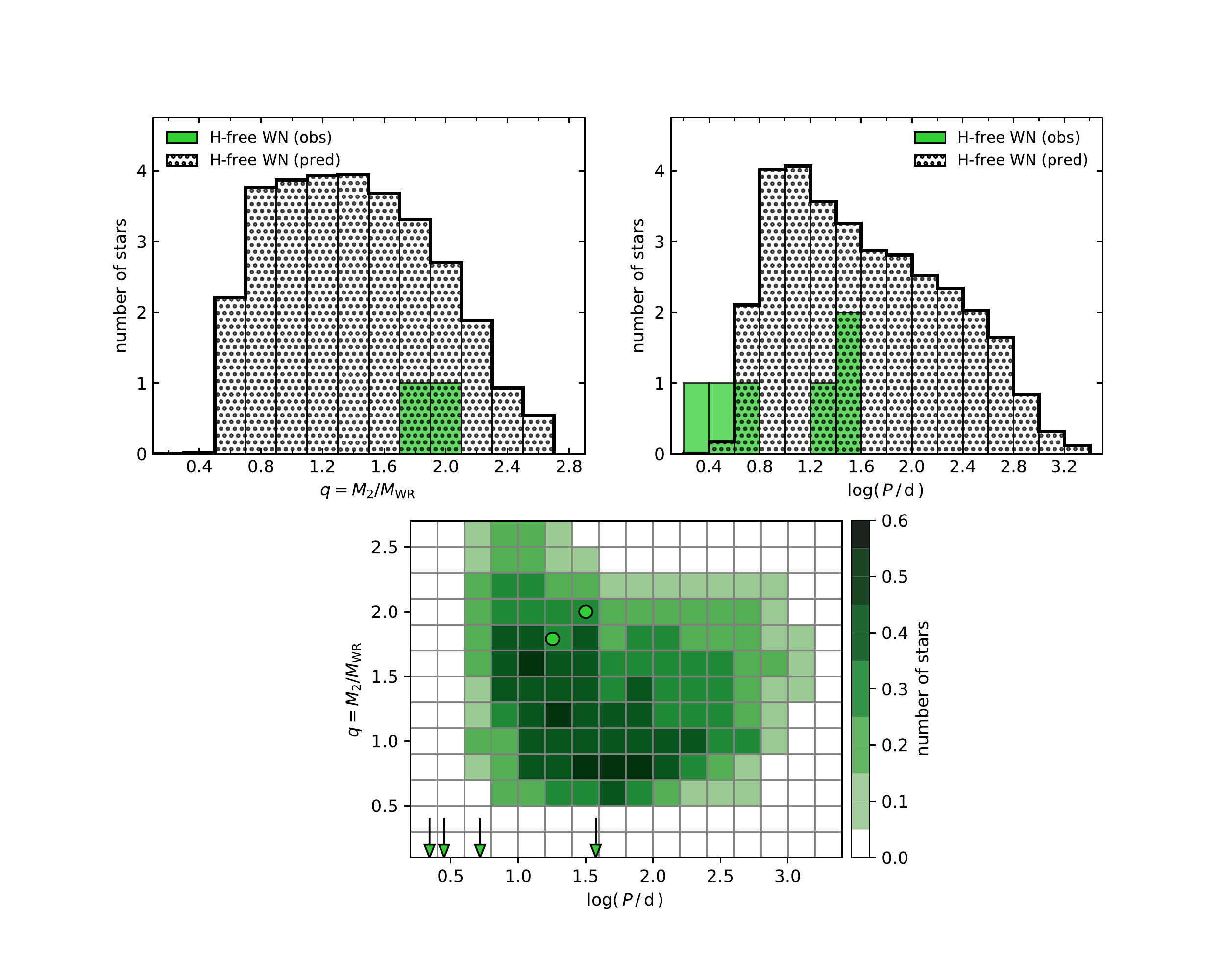}};
            \draw[green!40!black] (5.86,10.11) node {\tiny 2.0};
            \draw[black!80!gray] (5.8,9.83) node {\tiny 32.3};
            \draw[green!40!black] (7.91,10.11) node {\tiny 6.0};
            \draw[black!80!gray] (7.85,9.83) node {\tiny 32.3};
        \end{tikzpicture}
        \caption{Same as Fig.\,\ref{fig:H-rich-WN-PQ}, but for H-free WN stars. In the lower panel, in the 2D histogram, systems with unknown mass ratio but known orbital period are indicated by arrows. The observational data is take from \citet{She1:19}.}
        \label{fig:H-free-WN-PQ}
    \end{figure*}
    
    \begin{figure*}[t]
        \centering
        \begin{tikzpicture}
            \node [anchor=south west] at (0,0)
            {\includegraphics[trim= 2.3cm 1.5cm 2.5cm 2.2cm ,clip ,width=0.7\textwidth]{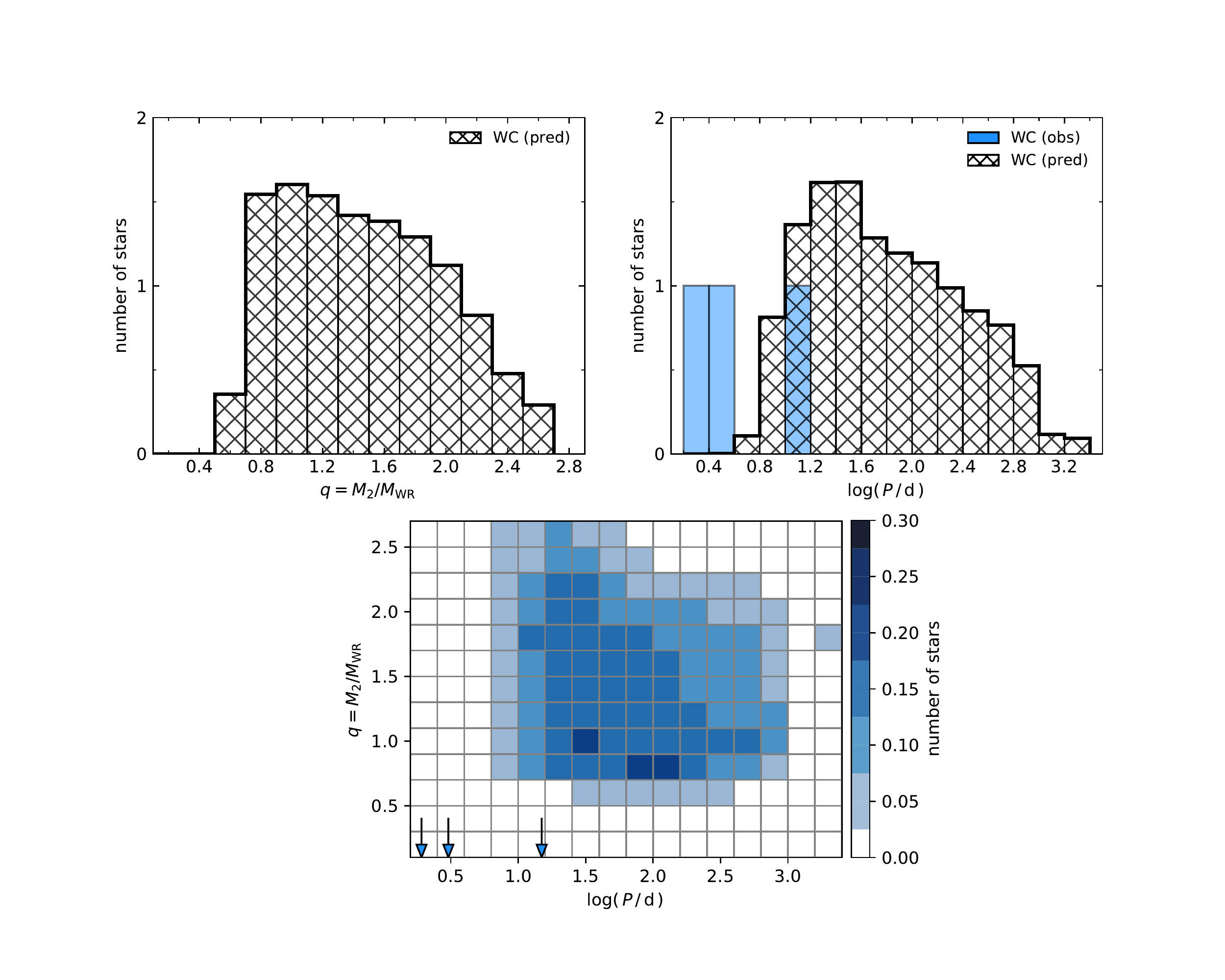}};
            \draw[black!80!gray] (1.12,10.11) node {\tiny 12.8};
            \draw[blue!60!gray] (7.91,10.11) node {\tiny 3.0};
            \draw[black!80!gray] (7.85,9.83) node {\tiny 12.8};
        \end{tikzpicture}
        \caption{Same as Fig.\,\ref{fig:H-rich-WN-PQ}, but for WC stars. The observational data is take from \citet{bar2:01}}
        \label{fig:WC-PQ}
    \end{figure*}
    
\subsubsection{H-poor WN stars}
\label{sec:orbtial_periods_and_mass_ratios_H-rich_WN}

    The orbital period distribution of the observed and predicted H-poor WN population in the LMC can be seen in the upper right panel of Fig.\,\ref{fig:H-rich-WN-PQ}. Most of the observed periods of the H-poor WN binaries are below $P\lesssim \SI{10}{d}$. This is linked to an observational bias towards smaller orbital periods as the companion in these systems has a stronger impact on the appearance of the spectrum, namely by a stronger Doppler-shift of the emission and absorption lines of each stellar component. Due to this observational bias, it is reasonable to assume that the sample is nearly complete for the shortest periods and nearly incomplete for long periods. 
    In the sample of observed periods there are two binary systems with orbital periods above $P\gtrsim \SI{100}{d}$ which are brighter than our most luminous models ($\log\,L/\lsun \geq 6.35$). Both binaries show X-ray emission, most probably originating from a colliding wind, which is an additional indicator that these stars have a binary companion. 
    
    By comparing the small sample of observed orbital periods to the predicted distribution of the H-poor WN binaries several similarities become apparent as well as some discrepancy. Our models with optical depths $\tau\geq1.5$ are under-predicting the number of binaries with orbital periods of about $P\approx \SI{2}{d}$, hinting towards a unconsidered formation channels. Possible formation channels are discussed in Sect.\,\ref{sec:disc}. However, our models with $\tau\geq0.5$ can reproduce the short period binaries and predict that they are still core hydrogen burning stars. Another discrepancy is that our models predict a large amount of systems with long period ($P\gtrsim\SI{100}{d}$) that are barely observed. This is most likely linked to the aforementioned observational bias. 
    
    The mass ratio distributions, found in the upper left panel of Fig.\,\ref{fig:H-rich-WN-PQ}, of the observed H-poor WN population is as small as the sample of the orbital periods. The sample is spread over a wide range of mass ratios and with respect to the small sample size is in good agreement with the predicted distribution, as there are no outliers as in the case of the orbital period distribution.
    
    In the lower panels of Fig.\,\ref{fig:H-rich-WN-PQ} 2D histograms of the combination of the period and mass ratio distributions with the two different optical depth criteria are shown. The predicted distributions show a wide variety of the mass ratio $q=\numrange{0.4}{2.4}$ at small orbital periods $P\lesssim \SI{1.0}{d}$. When going to larger orbital periods the mass ratio distribution gets shallower and closer to $q=1$, which is linked to the efficiency of mass transfer. It appears that the predicted distribution of mass ratios and orbital periods is in rough agreement with the observed distribution with respect to the small available sample size and their respective uncertainties.

\subsubsection{H-free WN stars}
\label{sec:orbtial_periods_and_mass_ratios_H-free_WN}

    Figure\,\ref{fig:H-free-WN-PQ} contains the observed and predicted period and mass ratio distributions of the H-free WN binaries. Similar to the orbital period distribution of the H-poor WN binaries, the predicted distribution under-estimates the amount of short period WR stars and over-estimates the binaries with long orbital periods. This can also be a consequence of observational biases. The under-prediction of the short period binaries hints at an unconsidered formation channel which seems to be independent of the WR subtype (see Sect.\,\ref{sec:scenarios_of_short_period_binaries}). 
    
    Compared to the predicted orbital period distribution of the H-poor WN binaries, the orbital period distribution of the H-free WN binaries is shifted towards longer orbital periods. This shift can be explained by the mass that is lost from the system during the WR phase, leading to a change in the orbital angular momentum and, therefore, to an longer orbital period.
    
    The observed mass ratio distribution is too small to get any useful information, but at least none of the observed mass ratios disagrees with the predicted mass ratio distribution. Compared to the predicted mass ratio distribution of the H-poor WN binaries the mass ratio distribution of the H-free WN binaries is shifted towards more extreme mass ratios $q>1$. This shift is also connected to the strong  mass-loss rates during the WR phase, which drastically alters the mass of the WR star on a timescale which is short compared to the main-sequence lifetime of the companion.
    
    Inspecting the 2D histogram of the predicted mass ratio and orbital period distribution of the H-free WN binaries in the lower panel of Fig.\,\ref{fig:H-free-WN-PQ} one can see that this distribution now spreads over a large range of mass ratios and orbital periods with a small tendency towards shorter orbital periods $P\approx \SI{20}{d}$. Comparing this with the small sample of observed H-free WN binaries (except of the binaries with the shortest periods) no discrepancies can be found between predictions and observations.

\subsubsection{WC stars}
\label{sec:orbtial_periods_and_mass_ratios_WC}

    The orbital period and mass ratio distributions of stars in the WC phase are shown in Fig.\,\ref{fig:WC-PQ}. The sample of observed WC binaries contains three short period binaries, of which two (those with the the shortest periods) cannot be explained by our models. However, there is a lack of intermediate orbital periods around $\sim \SI{10}{d}$ which is also found by \citet{dsi:20} for the galactic WC binaries.
    
    By comparing the predicted orbital period distribution with the ones of the H-free and H-poor WN binaries, a shift towards even longer orbits can be seen. As in the case of the H-free WN, we explain this shift by the strong mass-loss rates that lead to a widening of the orbit. Furthermore, it becomes evident that the predicted number of WC binaries is much smaller than the number of WN binaries. This is linked to the evolution of the stars themselves, as not every WN star will evolve into a WC star because the helium rich layers will not be removed for all WN stars and they spend less time as WCs than as WNs in this mass and metallicity range \citep[see][]{agu1:21}.
    
    Unfortunately the mass ratios of the WC binaries have not been measured yet. Therefore, we can only show the predicted distribution and compare it with the ones of the other WR subtypes. As in the case of the H-free WN binaries, the distribution of the mass ratios is again shifted towards more extreme mass ratios $q>1$ as the donor star looses several solar masses via its strong WR wind.
    
    The 2D histogram in the lower panel of Fig.\,\ref{fig:WC-PQ} shows that the expected distribution of mass ratios and orbital periods is rather uniform for orbital periods in the range of ${P=\SIrange{10}{1000}{d}}$ and for mass ratios in the range of ${q=\numrange{0.8}{2.0}}$. The distribution clearly lacks of short orbital periods $P\lesssim \SI{10}{d}$ and mass ratios below $q\lesssim 0.8$.
 
\section{Discussion}
\label{sec:disc}

     In the previous section we found that the observed WR population of the LMC with luminosities below $10^6 \lsun$ cannot be explained by our singe star models. Therefore, we compare our findings with the most commonly used single and binary evolution models to see whether they lead to similar conclusions or not. In this Section we  briefly summarize the predictions of the WR population of the different stellar evolutionary models. A more detailed comparison can be found in Appendix\,\ref{sec:comparision_with_previous_works}. Moreover, we discuss possible formation channels which are not included in our analysis that can explain the discrepancies between the predicted and observed WR population.

\subsection{Comparison with previous single and binary star evolution models}    
\label{sec:comp}

    We first focus on the similarities and differences of non-rotating stellar evolution models. Commonly used are the so called Geneva models \citep[e.g.][]{eks1:12,geo1:12,geo1:13,egg1:21}. These models are available for Galactic (MW), LMC and SMC ($Z=0.014$, $0.006$ and $0.002$) metallicity which makes it possible to compare these models with our models and to use them as a proxy for  other model grids that do not cover LMC metallicity. Their lower WR luminosity limits are $\log(L\,/\lsun) \approx 5.4$, $\log(L\,/\lsun) \approx 5.75$ and $\log(L\,/\lsun) \gtrsim 6.4$  (see Fig.\,\ref{fig:WR_single_literature}) for MW, LMC and SMC models, respectively. Their $40\,\msun$ LMC model spends a very short time in a H-free WR phase, which is not predicted by our grid. Their $60\,\msun$ LMC model predicts similar WR phases, however, their model is less luminous. The $85\,\msun$ model is predicted to evolve similarly like our $80\,\msun$ model. Our models seem to fit the general trend, although the WR luminosity limit of the Geneva models for the LMC is lower than our prediction which is linked to the more efficient red supergiant (RSG, $T_\mathrm{eff}\leq\SI{4800}{K}$) winds in their models. 
    
    The MIST \citep{cho1:16} stellar evolutionary models are also calculated with MESA but use less efficient overshooting and semiconvective mixing. It is a known issue, that the choice of the mixing parameters can have a deep effect on the resulting stellar populations, particularly after the main-sequence \citep[e.g][]{sch1:19,gil1:21,kle1:21}. Their LMC models (${[\mathrm{Fe/H}]=-0.5}$) only produce WR stars with luminosities above $\log(L\,/\lsun) \gtrsim6.4$ (see Fig.\,\ref{fig:WR_single_literature}) as their models spend most of their time in the RSG phase.
    
    Finally, the FRANEC models \citep{lim1:18} are only calculated for MW and SMC  metallicity (${[\mathrm{Fe/H}]=0}$ and ${[\mathrm{Fe/H}]=-1}$ ). Their lower WR luminosity limits are ${\log(L\,/\lsun) \approx5.4}$ and ${\log(L\,/\lsun)\approx 6.2}$ for MW and SMC, respectively. As for the Geneva models, our models seem to fit the general trend. Even though we are unable to directly compare the models, their predicted H-poor, H-free WN and WC phases appear to agree better with our models (cf., Fig.\,\ref{fig:WR_single_literature}).
    
    Comparing the different slow and the fast rotating models of the different codes, presented in Fig.\,\ref{fig:WR_single_literature}, one can see two major effects: First, rotation may extend the time spent in the WR phase significantly as seen from the Geneva models, but this needs not to be the case, as the FRANEC models show. Second, rotation leads to more efficient mixing in the stars, which then have a shortened H-free WN phase.
    
    From the MIST models one can conclude that the choice of mixing and the stronger RSG mass-loss rates can affect the appearance of the WR population drastically. Even though our models spend less than a quarter of their core helium burning time in the cold supergiant phase ($T_\mathrm{eff}\leq\SI{12500}{K}$) and are in relative good agreement with the other single star models, we are still overestimating the RSG population. \citet{gil1:21} studied this phenomenon in more detail using MESA models, and show that more efficient mixing can shorten the RSG phase, and avoid an overprediction of cool supergiants. 
    
    We conclude that single star models with enhanced mixing, and/or with enhanced
    mass loss in the cool supergiant regime, may predict WR stars down to luminosities
    of about $\log(L\,/\lsun)\sim 5.6$, at LMC metallicity, but have difficulties to
    reproduce the observed fainter population of WR stars in the LMC \citep[see also][]{She1:20}.

    For the binary models we compare our models with those from \citet{eld1:17}. Their binary models are calculated for LMC metallicity ($Z=0.004$) and are based on detailed stellar evolution models. We note, that typically a metallicity of $Z=0.006$ is used for the LMC, but as already argued in Sect.\,\ref{sec:binary_model_grids}, observations support a lower iron abundance which is taken into account in our model grid. Thus for a comparison to our stellar evolutionary models we prefer to use the BPASS models calculated at the lower metallicity of $Z=0.004$.
    We find, that most of their models spend similar times in the individual WR phases and have similar luminosities compared to our models (c.f. Fig.~\ref{fig:WR_binary_literature}). 
    We find that some of their models that undergo a Case A mass-transfer are predicted to produce WR stars that are by far more luminous than our models predict. This is because their models enter a common envelope phase, during which they merge and, therefore, leading to the formation of more massive and luminous WR stars.
    Besides that we find that their models spend to some extend different amount of time in the individual WR phases.
    The binary models of \citet{eld1:17} differ to our models by four assumptions: First, in BPASS the WR mass-loss recipe of \citet{nug1:00} is used only when $X_\mathrm{H}<0.4$ and $T_\mathrm{eff}>\SI{10}{kK}$. Second, RLOF is modeled using a simplified formula to ensure numerical stability \citep[][their equations 2 and 3]{eld1:17}. Third, the secondary is approximated by the single star stellar evolution equations of \citet{hur1:00} and only retrospectively modeled in detail.
    Fourth, their models do not include rotation.
    We find that the different physical assumptions can well explain the differences in the time a stellar model spends in the individual WR phases.
    
\subsection{Short period WR binaries}
\label{sec:scenarios_of_short_period_binaries}

    In Sect.\,\ref{sec:orbtial_periods_and_mass_ratios} it was shown that the observed shortest-period WR binaries, with orbital periods below $P_{\rm orb} < 3\,$d are not explained by our binary models. Below, we discuss three possible implications of this finding.
    
    \textbf{Common envelope evolution.} The initially widest interacting binaries in our model grid are expected to undergo a common envelope evolution, which we do not model. In this scenario, the binary goes through an unstable mass-transfer phase, during which the primary's envelope engulfs the secondary star. This is followed by an spiral-in process during which the orbital energy of the secondary is deposited in the envelope 
    of the primary. This might lead to the ejection of the envelope and the emergence of a compact binary, consisting of the core of the primary (now a WR star) and the mostly unaffected secondary \citep{iva1:13}. While we can not exclude that the observed shortest-period WR binaries evolved through this path, simplified one-dimensional estimates which compare the binding energy of the primary's envelope with the released orbital energy indicate that for massive stars, a successful common envelope ejection appears difficult as long as the companion star is a main sequence star \citep[e.g.][]{kru1:16}. \citet{pau1:20} investigated the binary models presented in this paper, and found that none of them with an initial primary mass above $M_\mathrm{1,\,i}\gtrsim28\msun$ is expected to eject the envelope. However, it is possible that there are unconsidered phenomena that contribute to a successful ejection, for example pulsations during the supergiant phase or LBV eruptions \citep[e.g.][]{lan1:20}, such that we can not exclude this formation channel.
            
    \textbf{Chemically homogeneous evolution.} Assuming that close binaries at the ZAMS are tidally locked leads to rapidly rotating stars in the binaries with the shortest orbital periods. This scenario has been suggested by \citet{dem1:09} to lead to chemically homogeneously evolving binary components, and investigated with MESA models using a similar input physics as our models for a large initial parameter space by \citet{mar1:16} and \citet{has1:20}. 
    Their conclusion was that this scenario applies only to very metal poor binaries,
    and only to the most massive systems \citep[e.g., possible to the extremely massive SMC binary HD\,5980;][]{koe2:14}. 
    Our binary models with initial orbital periods below $\lesssim\SI{2}{d}$ all result in a merger during their contact phase, and binary systems with larger initial orbital period cannot explain the observed WR binaries, which renders this scenario unlikely.
 
    \textbf{Angular momentum loss by winds.} When wind particles leave a massive star and the binary system, they may still interact gravitationally with both stellar components and thereby enhance or reduce the orbital angular momentum of the binary.  This mechanism, which is not included in our models, may be of relevance when the wind speed is comparable to the orbital velocities of the two stars, which may be particularly applicable to the tightest binaries \citep{bro1:93,dem1:16}. \citet{leo1:20} show that in binaries with identical components a strong orbital shrinkage may occur. Whether this scenario can explain the shortest period WR binaries remains unclear. However, 
    since the terminal wind velocities of WR stars are typically larger than $\varv_\infty \gtrsim \SI{2000}{km\,s^{-1}}$, while the orbital velocities in our short-period models rarely exceed $\SI{400}{km\,s^{-1}}$ we assume this effect to be small. 

    In the end, we can not rule out any of the three channels, and conclude that
    all three deserve deeper study. An understanding of the shortest period WR binaries my be of significant relevance for a reliable prediction of the population of black hole mergers in the universe.

\subsection{Disregarded binary products}
\label{sec:unconsidered_binary_outcomes}

    Due to the limitations of our method, we need to disregard various binary evolution products in our synthetic WR population. Here, we briefly discuss the various neglected channels, and give an estimate of their relevance.
    
    As shown by our example plots giving an overview of the fates of individual binary models for a given primary mass (cf., Appendix\,\ref{app:phase_diagramm}), we expect our binaries to merge in various corners of the parameter space spanned by the initial orbital period and the initial mass ratio. In particular, very short and very large initial periods, as well as extreme mass ratios are found to lead to merger conditions. Overall, we see from these plots that perhaps one quarter of all computed binaries are thought to have this fate, which, for flat initial distributions in log period and mass ratio might represent the corresponding fraction in a binary population.
    
    As the merger product is a single star, the mergers will enhance the single star population. Thus, to first approximation, neglecting the merger products may be simply compensated by adopting a smaller initial binary fraction --- which is still uncertain anyways. In case one of the merging components already undergoes core helium burning,
    the merger product may strongly differ from a single star. However, the fraction of
    these cases in a given population is estimated be small, due to the small lifetime
    of these merger products \citep{jus1:14}.

    In our binary evolutionary models the secondary star is evolved as a single star after the primary has depleted helium in its core. This might be in general a good approach  when assuming that the BH formation is associated with a birth kick as that of neutron stars \citep{jan1:13} and be representative for what happens in a sizable fraction of the systems. However, this might not be true for all systems in which the primary forms a BH. Some systems might not get disrupted, even though the orbit could widen significantly. For those systems, we neglect in our final population the potential WR stars that would be formed when the secondary interacts with compact companion via RLOF.

    The number of WR stars formed in this way will be smaller than the number of WR
    stars formed from the first mass transfer, because binaries can break up when
    the compact object forms, or merge upon mass-transfer from the initial secondary
    to the compact object. In fact, in case the formed compact object is a neutron
    star, both processes are very likely, such that WR stars are not expected.
    However, when the compact object is a BH --- which is more likely in the mass
    range we consider here --- the first process depends on the unknown magnitude of
    BH formation kicks \citep{jan1:13,mir1:17,cha1:18}.
    And the second process depends on the weakly constrained rate at
    which the BH can expel the transferred matter from the binary \citep{kin1:99,kin1:00}. It is therefore difficult to quantify the number of
    WR stars formed from the inverse mass transfer process. Currently, with Cyg\,X-3, there is only one candidate WR+BH binary in the Milky Way \citep{van1:92}.

    While we are not able to quantify the fraction of WR stars originating from 
    secondaries via stable RLOF, the above arguments substantiate that the WR stars originating from primaries are dominating in any given population. We note that, while this implies that the comparison of our model results with the observed LMC WR population is still meaningful, it does not preclude that a fraction of the LMC WR stars hosts
    a compact object, as speculated by \citet{van1:17}.

\subsection{WR mass loss rates}
\label{sec:reliability}

    Stellar wind mass-loss of massive stars is still poorly understood, especially those of the later evolutionary stages. In the case of WR stars, empirical mass loss rates have been proposed, but the true mass-loss rates strongly depend on the assumed physics, like wind clumping. In our models we use a clumping factor of $D=4$ as suggested by \citet{yoo1:17} in order to be able to reproduce the luminosity range of the observed WC stars in the LMC. In the literature, for a long time a clumping factor of $D=10$ was used for spectral analyses of WR winds, but this value is not accurately constrained \citep{nug1:98,hai1:14}. However, increasing the clumping factor from $D=4$ to $D=10$ would imply that the mass-loss rate drops by a factor of $\approx1.58$. This would imply that many of the WC stars predicted by our synthetic WR population could not have been formed. If one would use an unclumped wind with $D=1$, the mass-loss rates would increase by a factor of 2 compared to our case, which would lead to the formation of too faint WC stars. Evidently, in particular number and properties of WC stars depends sensitively on the adopted WR mass loss recipe. 
    We note that while using the one proposed by \citet{yoo1:17} allows us to recover
    the lower luminosity limit of LMC WC stars, their absolute number and luminosity function are not prescribed by selecting this recipe. While a better understanding of the winds from WR stars is highly desirable, we note that the recipe used here
    is in good agreement with the recent findings of \citet{san1:20}.

\subsection{Single star contribution}
\label{sec:ss}
    While our predicted WR luminosity functions represent the corresponding observed luminosities 
    rather well, the high-luminosity peaks produced by the single star contributions appear to be too large (see Fig.\,\ref{fig:WR_both_0.5}). The largest discrepancies occur for the H-free WN stars and for the WC stars, for which the single star models provide large contributions above $\gtrsim 10^6\,{\rm L}_\odot$ which are not reflected in the observed WR star population. 
    A simple solution is to reduce the single star fraction would be to increase the binary fraction.
    As Fig.\,\ref{fig:WR_both_0.5} shows, reducing the single star contribution by a factor of two,
    i.e., assuming a binary contribution of 75\%, would yield a much better agreement. 
    
    A larger binary fraction would also help to alleviate another problem of our single star models.
    In our synthetic population, we predict 3 red supergiants ($T_\mathrm{eff}\leq\SI{4800}{K}$) and 5 yellow supergiants ($T_\mathrm{eff}\lesssim\SI{7500}{K}$) with luminosities above $\log(L\,/\lsun) = 5.6$. \citet{dav1:18} and \citet{gil1:21} find that the number of observed stars of this types is smaller by roughly 50\%. While  \citet{gil1:21} conclude that stronger mixing in stars can reduce the number of luminous cool supergiants, they do not investigate which effect this has on the WR luminosity functions. From their figure\,6, we expect that they would indeed overproduce very luminous WR stars ($\log(L\,/\lsun) \geq 6$). The same is true for models with strong rotationally induced mixing (cf., Fig.\,\ref{fig:WR_single_literature}). 
    
    Our results indicate that for the most luminous stars, with initial masses above $\sim 40\,{\rm M}_{\odot}$, a binary fraction larger than >50\%  might resolve the two issues discussed above. An increased initial binary fraction would be in agreement with the results from \citet{moe1:17} who find for stars in our Galaxy with initial masses above $>16\,\msun$ an intrinsic close binary ($\log(P/\mathrm{d})\leq3.7$ and $q>0.1$) frequency  of $f_\mathrm{bin}=1.0\pm0.2$. On the other hand, we can not exclude the possibility that in the large parameter space of mass  loss and mixing, there may be single star models which also overcome the mentioned problems. In any case, our results show that the contribution of binary stars to our understanding of the populations of the various types of massive stars is essential, and that any calibration of the uncertain parameters in massive star evolution based on comparing only single star models with observations may lead into the wrong direction. 

\subsection{The WR binary fraction}
\label{sec:wr-bin}
    Several efforts to determine the fraction of WR stars which reside in binary systems
    are documented in the literature. \citet{foe1:03} and \citet{sch2:08} investigated the WN stars in the LMC and found a binary fraction of the order of 30\%. More recently, \citet{hai1:14} and \citet{She1:19} derive a WN binary fraction between 16 and 36\%. In the following, we refrain from discussing the binary fraction of WN stars with hydrogen, because in the corresponding observational samples we find many objects which may not be genuine WN stars, but rather fall into the Of category.  From the 38 H-free LMC WR stars listed in Hainich et al. we find 5 certain and 6 possible binaries, leading to a binary fraction in the range $\SIrange{13}{29}{\percent}$. \citet{bar1:01} investigated the WC stars in the LMC
    and found binary induced radial velocity variations in 13\% of their sample, but signatures of
    an O star in the spectrum for 65\%. 
     
    The numbers appear difficult to reconcile with to our predicted WR binary fractions. In our models, assuming a single star fraction of 50\%, we find an overall WR binary fraction of $\sim 80$\%, while
    for WR stars with luminosities below $\lesssim 10^6\,$L$_{\odot}$ we expect nearly all WR stars
    to reside in binary systems. However, the quoted numbers are detection fractions, and bias corrections are difficult. When considering the measured orbital periods of WR binaries (cf., Figs.\,\ref{fig:H-free-WN-PQ} and\,\ref{fig:WC-PQ}), we find an average of 13\,d for the H-free WN stars, and 6.6\,d for the WC stars. In contrast, average predicted orbital periods are above 50\,d
    for both groups. Since radial velocity variations are larger for short-period binaries, conceivably, many long period WR binaries may have escaped detection so far. Similarly, though based on even less data (Figs.\,\ref{fig:H-free-WN-PQ}), the two mass ratios measured in H-free WN binaries indicate O\,star companions with nearly twice the mass of the WR star, while our models predict many lighter companions, with spectral types well into the B star regime \citep[see also][]{lan1:20}. 
    
    We conclude that our binary models can not be ruled out based on the mismatch between predicted and detected WR binary fractions. Notably, in a recent study of northern Galactic WC stars, \citet{dsi:20} find a bias corrected minimum binary fraction of 72\%.

\section{Conclusions}
\label{sec:conclusions}

    In this work, a large grid of detailed single star and binary evolution models at LMC metallicity is used to create a synthetic WR star population. These models are calculated with the MESA code and include the physics of mass-loss, differential rotation, angular momentum transport by magnetic fields, inflation, tides, and binary mass and angular momentum transfer. The grid covers initial primary masses in the range of ${M_{1,\mathrm{i}}\simeq\SIrange{28}{89}{M_\odot}}$. We compare our results with the observed LMC WR population, which is thought to be complete.  
    
    The WR stars in our the synthetic population reproduce the number of observed WR stars of different subtype (WN with hydrogen, WN without hydrogen, WC) and their luminosity distributions well (see Fig.\,\ref{fig:WR_both_0.5}). Similarly, we find a good overall agreement of the observed and predicted hydrogen mass fractions in WN stars (see Fig.\,\ref{fig:H}). This is remarkable, since in contrast to single star models, the properties of WR star models produced by mass transfer are much less affected by the uncertain physics of stellar wind mass loss. Based on the observed high fraction of main sequence binaries, a large impact of  binary mass transfer on the predicted WR population is expected and confirmed. This renders predictions for massive star populations which are solely based on single star models questionable.
    
    Our results also raise several new questions. When comparing the distributions of our
    WR model binaries, in particular orbital periods and mass ratios, with the (sparse!) available observations (Figs.\,\ref{fig:H-free-WN-PQ} and\,\ref{fig:WC-PQ}), we find that our models do not reproduce the few shortest-period WR binaries ($P_{\rm orb} < 3\,$d). We discuss several possible reasons for this (Sect.\,\ref{sec:scenarios_of_short_period_binaries}), and point out that discriminating these may be important for the predictions of BH mergers. 
    
    When adopting an initial binary fraction of 50\%, we find that the large impact of single stars on
    the predicted number of very luminous WR stars ($> 10^6\,$L$_{\odot}$) is not seen in the
    observational data (Fig.\,\ref{fig:WR_both_0.5}), while halving the initial single star fraction for
    stars above $\sim 40\,$M$_{\odot}$) can remedy this, and at the same time prevents an overproduction
    of luminous red and yellow supergiants (Sect.\,\ref{sec:ss}). 
    
    A large initial binary fraction leads to a large expected binary fraction of the WR stars.
    Our models predict an overall binary fraction of $\sim 80$\%. While this may seem to be at odds with rather modest binary detection fractions in observed WR populations, our models imply that the bias correction may be large. This view is supported by a recent study of Galactic WC stars, where a bias corrected minimum binary fraction of 72\% is derived, while based on actual detections, a binary fraction of 13\% has been suggested for the LMV WC stars (cf., Sect.\,\ref{sec:wr-bin}).
    
    Clearly, our work is just a small step to assess the impact of binary evolution on the formation of WR stars, and on massive star evolution as a whole. The WR stars, due to their outstanding spectra, appear suitable to provide well defined populations for robust tests of our models.
    However, in the end, we need to compare the same models with populations of many other types of massive stars (contact systems, massive Algols, X-ray binaries, BH-merger, etc.), and at different metallicities, before robust conclusions can be drawn. 
    
\begin{acknowledgements}
    The authors thank Koushik Sen for his useful discussions and advise when analyzing the binary model grid.
    DP acknowledges financial support by the Deutsches Zentrum f\"ur Luft und Raumfahrt (DLR) grant FKZ 50 OR 2005. D.R.A-D. is supported by the Stavros Niarchos Foundation (SNF) and the Hellenic Foundation for Research and Innovation (H.F.R.I.) under the 2nd Call of ``Science and Society’' Action Always strive for excellence – ``Theodoros Papazoglou’' (Project Number: 01431). PM acknowledges support from the FWO junior postdoctoral fellowship No. 12ZY520N

\end{acknowledgements}

% BibTex Commands:                                                              
\bibliographystyle{aa}                                                         
\bibliography{astro.bib}           

\begin{thebibliography}{152}
\expandafter\ifx\csname natexlab\endcsname\relax\def\natexlab#1{#1}\fi

\bibitem[{{Aadland} {et~al.}(2022){Aadland}, {Massey}, {John Hillier},
  {Morrell}, {Neugent}, \& {Eldridge}}]{aad1:22}
{Aadland}, E., {Massey}, P., {John Hillier}, D., {et~al.} 2022, \apj, 931, 157

\bibitem[{{Abbott} \& {Conti}(1987)}]{abb1:87}
{Abbott}, D.~C. \& {Conti}, P.~S. 1987, \araa, 25, 113

\bibitem[{{Aguilera-Dena} {et~al.}(2020){Aguilera-Dena}, {Langer},
  {Antoniadis}, \& {M{\"u}ller}}]{agu1:20}
{Aguilera-Dena}, D.~R., {Langer}, N., {Antoniadis}, J., \& {M{\"u}ller}, B.
  2020, \apj, 901, 114

\bibitem[{{Aguilera-Dena} {et~al.}(2022{\natexlab{a}}){Aguilera-Dena},
  {Langer}, {Antoniadis}, {Pauli}, {Dessart}, {Vigna-G{\'o}mez},
  {Gr{\"a}fener}, \& {Yoon}}]{agu1:21}
{Aguilera-Dena}, D.~R., {Langer}, N., {Antoniadis}, J., {et~al.}
  2022{\natexlab{a}}, \aap, 661, A60

\bibitem[{{Aguilera-Dena} {et~al.}(2018){Aguilera-Dena}, {Langer}, {Moriya}, \&
  {Schootemeijer}}]{agu1:18}
{Aguilera-Dena}, D.~R., {Langer}, N., {Moriya}, T.~J., \& {Schootemeijer}, A.
  2018, \apj, 858, 115

\bibitem[{{Aguilera-Dena} {et~al.}(2022{\natexlab{b}}){Aguilera-Dena},
  {M{\"u}ller}, {Antoniadis}, {Langer}, {Dessart}, {Vigna-G{\'o}mez}, \&
  {Yoon}}]{agu1:22}
{Aguilera-Dena}, D.~R., {M{\"u}ller}, B., {Antoniadis}, J., {et~al.}
  2022{\natexlab{b}}, arXiv e-prints, arXiv:2204.00025

\bibitem[{{Asplund} {et~al.}(2009){Asplund}, {Grevesse}, {Sauval}, \&
  {Scott}}]{asp1:09}
{Asplund}, M., {Grevesse}, N., {Sauval}, A.~J., \& {Scott}, P. 2009, \araa, 47,
  481

\bibitem[{{Bartzakos} {et~al.}(2001{\natexlab{a}}){Bartzakos}, {Moffat}, \&
  {Niemela}}]{bar1:01}
{Bartzakos}, P., {Moffat}, A.~F.~J., \& {Niemela}, V.~S. 2001{\natexlab{a}},
  \mnras, 324, 18

\bibitem[{{Bartzakos} {et~al.}(2001{\natexlab{b}}){Bartzakos}, {Moffat}, \&
  {Niemela}}]{bar2:01}
{Bartzakos}, P., {Moffat}, A.~F.~J., \& {Niemela}, V.~S. 2001{\natexlab{b}},
  \mnras, 324, 33

\bibitem[{{Bestenlehner} {et~al.}(2014){Bestenlehner}, {Gr{\"a}fener}, {Vink},
  {Najarro}, {de Koter}, {Sana}, {Evans}, {Crowther}, {H{\'e}nault-Brunet},
  {Herrero}, {Langer}, {Schneider}, {Sim{\'o}n-D{\'{\i}}az}, {Taylor}, \&
  {Walborn}}]{bes1:14}
{Bestenlehner}, J.~M., {Gr{\"a}fener}, G., {Vink}, J.~S., {et~al.} 2014, \aap,
  570, A38

\bibitem[{{B{\"o}hm-Vitense}(1958)}]{boe1:58}
{B{\"o}hm-Vitense}, E. 1958, \zap, 46, 108

\bibitem[{{Breysacher} {et~al.}(1999){Breysacher}, {Azzopardi}, \&
  {Testor}}]{bre1:99}
{Breysacher}, J., {Azzopardi}, M., \& {Testor}, G. 1999, \aaps, 137, 117

\bibitem[{{Brinchmann} {et~al.}(2008){Brinchmann}, {Kunth}, \&
  {Durret}}]{bri2:08}
{Brinchmann}, J., {Kunth}, D., \& {Durret}, F. 2008, \aap, 485, 657

\bibitem[{{Brookshaw} \& {Tavani}(1993)}]{bro1:93}
{Brookshaw}, L. \& {Tavani}, M. 1993, \apj, 410, 719

\bibitem[{{Brott} {et~al.}(2011){Brott}, {de Mink}, {Cantiello}, {Langer}, {de
  Koter}, {Evans}, {Hunter}, {Trundle}, \& {Vink}}]{bro1:11}
{Brott}, I., {de Mink}, S.~E., {Cantiello}, M., {et~al.} 2011, \aap, 530, A115

\bibitem[{{Chan} {et~al.}(2018){Chan}, {M{\"u}ller}, {Heger}, {Pakmor}, \&
  {Springel}}]{cha1:18}
{Chan}, C., {M{\"u}ller}, B., {Heger}, A., {Pakmor}, R., \& {Springel}, V.
  2018, \apjl, 852, L19

\bibitem[{{Cherchneff} {et~al.}(2000){Cherchneff}, {Le Teuff}, {Williams}, \&
  {Tielens}}]{che1:00}
{Cherchneff}, I., {Le Teuff}, Y.~H., {Williams}, P.~M., \& {Tielens},
  A.~G.~G.~M. 2000, \aap, 357, 572

\bibitem[{{Choi} {et~al.}(2016){Choi}, {Dotter}, {Conroy}, {Cantiello},
  {Paxton}, \& {Johnson}}]{cho1:16}
{Choi}, J., {Dotter}, A., {Conroy}, C., {et~al.} 2016, \apj, 823, 102

\bibitem[{{Chruslinska} \& {Nelemans}(2019)}]{mar1:19}
{Chruslinska}, M. \& {Nelemans}, G. 2019, \mnras, 488, 5300

\bibitem[{{Conti}(1991)}]{con1:91}
{Conti}, P.~S. 1991, \apj, 377, 115

\bibitem[{{Crowther}(2007)}]{cro1:07}
{Crowther}, P.~A. 2007, \araa, 45, 177

\bibitem[{{Crowther}(2019)}]{cro1:19}
{Crowther}, P.~A. 2019, Galaxies, 7, 88

\bibitem[{{Crowther} {et~al.}(2002){Crowther}, {Dessart}, {Hillier}, {Abbott},
  \& {Fullerton}}]{cro1:02}
{Crowther}, P.~A., {Dessart}, L., {Hillier}, D.~J., {Abbott}, J.~B., \&
  {Fullerton}, A.~W. 2002, \aap, 392, 653

\bibitem[{{Davies} {et~al.}(2018){Davies}, {Crowther}, \& {Beasor}}]{dav1:18}
{Davies}, B., {Crowther}, P.~A., \& {Beasor}, E.~R. 2018, \mnras, 478, 3138

\bibitem[{{de Jager} {et~al.}(1988){de Jager}, {Nieuwenhuijzen}, \& {van der
  Hucht}}]{jag1:88}
{de Jager}, C., {Nieuwenhuijzen}, H., \& {van der Hucht}, K.~A. 1988, \aaps,
  72, 259

\bibitem[{{De Loore} {et~al.}(1978){De Loore}, {De Greve}, \&
  {Vanbeveren}}]{loo1:78}
{De Loore}, C., {De Greve}, J.~P., \& {Vanbeveren}, D. 1978, \aap, 67, 373

\bibitem[{{de Mink} {et~al.}(2009){de Mink}, {Cantiello}, {Langer}, {Pols},
  {Brott}, \& {Yoon}}]{dem1:09}
{de Mink}, S.~E., {Cantiello}, M., {Langer}, N., {et~al.} 2009, \aap, 497, 243

\bibitem[{{de Mink} {et~al.}(2013){de Mink}, {Langer}, {Izzard}, {Sana}, \& {de
  Koter}}]{dem2:13}
{de Mink}, S.~E., {Langer}, N., {Izzard}, R.~G., {Sana}, H., \& {de Koter}, A.
  2013, \apj, 764, 166

\bibitem[{{de Mink} \& {Mandel}(2016)}]{dem1:16}
{de Mink}, S.~E. \& {Mandel}, I. 2016, \mnras, 460, 3545

\bibitem[{{Dessart} {et~al.}(2011){Dessart}, {Hillier}, {Livne}, {Yoon},
  {Woosley}, {Waldman}, \& {Langer}}]{des1:11}
{Dessart}, L., {Hillier}, D.~J., {Livne}, E., {et~al.} 2011, \mnras, 414, 2985

\bibitem[{{Dessart} {et~al.}(2020){Dessart}, {Yoon}, {Aguilera-Dena}, \&
  {Langer}}]{des1:20}
{Dessart}, L., {Yoon}, S.-C., {Aguilera-Dena}, D.~R., \& {Langer}, N. 2020,
  \aap, 642, A106

\bibitem[{{Dray} {et~al.}(2003){Dray}, {Tout}, {Karakas}, \&
  {Lattanzio}}]{dra1:03}
{Dray}, L.~M., {Tout}, C.~A., {Karakas}, A.~I., \& {Lattanzio}, J.~C. 2003,
  \mnras, 338, 973

\bibitem[{{Dsilva} {et~al.}(2020){Dsilva}, {Shenar}, {Sana}, \&
  {Marchant}}]{dsi:20}
{Dsilva}, K., {Shenar}, T., {Sana}, H., \& {Marchant}, P. 2020, \aap, 641, A26

\bibitem[{{Eggenberger} {et~al.}(2021){Eggenberger}, {Ekstr{\"o}m}, {Georgy},
  {Martinet}, {Pezzotti}, {Nandal}, {Meynet}, {Buldgen}, {Salmon},
  {Haemmerl{\'e}}, {Maeder}, {Hirschi}, {Yusof}, {Groh}, {Farrell}, {Murphy},
  \& {Choplin}}]{egg1:21}
{Eggenberger}, P., {Ekstr{\"o}m}, S., {Georgy}, C., {et~al.} 2021, \aap, 652,
  A137

\bibitem[{{Ekstr{\"o}m} {et~al.}(2012){Ekstr{\"o}m}, {Georgy}, {Eggenberger},
  {Meynet}, {Mowlavi}, {Wyttenbach}, {Granada}, {Decressin}, {Hirschi},
  {Frischknecht}, {Charbonnel}, \& {Maeder}}]{eks1:12}
{Ekstr{\"o}m}, S., {Georgy}, C., {Eggenberger}, P., {et~al.} 2012, \aap, 537,
  A146

\bibitem[{{Eldridge} {et~al.}(2008){Eldridge}, {Izzard}, \& {Tout}}]{eld1:08}
{Eldridge}, J.~J., {Izzard}, R.~G., \& {Tout}, C.~A. 2008, \mnras, 384, 1109

\bibitem[{{Eldridge} {et~al.}(2017){Eldridge}, {Stanway}, {Xiao}, {McClelland},
  {Taylor}, {Ng}, {Greis}, \& {Bray}}]{eld1:17}
{Eldridge}, J.~J., {Stanway}, E.~R., {Xiao}, L., {et~al.} 2017, \pasa, 34, e058

\bibitem[{{Ferraro} {et~al.}(2006){Ferraro}, {Mucciarelli}, {Carretta}, \&
  {Origlia}}]{fer1:06}
{Ferraro}, F.~R., {Mucciarelli}, A., {Carretta}, E., \& {Origlia}, L. 2006,
  \apjl, 645, L33

\bibitem[{{Foellmi} {et~al.}(2003){Foellmi}, {Moffat}, \& {Guerrero}}]{foe1:03}
{Foellmi}, C., {Moffat}, A.~F.~J., \& {Guerrero}, M.~A. 2003, \mnras, 338, 1025

\bibitem[{{Garcia-Segura} {et~al.}(1996{\natexlab{a}}){Garcia-Segura},
  {Langer}, \& {Mac Low}}]{gar2:96}
{Garcia-Segura}, G., {Langer}, N., \& {Mac Low}, M.~M. 1996{\natexlab{a}},
  \aap, 316, 133

\bibitem[{{Garcia-Segura} {et~al.}(1996{\natexlab{b}}){Garcia-Segura}, {Mac
  Low}, \& {Langer}}]{gar1:96}
{Garcia-Segura}, G., {Mac Low}, M.~M., \& {Langer}, N. 1996{\natexlab{b}},
  \aap, 305, 229

\bibitem[{{Georgy} {et~al.}(2013){Georgy}, {Ekstr{\"o}m}, {Eggenberger},
  {Meynet}, {Haemmerl{\'e}}, {Maeder}, {Granada}, {Groh}, {Hirschi}, {Mowlavi},
  {Yusof}, {Charbonnel}, {Decressin}, \& {Barblan}}]{geo1:13}
{Georgy}, C., {Ekstr{\"o}m}, S., {Eggenberger}, P., {et~al.} 2013, \aap, 558,
  A103

\bibitem[{{Georgy} {et~al.}(2012){Georgy}, {Ekstr{\"o}m}, {Meynet}, {Massey},
  {Levesque}, {Hirschi}, {Eggenberger}, \& {Maeder}}]{geo1:12}
{Georgy}, C., {Ekstr{\"o}m}, S., {Meynet}, G., {et~al.} 2012, \aap, 542, A29

\bibitem[{{Gilkis} {et~al.}(2021){Gilkis}, {Shenar}, {Ramachandran}, {Jermyn},
  {Mahy}, {Oskinova}, {Arcavi}, \& {Sana}}]{gil1:21}
{Gilkis}, A., {Shenar}, T., {Ramachandran}, V., {et~al.} 2021, \mnras
  [\eprint[arXiv]{2102.03102}]

\bibitem[{{G{\"o}tberg} {et~al.}(2018){G{\"o}tberg}, {de Mink}, {Groh},
  {Kupfer}, {Crowther}, {Zapartas}, \& {Renzo}}]{goe1:18}
{G{\"o}tberg}, Y., {de Mink}, S.~E., {Groh}, J.~H., {et~al.} 2018, \aap, 615,
  A78

\bibitem[{{Gr{\"a}fener} {et~al.}(2017){Gr{\"a}fener}, {Owocki}, {Grassitelli},
  \& {Langer}}]{gra1:17}
{Gr{\"a}fener}, G., {Owocki}, S.~P., {Grassitelli}, L., \& {Langer}, N. 2017,
  \aap, 608, A34

\bibitem[{{Gr{\"a}fener} {et~al.}(2012){Gr{\"a}fener}, {Owocki}, \&
  {Vink}}]{gra1:12}
{Gr{\"a}fener}, G., {Owocki}, S.~P., \& {Vink}, J.~S. 2012, \aap, 538, A40

\bibitem[{Gr\"afener {et~al.}(2013)Gr\"afener, Vink, Harries, \&
  Langer}]{gra1:13}
Gr\"afener, G., Vink, J.~S., Harries, T.~J., \& Langer, N. 2013, VizieR Online
  Data Catalog, 354, 79083

\bibitem[{{Grassitelli} {et~al.}(2021){Grassitelli}, {Langer}, {Mackey},
  {Gr{\"a}fener}, {Grin}, {Sander}, \& {Vink}}]{gra1:21}
{Grassitelli}, L., {Langer}, N., {Mackey}, J., {et~al.} 2021, \aap, 647, A99

\bibitem[{{Grin} {et~al.}(2017){Grin}, {Ram{\'{\i}}rez-Agudelo}, {de Koter},
  {Sana}, {Puls}, {Brott}, {Crowther}, {Dufton}, {Evans}, {Gr{\"a}fener},
  {Herrero}, {Langer}, {Lennon}, {van Loon}, {Markova}, {de Mink}, {Najarro},
  {Schneider}, {Taylor}, {Tramper}, {Vink}, \& {Walborn}}]{gri1:17}
{Grin}, N.~J., {Ram{\'{\i}}rez-Agudelo}, O.~H., {de Koter}, A., {et~al.} 2017,
  \aap, 600, A82

\bibitem[{{Groh} {et~al.}(2013){Groh}, {Georgy}, \& {Ekstr{\"o}m}}]{gro3:13}
{Groh}, J.~H., {Georgy}, C., \& {Ekstr{\"o}m}, S. 2013, \aap, 558, L1

\bibitem[{{Hainich} {et~al.}(2014){Hainich}, {R{\"u}hling}, {Todt}, {Oskinova},
  {Liermann}, {Gr{\"a}fener}, {Foellmi}, {Schnurr}, \& {Hamann}}]{hai1:14}
{Hainich}, R., {R{\"u}hling}, U., {Todt}, H., {et~al.} 2014, \aap, 565, A27

\bibitem[{{Hamann} {et~al.}(2019){Hamann}, {Gr{\"a}fener}, {Liermann},
  {Hainich}, {Sander}, {Shenar}, {Ramachandran}, {Todt}, \&
  {Oskinova}}]{ham1:19}
{Hamann}, W.-R., {Gr{\"a}fener}, G., {Liermann}, A., {et~al.} 2019, \aap, 625,
  A57

\bibitem[{{Harris} \& {Zaritsky}(2009)}]{har1:09}
{Harris}, J. \& {Zaritsky}, D. 2009, \aj, 138, 1243

\bibitem[{{Hastings} {et~al.}(2020){Hastings}, {Langer}, \&
  {Koenigsberger}}]{has1:20}
{Hastings}, B., {Langer}, N., \& {Koenigsberger}, G. 2020, \aap, 641, A86

\bibitem[{{Heger} {et~al.}(2000){Heger}, {Langer}, \& {Woosley}}]{heg1:00}
{Heger}, A., {Langer}, N., \& {Woosley}, S.~E. 2000, \apj, 528, 368

\bibitem[{{Hillier} {et~al.}(2021){Hillier}, {Aadland}, {Massey}, \&
  {Morrell}}]{hil1:21}
{Hillier}, D.~J., {Aadland}, E., {Massey}, P., \& {Morrell}, N. 2021, \mnras,
  503, 2726

\bibitem[{{Hunter} {et~al.}(2009){Hunter}, {Brott}, {Langer}, {Lennon},
  {Dufton}, {Howarth}, {Ryans}, {Trundle}, {Evans}, {de Koter}, \&
  {Smartt}}]{hun1:09}
{Hunter}, I., {Brott}, I., {Langer}, N., {et~al.} 2009, \aap, 496, 841

\bibitem[{{Hunter} {et~al.}(2008){Hunter}, {Brott}, {Lennon}, {Langer},
  {Dufton}, {Trundle}, {Smartt}, {de Koter}, {Evans}, \& {Ryans}}]{hun1:08}
{Hunter}, I., {Brott}, I., {Lennon}, D.~J., {et~al.} 2008, \apjl, 676, L29

\bibitem[{{Hunter} {et~al.}(2007){Hunter}, {Dufton}, {Smartt}, {Ryans},
  {Evans}, {Lennon}, {Trundle}, {Hubeny}, \& {Lanz}}]{hun1:07}
{Hunter}, I., {Dufton}, P.~L., {Smartt}, S.~J., {et~al.} 2007, \aap, 466, 277

\bibitem[{{Hurley} {et~al.}(2000){Hurley}, {Pols}, \& {Tout}}]{hur1:00}
{Hurley}, J.~R., {Pols}, O.~R., \& {Tout}, C.~A. 2000, \mnras, 315, 543

\bibitem[{{Inserra} {et~al.}(2013){Inserra}, {Smartt}, {Jerkstrand}, {Valenti},
  {Fraser}, {Wright}, {Smith}, {Chen}, {Kotak}, {Pastorello}, {Nicholl},
  {Bresolin}, {Kudritzki}, {Benetti}, {Botticella}, {Burgett}, {Chambers},
  {Ergon}, {Flewelling}, {Fynbo}, {Geier}, {Hodapp}, {Howell}, {Huber},
  {Kaiser}, {Leloudas}, {Magill}, {Magnier}, {McCrum}, {Metcalfe}, {Price},
  {Rest}, {Sollerman}, {Sweeney}, {Taddia}, {Taubenberger}, {Tonry},
  {Wainscoat}, {Waters}, \& {Young}}]{ins1:13}
{Inserra}, C., {Smartt}, S.~J., {Jerkstrand}, A., {et~al.} 2013, \apj, 770, 128

\bibitem[{{Ivanova} {et~al.}(2013){Ivanova}, {Justham}, {Chen}, {De Marco},
  {Fryer}, {Gaburov}, {Ge}, {Glebbeek}, {Han}, {Li}, {Lu}, {Marsh},
  {Podsiadlowski}, {Potter}, {Soker}, {Taam}, {Tauris}, {van den Heuvel}, \&
  {Webbink}}]{iva1:13}
{Ivanova}, N., {Justham}, S., {Chen}, X., {et~al.} 2013, \aapr, 21, 59

\bibitem[{{Ivanova} {et~al.}(2020){Ivanova}, {Justham}, \& {Ricker}}]{iva1:20}
{Ivanova}, N., {Justham}, S., \& {Ricker}, P. 2020, {Common Envelope Evolution}

\bibitem[{{Janka}(2013)}]{jan1:13}
{Janka}, H.-T. 2013, \mnras, 434, 1355

\bibitem[{{Justham} {et~al.}(2014){Justham}, {Podsiadlowski}, \&
  {Vink}}]{jus1:14}
{Justham}, S., {Podsiadlowski}, P., \& {Vink}, J.~S. 2014, \apj, 796, 121

\bibitem[{{Kalari} {et~al.}(2018){Kalari}, {Vink}, {Dufton}, \&
  {Fraser}}]{kal1:18}
{Kalari}, V.~M., {Vink}, J.~S., {Dufton}, P.~L., \& {Fraser}, M. 2018, \aap,
  618, A17

\bibitem[{{Kee} {et~al.}(2021){Kee}, {Sundqvist}, {Decin}, {de Koter}, \&
  {Sana}}]{kee1:21}
{Kee}, N.~D., {Sundqvist}, J.~O., {Decin}, L., {de Koter}, A., \& {Sana}, H.
  2021, \aap, 646, A180

\bibitem[{{King} \& {Begelman}(1999)}]{kin1:99}
{King}, A.~R. \& {Begelman}, M.~C. 1999, \apjl, 519, L169

\bibitem[{{King} {et~al.}(2000){King}, {Taam}, \& {Begelman}}]{kin1:00}
{King}, A.~R., {Taam}, R.~E., \& {Begelman}, M.~C. 2000, \apjl, 530, L25

\bibitem[{{Kippenhahn} {et~al.}(1980){Kippenhahn}, {Ruschenplatt}, \&
  {Thomas}}]{kip1:80}
{Kippenhahn}, R., {Ruschenplatt}, G., \& {Thomas}, H.~C. 1980, \aap, 91, 175

\bibitem[{{Klencki} {et~al.}(2021){Klencki}, {Istrate}, {Nelemans}, \&
  {Pols}}]{kle1:21}
{Klencki}, J., {Istrate}, A.~G., {Nelemans}, G., \& {Pols}, O. 2021, arXiv
  e-prints, arXiv:2111.10271

\bibitem[{{Koenigsberger} {et~al.}(2014){Koenigsberger}, {Morrell}, {Hillier},
  {Gamen}, {Schneider}, {Gonz{\'a}lez-Jim{\'e}nez}, {Langer}, \&
  {Barb{\'a}}}]{koe2:14}
{Koenigsberger}, G., {Morrell}, N., {Hillier}, D.~J., {et~al.} 2014, \aj, 148,
  62

\bibitem[{{Koesterke} \& {Hamann}(1995)}]{koe1:95}
{Koesterke}, L. \& {Hamann}, W.-R. 1995, \aap, 299, 503

\bibitem[{{K{\"o}hler} {et~al.}(2015){K{\"o}hler}, {Langer}, {de Koter}, {de
  Mink}, {Crowther}, {Evans}, {Gr{\"a}fener}, {Sana}, {Sanyal}, {Schneider}, \&
  {Vink}}]{koe1:15}
{K{\"o}hler}, K., {Langer}, N., {de Koter}, A., {et~al.} 2015, \aap, 573, A71

\bibitem[{{Kruckow} {et~al.}(2016){Kruckow}, {Tauris}, {Langer}, {Sz{\'e}csi},
  {Marchant}, \& {Podsiadlowski}}]{kru1:16}
{Kruckow}, M.~U., {Tauris}, T.~M., {Langer}, N., {et~al.} 2016, \aap, 596, A58

\bibitem[{{Kurt} \& {Dufour}(1998)}]{kur1:98}
{Kurt}, C.~M. \& {Dufour}, R.~J. 1998, in Revista Mexicana de Astronomia y
  Astrofisica Conference Series, Vol.~7, Revista Mexicana de Astronomia y
  Astrofisica Conference Series, ed. R.~J. {Dufour} \& S.~{Torres-Peimbert},
  202

\bibitem[{{Langer}(1987)}]{lan1:87}
{Langer}, N. 1987, \aap, 171, L1

\bibitem[{{Langer}(1989)}]{lan1:89}
{Langer}, N. 1989, \aap, 210, 93

\bibitem[{{Langer}(2012)}]{lan1:12}
{Langer}, N. 2012, \araa, 50, 107

\bibitem[{{Langer} {et~al.}(1983){Langer}, {Fricke}, \& {Sugimoto}}]{lan1:83}
{Langer}, N., {Fricke}, K.~J., \& {Sugimoto}, D. 1983, \aap, 126, 207

\bibitem[{{Langer} {et~al.}(1994){Langer}, {Hamann}, {Lennon}, {Najarro},
  {Pauldrach}, \& {Puls}}]{lan1:94}
{Langer}, N., {Hamann}, W.-R., {Lennon}, M., {et~al.} 1994, \aap, 290, 819

\bibitem[{{Langer} {et~al.}(2020){Langer}, {Sch{\"u}rmann}, {Stoll},
  {Marchant}, {Lennon}, {Mahy}, {de Mink}, {Quast}, {Riedel}, {Sana},
  {Schneider}, {Schootemeijer}, {Wang}, {Almeida}, {Bestenlehner},
  {Bodensteiner}, {Castro}, {Clark}, {Crowther}, {Dufton}, {Evans}, {Fossati},
  {Gr{\"a}fener}, {Grassitelli}, {Grin}, {Hastings}, {Herrero}, {de Koter},
  {Menon}, {Patrick}, {Puls}, {Renzo}, {Sander}, {Schneider}, {Sen}, {Shenar},
  {Sim{\'o}n-D{\'\i}as}, {Tauris}, {Tramper}, {Vink}, \& {Xu}}]{lan1:20}
{Langer}, N., {Sch{\"u}rmann}, C., {Stoll}, K., {et~al.} 2020, \aap, 638, A39

\bibitem[{{Limongi} \& {Chieffi}(2018)}]{lim1:18}
{Limongi}, M. \& {Chieffi}, A. 2018, \apjs, 237, 13

\bibitem[{{MacLeod} \& {Loeb}(2020)}]{leo1:20}
{MacLeod}, M. \& {Loeb}, A. 2020, \apj, 902, 85

\bibitem[{{Maeder}(1983)}]{mae1:83}
{Maeder}, A. 1983, \aap, 120, 113

\bibitem[{{Maeder}(1987)}]{mae1:87}
{Maeder}, A. 1987, \aap, 178, 159

\bibitem[{{Maeder} \& {Meynet}(1987)}]{mae2:87}
{Maeder}, A. \& {Meynet}, G. 1987, \aap, 182, 243

\bibitem[{{Maeder} \& {Meynet}(2012)}]{mae1:12}
{Maeder}, A. \& {Meynet}, G. 2012, Reviews of Modern Physics, 84, 25

\bibitem[{{Marchant}(2016)}]{mar2:16}
{Marchant}, P. 2016, PhD thesis, Rheinischen Friedrich-Wilhelms-Universität
  Bonn

\bibitem[{{Marchant} {et~al.}(2016){Marchant}, {Langer}, {Podsiadlowski},
  {Tauris}, \& {Moriya}}]{mar1:16}
{Marchant}, P., {Langer}, N., {Podsiadlowski}, P., {Tauris}, T.~M., \&
  {Moriya}, T.~J. 2016, \aap, 588, A50

\bibitem[{{Markova} {et~al.}(2018){Markova}, {Puls}, \& {Langer}}]{mar1:18}
{Markova}, N., {Puls}, J., \& {Langer}, N. 2018, \aap, 613, A12

\bibitem[{{Massey} {et~al.}(2021){Massey}, {Neugent}, {Dorn-Wallenstein},
  {Eldridge}, {Stanway}, \& {Levesque}}]{mas1:21}
{Massey}, P., {Neugent}, K.~F., {Dorn-Wallenstein}, T.~Z., {et~al.} 2021, \apj,
  922, 177

\bibitem[{{Mauron} \& {Josselin}(2011)}]{mau1:11}
{Mauron}, N. \& {Josselin}, E. 2011, \aap, 526, A156

\bibitem[{{Menon} {et~al.}(2021){Menon}, {Langer}, {de Mink}, {Justham}, {Sen},
  {Sz{\'e}csi}, {de Koter}, {Abdul-Masih}, {Sana}, {Mahy}, \&
  {Marchant}}]{men1:21}
{Menon}, A., {Langer}, N., {de Mink}, S.~E., {et~al.} 2021, \mnras, 507, 5013

\bibitem[{{Meynet} \& {Maeder}(2003)}]{mey1:03}
{Meynet}, G. \& {Maeder}, A. 2003, \aap, 404, 975

\bibitem[{{Meynet} \& {Maeder}(2005)}]{mey1:05}
{Meynet}, G. \& {Maeder}, A. 2005, \aap, 429, 581

\bibitem[{{Mirabel}(2017)}]{mir1:17}
{Mirabel}, I.~F. 2017, in New Frontiers in Black Hole Astrophysics, ed.
  A.~{Gomboc}, Vol. 324, 303--306

\bibitem[{{Moe} \& {Di Stefano}(2017)}]{moe1:17}
{Moe}, M. \& {Di Stefano}, R. 2017, \apjs, 230, 15

\bibitem[{{Mokiem} {et~al.}(2007){Mokiem}, {de Koter}, {Vink}, {Puls}, {Evans},
  {Smartt}, {Crowther}, {Herrero}, {Langer}, {Lennon}, {Najarro}, \&
  {Villamariz}}]{mok2:07}
{Mokiem}, M.~R., {de Koter}, A., {Vink}, J.~S., {et~al.} 2007, \aap, 473, 603

\bibitem[{{Neugent} {et~al.}(2018){Neugent}, {Massey}, \& {Morrell}}]{neu1:18}
{Neugent}, K.~F., {Massey}, P., \& {Morrell}, N. 2018, \apj, 863, 181

\bibitem[{{Nieuwenhuijzen} \& {de Jager}(1990)}]{nie1:90}
{Nieuwenhuijzen}, H. \& {de Jager}, C. 1990, \aap, 231, 134

\bibitem[{{Nugis} {et~al.}(1998){Nugis}, {Crowther}, \& {Willis}}]{nug1:98}
{Nugis}, T., {Crowther}, P.~A., \& {Willis}, A.~J. 1998, \aap, 333, 956

\bibitem[{{Nugis} \& {Lamers}(2000)}]{nug1:00}
{Nugis}, T. \& {Lamers}, H.~J.~G.~L.~M. 2000, \aap, 360, 227

\bibitem[{{Paczy{\'n}ski}(1967)}]{pac1:67}
{Paczy{\'n}ski}, B. 1967, \actaa, 17, 355

\bibitem[{{Pauli}(2020)}]{pau1:20}
{Pauli}, D. 2020, Master's thesis, Rheinischen Friedrich-Wilhelms-Universität
  Bonn

\bibitem[{{Paxton} {et~al.}(2011){Paxton}, {Bildsten}, {Dotter}, {Herwig},
  {Lesaffre}, \& {Timmes}}]{pax1:11}
{Paxton}, B., {Bildsten}, L., {Dotter}, A., {et~al.} 2011, \apjs, 192, 3

\bibitem[{{Paxton} {et~al.}(2013){Paxton}, {Cantiello}, {Arras}, {Bildsten},
  {Brown}, {Dotter}, {Mankovich}, {Montgomery}, {Stello}, {Timmes}, \&
  {Townsend}}]{pax1:13}
{Paxton}, B., {Cantiello}, M., {Arras}, P., {et~al.} 2013, \apjs, 208, 4

\bibitem[{{Paxton} {et~al.}(2015){Paxton}, {Marchant}, {Schwab}, {Bauer},
  {Bildsten}, {Cantiello}, {Dessart}, {Farmer}, {Hu}, {Langer}, {Townsend},
  {Townsley}, \& {Timmes}}]{pax1:15}
{Paxton}, B., {Marchant}, P., {Schwab}, J., {et~al.} 2015, \apjs, 220, 15

\bibitem[{{Paxton} {et~al.}(2018){Paxton}, {Schwab}, {Bauer}, {Bildsten},
  {Blinnikov}, {Duffell}, {Farmer}, {Goldberg}, {Marchant}, {Sorokina},
  {Thoul}, {Townsend}, \& {Timmes}}]{pax1:18}
{Paxton}, B., {Schwab}, J., {Bauer}, E.~B., {et~al.} 2018, \apjs, 234, 34

\bibitem[{{Paxton} {et~al.}(2019){Paxton}, {Smolec}, {Schwab}, {Gautschy},
  {Bildsten}, {Cantiello}, {Dotter}, {Farmer}, {Goldberg}, {Jermyn}, {Kanbur},
  {Marchant}, {Thoul}, {Townsend}, {Wolf}, {Zhang}, \& {Timmes}}]{pax1:19}
{Paxton}, B., {Smolec}, R., {Schwab}, J., {et~al.} 2019, \apjs, 243, 10

\bibitem[{{Ramachandran} {et~al.}(2018){Ramachandran}, {Hamann}, {Hainich},
  {Oskinova}, {Shenar}, {Sander}, {Todt}, \& {Gallagher}}]{ram1:18}
{Ramachandran}, V., {Hamann}, W.~R., {Hainich}, R., {et~al.} 2018, \aap, 615,
  A40

\bibitem[{{Ram{\'{\i}}rez-Agudelo} {et~al.}(2017){Ram{\'{\i}}rez-Agudelo},
  {Sana}, {de Koter}, {Tramper}, {Grin}, {Schneider}, {Langer}, {Puls},
  {Markova}, {Bestenlehner}, {Castro}, {Crowther}, {Evans}, {Garc{\'{\i}}a},
  {Gr{\"a}fener}, {Herrero}, {van Kempen}, {Lennon}, {Ma{\'{\i}}z
  Apell{\'a}niz}, {Najarro}, {Sab{\'{\i}}n-Sanjuli{\'a}n},
  {Sim{\'o}n-D{\'{\i}}az}, {Taylor}, \& {Vink}}]{ram1:17}
{Ram{\'{\i}}rez-Agudelo}, O.~H., {Sana}, H., {de Koter}, A., {et~al.} 2017,
  \aap, 600, A81

\bibitem[{{Remillard} \& {McClintock}(2006)}]{rem1:06}
{Remillard}, R.~A. \& {McClintock}, J.~E. 2006, \araa, 44, 49

\bibitem[{{Sab{\'{\i}}n-Sanjuli{\'a}n}
  {et~al.}(2017){Sab{\'{\i}}n-Sanjuli{\'a}n}, {Sim{\'o}n-D{\'{\i}}az},
  {Herrero}, {Puls}, {Schneider}, {Evans}, {Garcia}, {Najarro}, {Brott},
  {Castro}, {Crowther}, {de Koter}, {de Mink}, {Gr{\"a}fener}, {Grin},
  {Holgado}, {Langer}, {Lennon}, {Ma{\'{\i}}z Apell{\'a}niz},
  {Ram{\'{\i}}rez-Agudelo}, {Sana}, {Taylor}, {Vink}, \& {Walborn}}]{sab1:17}
{Sab{\'{\i}}n-Sanjuli{\'a}n}, C., {Sim{\'o}n-D{\'{\i}}az}, S., {Herrero}, A.,
  {et~al.} 2017, \aap, 601, A79

\bibitem[{{Salpeter}(1955)}]{sal1:55}
{Salpeter}, E.~E. 1955, \apj, 121, 161

\bibitem[{{Sana} {et~al.}(2013){Sana}, {de Koter}, {de Mink}, {Dunstall},
  {Evans}, {H{\'e}nault-Brunet}, {Ma{\'{\i}}z Apell{\'a}niz},
  {Ram{\'{\i}}rez-Agudelo}, {Taylor}, {Walborn}, {Clark}, {Crowther},
  {Herrero}, {Gieles}, {Langer}, {Lennon}, \& {Vink}}]{san1:13}
{Sana}, H., {de Koter}, A., {de Mink}, S.~E., {et~al.} 2013, \aap, 550, A107

\bibitem[{{Sana} {et~al.}(2012){Sana}, {de Mink}, {de Koter}, {Langer},
  {Evans}, {Gieles}, {Gosset}, {Izzard}, {Le Bouquin}, \&
  {Schneider}}]{san2:12}
{Sana}, H., {de Mink}, S.~E., {de Koter}, A., {et~al.} 2012, Science, 337, 444

\bibitem[{{Sana} {et~al.}(2014){Sana}, {Le Bouquin}, {Lacour}, {Berger},
  {Duvert}, {Gauchet}, {Norris}, {Olofsson}, {Pickel}, {Zins}, {Absil}, {de
  Koter}, {Kratter}, {Schnurr}, \& {Zinnecker}}]{san1:14}
{Sana}, H., {Le Bouquin}, J.~B., {Lacour}, S., {et~al.} 2014, \apjs, 215, 15

\bibitem[{{Sander} {et~al.}(2012){Sander}, {Hamann}, \& {Todt}}]{san1:12}
{Sander}, A., {Hamann}, W.-R., \& {Todt}, H. 2012, \aap, 540, A144

\bibitem[{{Sander} {et~al.}(2019){Sander}, {Hamann}, {Todt}, {Hainich},
  {Shenar}, {Ramachandran}, \& {Oskinova}}]{san1:19}
{Sander}, A.~A.~C., {Hamann}, W.-R., {Todt}, H., {et~al.} 2019, \aap, 621, A92

\bibitem[{{Sander} \& {Vink}(2020)}]{san1:20}
{Sander}, A. A.~C. \& {Vink}, J.~S. 2020, \mnras, 499, 873

\bibitem[{{Sanyal} {et~al.}(2015){Sanyal}, {Grassitelli}, {Langer}, \&
  {Bestenlehner}}]{san1:15}
{Sanyal}, D., {Grassitelli}, L., {Langer}, N., \& {Bestenlehner}, J.~M. 2015,
  \aap, 580, A20

\bibitem[{{Schmutz} {et~al.}(1989){Schmutz}, {Hamann}, \&
  {Wessolowski}}]{sch1:89}
{Schmutz}, W., {Hamann}, W.-R., \& {Wessolowski}, U. 1989, \aap, 210, 236

\bibitem[{{Schneider} {et~al.}(2015){Schneider}, {Izzard}, {Langer}, \& {de
  Mink}}]{sch1:15}
{Schneider}, F.~R.~N., {Izzard}, R.~G., {Langer}, N., \& {de Mink}, S.~E. 2015,
  \apj, 805, 20

\bibitem[{{Schnurr} {et~al.}(2008){Schnurr}, {Moffat}, {St-Louis}, {Morrell},
  \& {Guerrero}}]{sch2:08}
{Schnurr}, O., {Moffat}, A.~F.~J., {St-Louis}, N., {Morrell}, N.~I., \&
  {Guerrero}, M.~A. 2008, \mnras, 389, 806

\bibitem[{{Schootemeijer} \& {Langer}(2018)}]{sch1:18}
{Schootemeijer}, A. \& {Langer}, N. 2018, \aap, 611, A75

\bibitem[{{Schootemeijer} {et~al.}(2019){Schootemeijer}, {Langer}, {Grin}, \&
  {Wang}}]{sch1:19}
{Schootemeijer}, A., {Langer}, N., {Grin}, N.~J., \& {Wang}, C. 2019, \aap,
  625, A132

\bibitem[{{Sen} {et~al.}(2022){Sen}, {Langer}, {Marchant}, {Menon}, {de Mink},
  {Schootemeijer}, {Sch{\"u}rmann}, {Mahy}, {Hastings}, {Nathaniel}, {Sana},
  {Wang}, \& {Xu}}]{sen1:22}
{Sen}, K., {Langer}, N., {Marchant}, P., {et~al.} 2022, \aap, 659, A98

\bibitem[{{Shenar} {et~al.}(2020){Shenar}, {Gilkis}, {Vink}, {Sana}, \& {Sand
  er}}]{She1:20}
{Shenar}, T., {Gilkis}, A., {Vink}, J.~S., {Sana}, H., \& {Sand er}, A.~A.~C.
  2020, \aap, 634, A79

\bibitem[{{Shenar} {et~al.}(2019){Shenar}, {Sablowski}, {Hainich}, {Todt},
  {Moffat}, {Oskinova}, {Ramachandran}, {Sana}, {Sander}, {Schnurr},
  {St-Louis}, {Vanbeveren}, {G{\"o}tberg}, \& {Hamann}}]{She1:19}
{Shenar}, T., {Sablowski}, D.~P., {Hainich}, R., {et~al.} 2019, \aap, 627, A151

\bibitem[{{Smith} {et~al.}(1994){Smith}, {Meynet}, \& {Mermilliod}}]{smi1:94}
{Smith}, L.~F., {Meynet}, G., \& {Mermilliod}, J.~C. 1994, \aap, 287, 835

\bibitem[{{Smith} \& {Owocki}(2006)}]{smi1:06}
{Smith}, N. \& {Owocki}, S.~P. 2006, \apjl, 645, L45

\bibitem[{{Spruit}(2002)}]{spr1:02}
{Spruit}, H.~C. 2002, \aap, 381, 923

\bibitem[{{Tauris} \& {van den Heuvel}(2006)}]{Tau:06}
{Tauris}, T.~M. \& {van den Heuvel}, E.~P.~J. 2006, {Formation and evolution of
  compact stellar X-ray sources}, Vol.~39, 623--665

\bibitem[{{Tramper} {et~al.}(2015){Tramper}, {Straal}, {Sanyal}, {Sana}, {de
  Koter}, {Gr{\"a}fener}, {Langer}, {Vink}, {de Mink}, \& {Kaper}}]{tra1:15}
{Tramper}, F., {Straal}, S.~M., {Sanyal}, D., {et~al.} 2015, \aap, 581, A110

\bibitem[{{Trundle} {et~al.}(2007){Trundle}, {Dufton}, {Hunter}, {Evans},
  {Lennon}, {Smartt}, \& {Ryans}}]{tru1:07}
{Trundle}, C., {Dufton}, P.~L., {Hunter}, I., {et~al.} 2007, {The VLT-FLAMES
  survey of massive stars: evolution of surface N abundances and effective
  temperature scales in the Galaxy and Magellanic Clouds}, Astronomy and
  Astrophysics, Volume 471, Issue 2, August IV 2007, pp.625-643

\bibitem[{{van den Heuvel} {et~al.}(2017){van den Heuvel}, {Portegies Zwart},
  \& {de Mink}}]{van1:17}
{van den Heuvel}, E.~P.~J., {Portegies Zwart}, S.~F., \& {de Mink}, S.~E. 2017,
  \mnras, 471, 4256

\bibitem[{{van Kerkwijk} {et~al.}(1992){van Kerkwijk}, {Charles}, {Geballe},
  {King}, {Miley}, {Molnar}, {van den Heuvel}, {van der Klis}, \& {van
  Paradijs}}]{van1:92}
{van Kerkwijk}, M.~H., {Charles}, P.~A., {Geballe}, T.~R., {et~al.} 1992, \nat,
  355, 703

\bibitem[{{Vanbeveren} \& {Conti}(1980)}]{van1:80}
{Vanbeveren}, D. \& {Conti}, P.~S. 1980, \aap, 88, 230

\bibitem[{{Vink}(2017)}]{vin1:17}
{Vink}, J.~S. 2017, \aap, 607, L8

\bibitem[{{Vink} {et~al.}(2001){Vink}, {de Koter}, \& {Lamers}}]{vin1:01}
{Vink}, J.~S., {de Koter}, A., \& {Lamers}, H.~J.~G.~L.~M. 2001, \aap, 369, 574

\bibitem[{{Walborn} {et~al.}(2004){Walborn}, {Morrell}, {Howarth}, {Crowther},
  {Lennon}, {Massey}, \& {Arias}}]{wal1:04}
{Walborn}, N.~R., {Morrell}, N.~I., {Howarth}, I.~D., {et~al.} 2004, \apj, 608,
  1028

\bibitem[{{Wang} {et~al.}(2020){Wang}, {Langer}, {Schootemeijer}, {Castro},
  {Adscheid}, {Marchant}, \& {Hastings}}]{wan1:20}
{Wang}, C., {Langer}, N., {Schootemeijer}, A., {et~al.} 2020, \apjl, 888, L12

\bibitem[{{Weaver} {et~al.}(1977){Weaver}, {McCray}, {Castor}, {Shapiro}, \&
  {Moore}}]{wea1:97}
{Weaver}, R., {McCray}, R., {Castor}, J., {Shapiro}, P., \& {Moore}, R. 1977,
  \apj, 218, 377

\bibitem[{{Weis} \& {Bomans}(2020)}]{wei1:20}
{Weis}, K. \& {Bomans}, D.~J. 2020, Galaxies, 8, 20

\bibitem[{{Wellstein} \& {Langer}(1999)}]{wel1:99}
{Wellstein}, S. \& {Langer}, N. 1999, \aap, 350, 148

\bibitem[{{Woosley} \& {Heger}(2006)}]{woo1:06}
{Woosley}, S. \& {Heger}, A. 2006, \apj, 637, 914

\bibitem[{{Woosley}(1993)}]{woo2:93}
{Woosley}, S.~E. 1993, \apj, 405, 273

\bibitem[{{Yoon} {et~al.}(2006){Yoon}, {Langer}, \& {Norman}}]{yoo1:06}
{Yoon}, S., {Langer}, N., \& {Norman}, C. 2006, \aap, 460, 199

\bibitem[{{Yoon}(2017)}]{yoo1:17}
{Yoon}, S.-C. 2017, \mnras, 470, 3970

\bibitem[{{Yoon} \& {Langer}(2005)}]{yoo1:05}
{Yoon}, S.-C. \& {Langer}, N. 2005, \aap, 443, 643

\end{thebibliography}

\begin{appendix}

\section{The stellar wind optical depth}
\label{app:optical_depth}
    In Sect.\,\ref{sec:optical_depth_of_WR_winds} we calculated the optical depth of the winds with the formula given in Eq.\,\ref{eq:optical_depth} for the main-sequence and post main-sequence evolution of our single and binary models. In the following we want to show exemplary phase diagrams of the  theoretical optical depth of the wind throughout the different evolutionary stages of each model would have. 
    
\begin{figure*}[tp]
        \centering
        \includegraphics[trim= 1.9cm 1.5cm 2.2cm 2cm ,clip ,height=0.37\textheight]{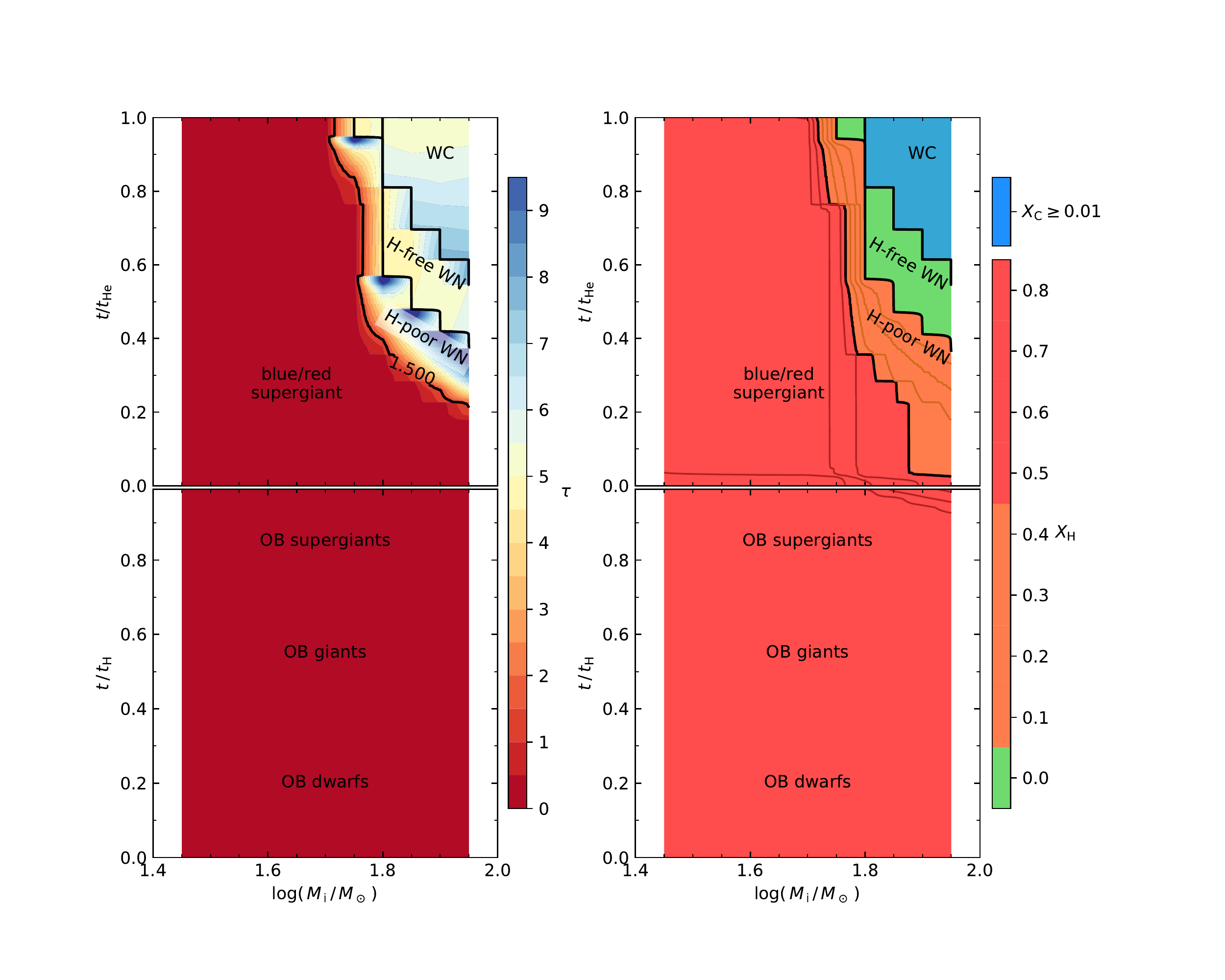}
        \caption{Optical depth (left, color coded) and surface hydrogen abundance (right) of our single star models as funcction of initial mass and normalized time, during their main-sequence (\textit{lower}) and core helium burning (\textit{upper}) evolution. The transition line from an optically thin to an optically thick wind is marked by a solid black line labeled with ``$1.500$''. The unlabeled solid black lines divide the different considered WR phases as labeled in each region. In the right plot, threshold values of the surface hydrogen abundances in steps of ${\Delta X_\mathrm{H}=0.1}$ are indicated
        by thin solid lines, which have dark red color for the range ${X_\mathrm{H}=\numrange{0.8}{0.5}}$, and brown color for  ${X_\mathrm{H}=\numrange{0.4}{0.1}}$. The borderlines of ${X_\mathrm{H}={0.4}}$ and ${X_\mathrm{H}={0.0}}$ are marked by solid black lines, as well as the borderline beyond which stellar models have a surface carbon mass fraction in excess of $X_\mathrm{C} = 0.01$, which is our threshold for defining WC stars.}
        \label{fig:Phase_Single}
        \centering
        \includegraphics[trim= 1.9cm 1.5cm 2.2cm 2cm ,clip ,height=0.37\textheight]{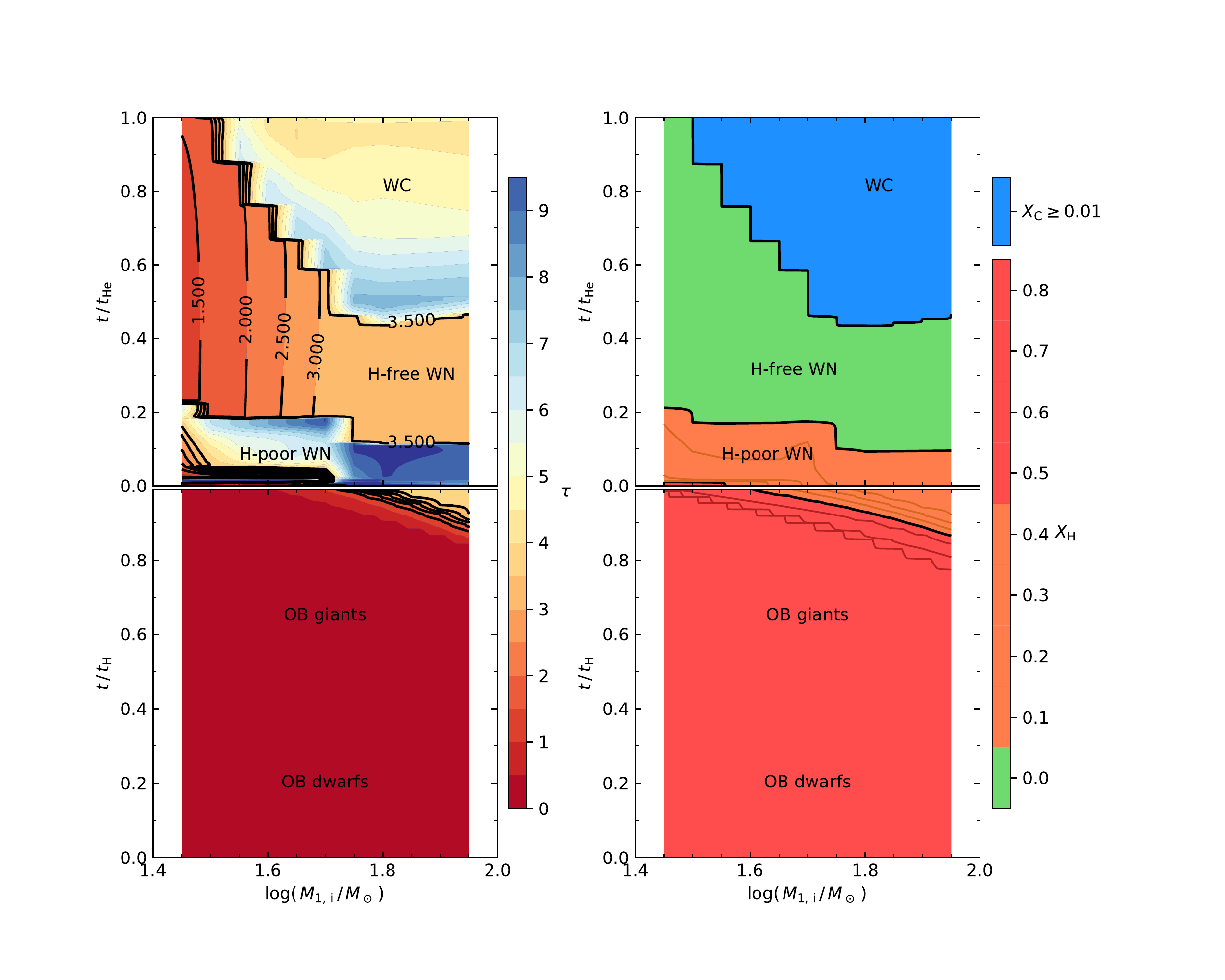}
        \caption{Same as in Fig.\,\ref{fig:Phase_Single}, but for our binary models with an initial orbital period of $P_\mathrm{\,i}=\SI{10}{d}$ and an initial mass ratio of $q_\mathrm{\,i}=0.5$. In contrast to Fig.\,\ref{fig:Phase_Single}, the optical depths here mark fixed values of the wind optical depths in the range from $\tau = \numrange{1.5}{3.5}$ and are labeled accordingly.}
        \label{fig:Phase_Binary}
    \end{figure*}

    Figure\,\ref{fig:Phase_Single} shows the calculated optical depth of the winds 
    of our single star models, and of the surface hydrogen abundance, as function of mass and normalised time during  the main-sequence and post main-sequence evolution. During the their post main-sequence evolution, models with initial mass $M_\mathrm{\,i} \gtrsim \SI{50}{M_\odot}$ are able to efficiently strip off their hydrogen-rich layers during the supergiant phase and go through a WR phase. During the transition from the supergiant stage to the WR phase, the optical depth of the wind rises quickly. The wind transitions from an optically thin to an optically thick wind occurs synchronously with the drop of the surface hydrogen abundance close to $X_\mathrm{H}\approx0.3$. Therefore, the produced helium stars are suggested to be observed as WR stars according to our criterion. Due to the quick transition from an optically thin to an optically thick wind a lower threshold of $\tau\geq0.5$ does not lead to a significant change on the contribution of the single stars to the H-poor WN population (see Fig.\,\ref{fig:WR_both_0.5}). 
    
    None of our single star models develops an optically thick wind while the surface hydrogen abundance is above $X_\mathrm{H}\gtrsim 0.3$. This means, that our models predict that all observed WR stars with surface hydrogen abundances above $X_\mathrm{H}\gtrsim 0.3$ are formed in binary systems. The quick transition from an optically thin to an optically thick wind in the models is linked to the switch from the \citet{vin1:01} OB wind scheme to the \citet{nug1:00} WR wind scheme when the surface hydrogen abundance drops from $X_\mathrm{H}={0.7}$ to $X_\mathrm{H}={0.4}$. Of course as discussed in Sects.\,\ref{sec:reliability} and \ref{sec:ss} this picture can change when using different mass-loss rates and mixing efficiencies.

    Figure\,\ref{fig:Phase_Binary} shows the same diagrams but now for the mass donors of our binary models with an initial orbital period ${P_\mathrm{\,i} = \SI{10}{d}}$ and an initial mass ratio $q_\mathrm{\,i}=0.5$. For these values of initial orbital period and mass ratio, all systems undergo stable mass-transfer and are calculated until core helium depletion. While the amount of mass that is transferred depends on both parameters, the dependence is weak in the sense that in all cases most parts of the hydrogen-rich envelope are removed during the mass-transfer phases.
    
    Similar to the single star models, the binary models are expected to have  an optically thin wind during most of their main-sequence lifetime. However, during a mass-transfer phase more than 70\% of the H-rich envelope is removed. The models shown have a mass-transfer phase during the main-sequence (Case\,A mass transfer).
    After the onset of mass transfer the surface hydrogen abundance drops rapidly from $X_\mathrm{H}=0.7$ to $X_\mathrm{H}=0.4$, and the wind of the models is predicted to become optically thick as soon as the surface hydrogen abundance drops below $X_\mathrm{H}\approx0.4$. 
    
    The most notable difference to the single star models is that for most massive primaries we expect them to develop an optically thick wind, and thus a WR-like spectrum, already during their late phases of core hydrogen burning. We like to note here, that here it can be seen that a lower threshold on the optical depth of $\tau\geq0.5$ is expected to strongly impact the H-poor WN population, as it leads to a sustainable longer  H-poor WN phase which starts during core hydrogen burning and is able to explain surface hydrogen abundances above $X_\mathrm{H}\geq0.4$ (see also Fig.\,\ref{fig:H}).
    
\clearpage
\section{Comparison with previous works}
\label{sec:comparision_with_previous_works}
    
    Evolutionary models of massive stars contain several weakly constrained physics parameters that can strongly affect their evolution and the fate. In the literature various stellar evolutionary models computed with different stellar evolution codes as well as different input parameters, in particular mass-loss recipes and mixing efficiencies, are known and used to explain observed population of stars at different metallicities. Here, we compare our single and binary star models with the predictions of other works. 
    
\subsection{Single star models}
\begin{figure*}[t]
        \centering
        \begin{tikzpicture}
            \node [anchor=south west] at (0,0)
            {\includegraphics[trim= 0cm 0cm 7cm 0cm ,clip,width=1.0\textwidth]{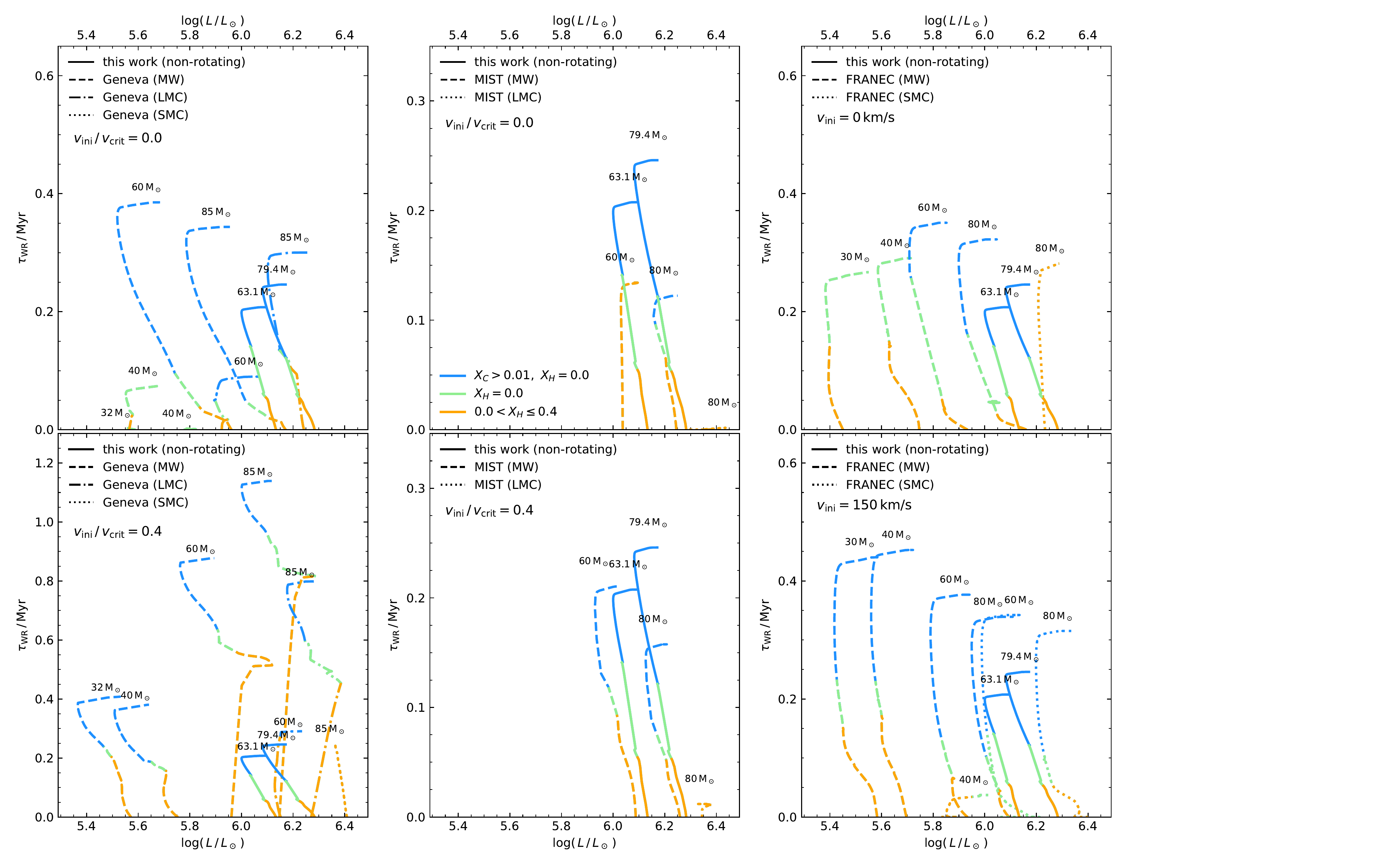}};
            
            \draw[black] (5.5,12.8) node {a)};
            \draw[black] (11.45,12.8) node {b)};
            \draw[black] (17.4,12.8) node {c)};
            \draw[black] (5.5,6.55) node {d)};
            \draw[black] (11.45,6.55) node {e)};
            \draw[black] (17.4,6.55) node {f)};
        \end{tikzpicture}
        \caption{Predicted time a model spends in the different WR phases $\tau_\mathrm{WR}$ as a function of luminosity for different initial parameters. The tracks taken from literature cover initial masses $M_\mathrm{\,i} = 30\msun$, $40\msun$, $60\msun$ and $80\msun$ (or $85\msun$ for the Geneva models). Due to the logarithmic spacing of our initial masses our models cover initial masses $M_\mathrm{\,i}= 32\msun$, $39.2\msun$, $63.1\msun$ and $79.4\msun$. The tail of each track is labeled with its initial mass. The tracks are colored in orange, green and blue according to their surface abundances which are used in the later analysis to respectively differentiate the H-poor WN, H-free WN and WC phase. In panel a) and d) the non-rotating and rotating Geneva tracks at galactic (dashed) and SMC (dotted) metallicity are compared to our model tracks (solid). Panels b) and e) show the non-rotating and rotating stellar evolution MIST tracks at galactic (dashed) and LMC (dotted) metallicity, in comparison to our tracks (solid). The last two panels c) and f) emphasize the differences of the galactic (dashed) and SMC (dotted) of the FRANEC models to our LMC models (solid).}
        \label{fig:WR_single_literature}
    \end{figure*}

    At first we compare our models to the Geneva \citep{eks1:12,geo1:12,geo1:13} stellar evolutionary models which cover a large range of masses and several metallicities. Unfortunately, the recent Geneva models are calculated for solar ($Z_\odot=0.014$). LMC ($Z=0.006$) and SMC ($Z=0.002$) metallicity. In Fig.\,\ref{fig:WR_single_literature} we show the predicted times a model spends as WR star as a function of luminosity, while the tracks are color coded by the different WR phases predicted by the models.
    
    The tracks of the non-rotating Geneva models are shown in panel a) of Fig.\,\ref{fig:WR_single_literature}. In our considered luminosity range their non-rotating SMC models avoid a WR phase at all, while their Galactic models can explain WR stars with luminosities above $\log(L/L_\odot) \gtrsim 5.4$. Their LMC models are a somewhat fainter than our models with luminosities above $\log(L/L_\odot) \gtrsim 5.7$. This is mainly because their adopted mass-loss rate during the RSG phase is larger. The Geneva models use the mass-loss rates of \citet{jag1:88}, which is similar to the one of \citet{nie1:90} used by us, except for luminosities above $\log(L/L_\odot)>5.5$, where the mass-loss rates of  \citet{jag1:88} become larger by a factor of 2 or more \citep{mau1:11}. Additionally, in the Geneva models the mass-loss rates during the RSG are increased by a factor of 3 if the luminosity in the envelope is five times higher than the Eddington luminosity in the convective envelope. This has a strong impact on the formation of WR stars at different luminosities. The Geneva models use the mass-loss recipe of \citet{nug1:00} with a clumping factor of $D=10$ for the WR phase, which makes their WR mass-loss less efficient than the one we adopt. This impacts the predicted times a stellar model can spend in a WR phase.
    
    Results from the Geneva models with rotation can be found in panel d) of Fig.\,\ref{fig:WR_single_literature}. These models drastically differ from the non-rotating case. First, they spend three times more time in a WR phase, most notably as H-poor WN. Second, all Galactic models can evolve now into WC stars with luminosities as low as $\log\,L/\lsun\approx5.4$. Third, their most massive SMC model with initial mass $85\,\msun$ now evolves into a H-poor WN type star.
    
    In contrast to the non-rotating models, these models 
    differ strongly from ours. As we argue in Sect.\,\ref{sec:ss}, their WR lifetimes are exceptionally high because of their efficient rotational mixing, which results in larger convective cores, which leads to an overestimate of the number of the most luminous WR stars.
    
    The second set of  stellar evolutionary models we compare to are the MESA Isochrones and Stellar Tracks (MIST) models from \citet{cho1:16}. As these models are also calculated with MESA like our models, one might expect relative similar results. However, their non-rotating Galactic and LMC models (panel b)) of Fig.\,\ref{fig:WR_single_literature}) produce a WR phase only for luminosities above $\log(L/L_\odot)= 6.3$. Their models are calculated with a smaller overshooting parameter of $0.2H_\mathrm{P}$ which leads to smaller He-cores, and their treatment of semiconvection is also less efficient than that in our models with $\alpha_\mathrm{sc}=0.1$. This inefficient mixing leave their models spend four times longer in the cool supergiant regime ($T_\mathrm{eff}\leq\SI{12500}{K}$) compared to our models, which explains the scarcity of predicted WR stars. 
    
    Their rotating LMC models show similar features, as they still can only explain WR stars with luminosities larger than $\log(L/L_\odot)\gtrsim 6.3$. It appears, that their Galactic models are in better agreement with our LMC models, as they match roughly the luminosity range, the times spend in the different WR phases.  
    
    The third set of single star models with which we want to compare our models to are the models calculated with the FRANEC code from \citet{lim1:18}. As in the case of the Geneva models the FRANEC models only cover Galactic ($[\mathrm{Fe/H}]=0$) and SMC ($[\mathrm{Fe/H}]=-1$) metallicity. In contrast to the Geneva models, the non-rotating FRANEC models at SMC metallicity, shown in panel c) of Fig.\,\ref{fig:WR_single_literature}, produce WR stars with luminosities above $\log(L/L_\odot)\gtrsim 6.2$. Their Galactic models have longer H-poor and H-free WN phases than those from the Geneva code. Overall these models seem to be in better agreement with our models, as they fit into the general trend of their models, while covering the expected luminosity range and WR phases. Contrary to the rotating Geneva models, the fast rotating FRANEC models spend only slightly more time in a WR phase, which appears to be in better agreement with the observations. 
    
    We emphasize that in case of the single stars, the predictions for the WR population depend sensitively on the adopted physics parameters. 
    
\subsection{Binary star models}

    \begin{figure*}[t]
        \centering
        \begin{tikzpicture}
        \node [anchor=south west] at (0,0) {\includegraphics[trim= 0.1cm 0cm 0.cm 0cm ,clip, width=1.0\textwidth]{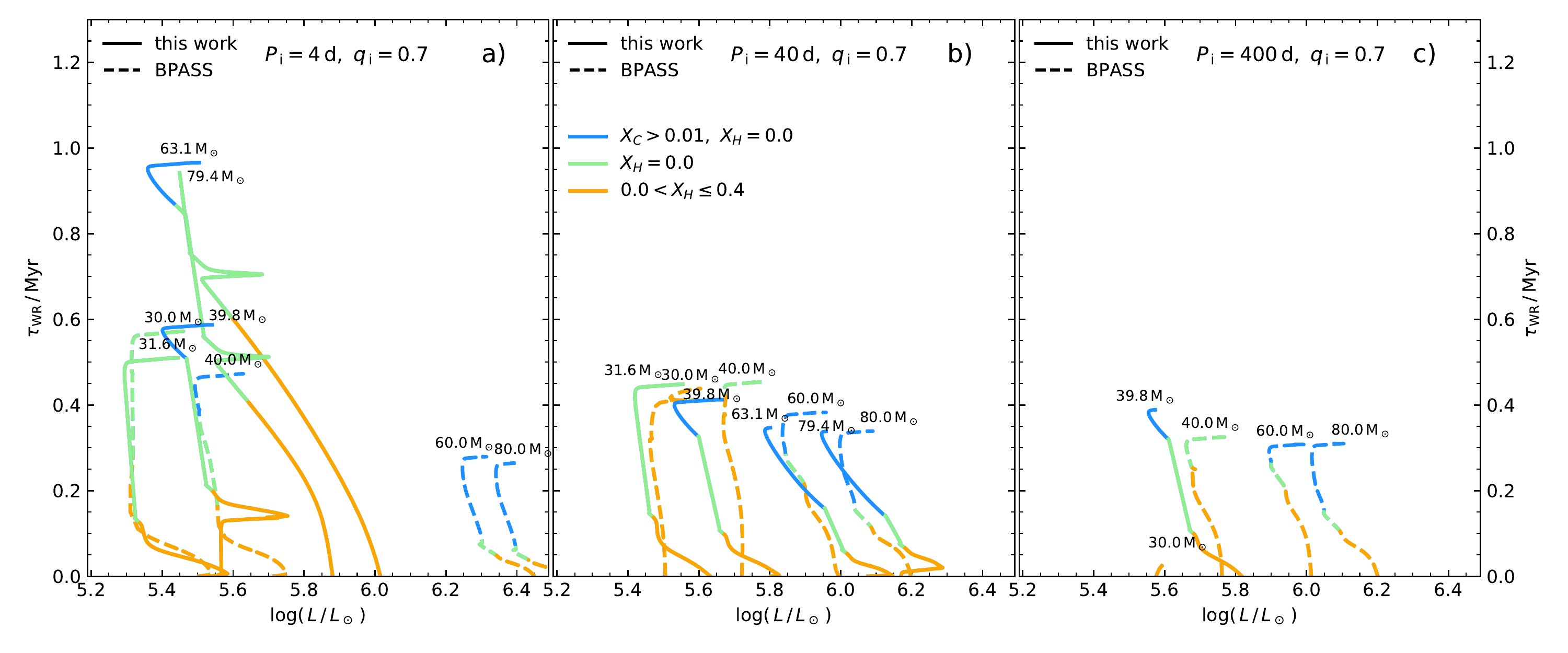}};
        \end{tikzpicture}
        \caption{Predicted time a model spends in the different WR phases $\tau_\mathrm{WR}$ as a function of luminosity for different initial masses and orbital periods of binary models at LMC metallicity. The BPASS tracks, depicted as dashed lines, cover initial donor masses ${M_\mathrm{1,\,i} = 30\msun}$, $40\msun$, $60\msun$ and $80\msun$. The solid lines correspond to our binary models which cover initial donor masses ${M_\mathrm{1,\,i}= 32\msun}$, $39.2\msun$, $63.1\msun$ and $79.4\msun$. Each track is labeled at its end with its initial mass. The tracks are colored in orange, green and blue according to their surface abundances which are used in the later analysis to respectively differentiate the H-poor WN, H-free WN and WC phase. All binary models have an initial mass-ratio of $q_\mathrm{\,i}=0.7$ and the tracks of initial orbital periods $P_\mathrm{\,i}=\SI{4}{d}$, $\SI{40}{d}$ and $\SI{400}{d}$ in panel a), b) and c), respectively.}
        \label{fig:WR_binary_literature}
    \end{figure*}
    
    In the literature, only few comprehensive sets of detailed binary evolution models are available, and the most commonly used ones are the Binary Population and Spectral Synthesis (BPASS) models from \citet{eld1:17}, covering wide ranges of masses, periods, mass-ratios and metallicities, including LMC metallicity. A comparison of selected BPASS tracks with our tracks for different initial orbital periods is shown in Fig.\,\ref{fig:WR_binary_literature}. 
    
    By comparing the binary models with initial periods of ${P_\mathrm{\,i}= \SI{4}{d}}$ shown in panel a) of Fig.\,\ref{fig:WR_binary_literature}, one can see  similarities and at the same time strong differences. The donor models with initial masses $30\msun$ and $40\msun$ evolve  similar. They show an agreement with luminosities and, therefore, masses, as well as in the time they spend in the individual WR phases. However, the predictions of the donor models with initial masses $60\msun$ and $80\msun$ differ strongly, with respect to their luminosity, lifetime and surface abundances. This can easily be explained, the BPASS models enter during the RLOF a common envelope phase which results in a merger. As a consequence, their models evolve into luminous RSGs ($\log(L/L_\odot)\geq6.5$), where a second mass-transfer phase is initiated and the remaining H-rich envelope is stripped, leading to the formation of the appearently ``overluminous'' WR stars.
    
    The BPASS models with initial periods of ${P_\mathrm{\,i}=\SI{40}{d}}$ (panel~b)) of Fig.\,\ref{fig:WR_binary_literature}) spend similar times as WR stars as for our models. However, the models predict different surface abundances and thus different WR phases at different luminosities. For instance, their model with initial mass $30\,\msun$ spends it entire WR phase as H-poor WN star, while our model with similar initial mass is able to full strip off its H-rich envelope and spends more than >50\% of its time as H-free WN star. This is likely linked to the effects of the different WR mass-loss recipes used. 
    
    A discrepancy for the different predicted WR subtypes can also be found for the longest period binaries, with initial periods $P_\mathrm{\,i}=\SI{400}{d}$  (panel c)) of Fig.\,\ref{fig:WR_binary_literature}). Here, the lack of very luminous WR stars in our calculations occurs due to mass-transfer rates exceeding the threshold beyond which a common envelope evolution is initiated. In BPASS there is a prescription to model systems undergoing unstable mass-transfer/common envelope evolution. Therefore, it is likely that these differences occur due to their different treatment of RLOF.
    
    We see that for close binaries, where BPASS assumes complete mixing during mass transfer, and for wide binaries, where BPASS appears to adopt a different merger criterion, the models which are not affected by these two issues agree
    well with our binary models.
    
 			\clearpage
\clearpage
\newpage
 			
\section{Model overview}
\label{app:phase_diagramm}
\begin{figure*}[p]
        \centering
        \includegraphics[height=1.3\textwidth]{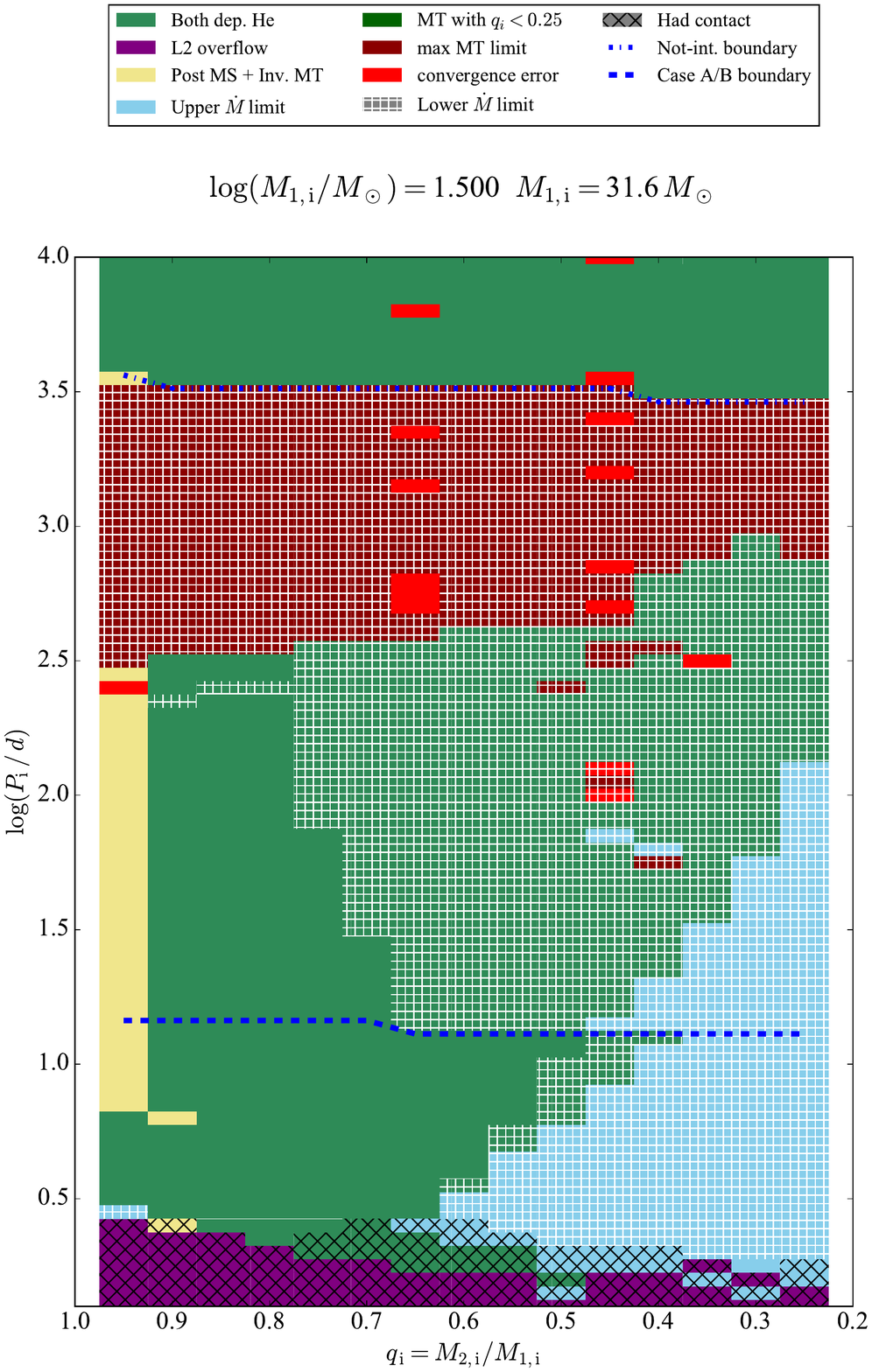}
        \caption{Phase diagram of binary models with fixed initial donor mass of $M_{1,\mathrm{i}}=31.6\msun$. The initial mass ratio~$q_\mathrm{\,i}$ is plotted versus the initial orbital period $P_\mathrm{\,i}$. The different phases are indicated in the legend and a more detailed description can be found in the text.}
        \label{fig:LMC_GRID_1.500}
    \end{figure*}

\begin{figure*}[p]
        \centering
        \includegraphics[height=1.3\textwidth]{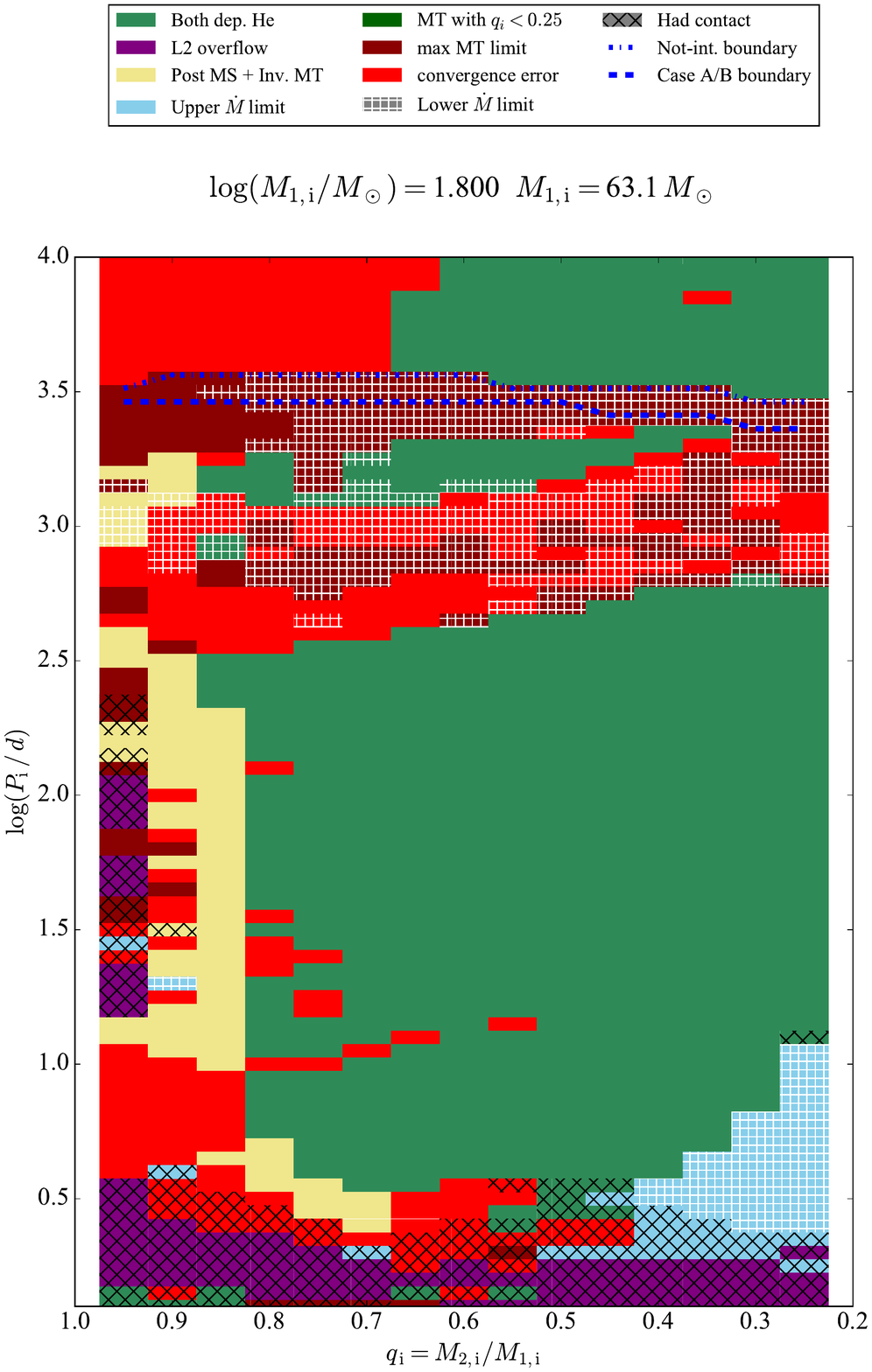}
        \caption{Same as Fig.\,\ref{fig:LMC_GRID_1.500} but for initial donor mass of $M_{1,\mathrm{i}}=63.1\msun$.}
        \label{fig:LMC_GRID_1.800}
    \end{figure*}

\begin{figure*}[p]
        \centering
        \includegraphics[height=1.3\textwidth]{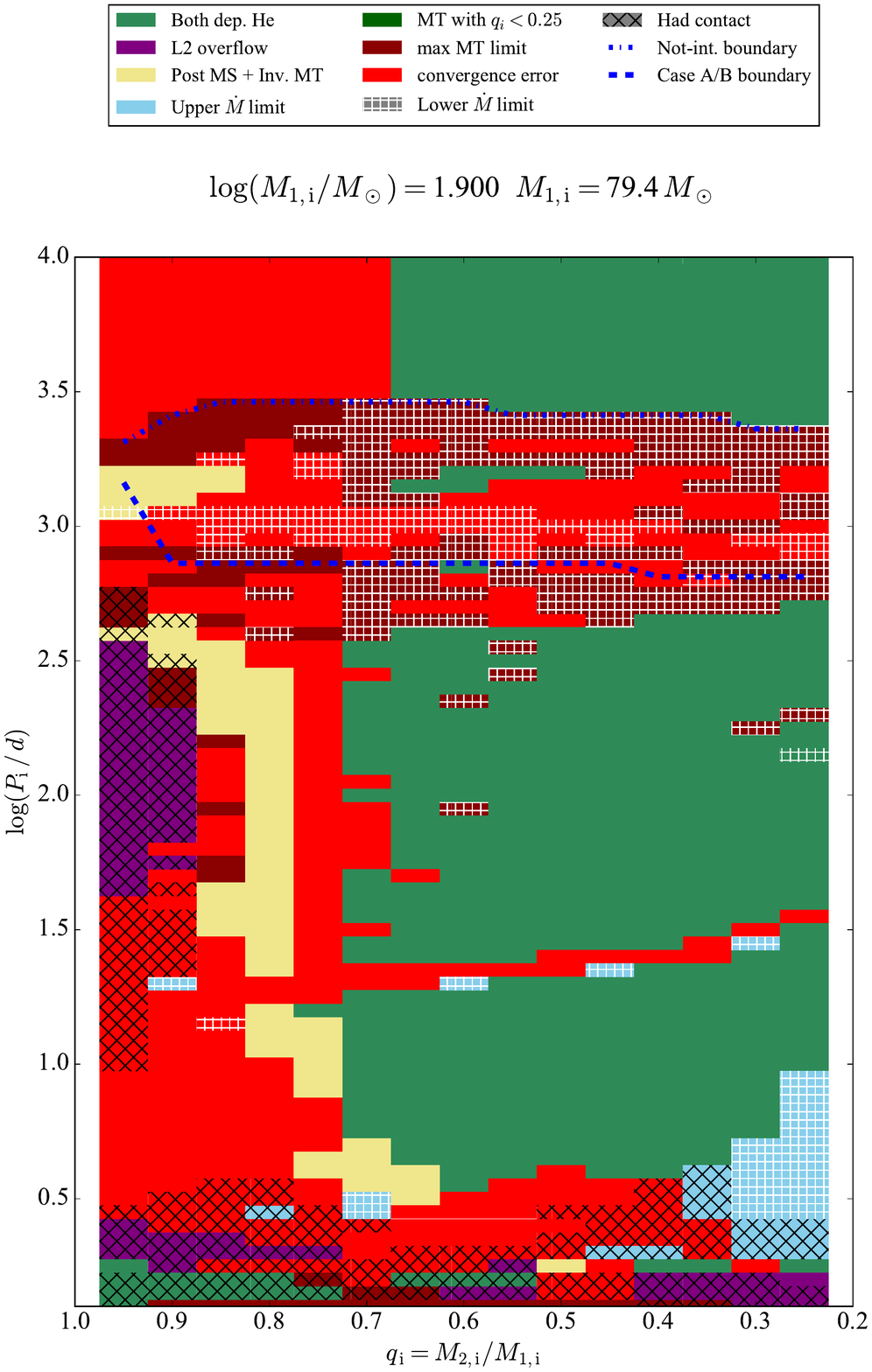}
        \caption{Same as Fig.\,\ref{fig:LMC_GRID_1.500} but for initial donor mass of $M_{1,\mathrm{i}}=79.4\msun$.}
        \label{fig:LMC_GRID_1.900}
    \end{figure*}

    Here, we provide three figures which allow to obtain an overview of the different final states of our binary models. For three exemplary initial primary masses (${M_{1\mathrm{,\,i}}=31.6\msun}$, $63.1\msun$ and $79.4\msun$), Figs.\,\ref{fig:LMC_GRID_1.500}, \ref{fig:LMC_GRID_1.800} and \ref{fig:LMC_GRID_1.900}
    show the covered initial parameter space in orbital period and mass ratio.
    Each pixel in these plots corresponds to one computed detailed binary evolution sequence, and its color and hatching gives information about its evolution (see below).
    They further indicate the borderline between Case\,A and Case\,B mass transfer (blue dashed line), and the borderline between interacting and non-interacting binaries (blue dotted line). The information coded in each pixel is as follows.

    \begin{itemize}
        \item[-] \textbf{Both dep. C}:\\
            Both stars are calculated until carbon depletion in their core. Due to convergence problems stellar models with helium core masses above $M_\mathrm{He}>14\msun$ are calculated to core helium depletion instead. When the primary depleted carbon the companion is modeled as a single star, ignoring the possibility of the formation of a binary with compact object due to the uncertainties of a supernova kick.

        \item[-] \textbf{L2 overflow}:\\
            During a contact phase both stars overfill their Roche lobe and mass is overflowing the second Lagrangian point $L_2$. The system is suspected to merge.
            
        \item[-] \textbf{Post MS + Inv. MT}:\\
            Inverse mass transfer from the secondary to a post main-sequence primary. These systems are suspected to have an unstable mass transfer and to merge, leading to the formation of a post main-sequence star.
            
        \item[-] \textbf{Upper $\mathbf{\dot{M}}$ limit}:\\
            The transferred material can neither be expelled nor accreted anymore (equation 5.3 with $R_\mathrm{RL,2}$ instead of $R_2$ in \citet{mar2:16} has been reached). The transferred material can neither be expelled nor accreted anymore. Therefore, the formation of a common envelope is assumed.

        \item[-] \textbf{Lower $\mathbf{\dot{M}}$ limit}:\\
            The transferred material might no-longer be able to be radiated away from the system (equation 5.3 in \citet{mar2:16} has been reached). Note that the models reaching this condition continue the calculation. This label has to be understood as a warning that radiation may not be able to drive a wind to expel the transferred material from the system.
    
        \item[-] \textbf{MT with $\mathbf{q_\mathrm{\,i}<0.25}$}:\\
            It is assumed that a mass transfer phase for systems with extreme mass ratios below $q_\mathrm{\,i}<0.25$ are will fill the Roche lobe of the secondary quickly leading to a common envelope phase and a potential merger.

        \item[-] \textbf{max MT limit}:\\
            This is the additional label mentioned above. Binary systems that trigger this condition have a mass transfer rate that exceeds $0.1\msunpyr$. These systems are suggested to form a common envelope.
    
        \item[-] \textbf{convergence error}:\\
            The calculation has stopped due to numerical issues and did not converge. This likely happens in the late phases of the evolution or late stages of mass transfer.
    
        \item[-] \textbf{Had contact}:\\
            Both stars had a contact phase. Note that this does not imply that this system results in a merger.
    
        \item[-] \textbf{Not-int. boundary}:\\
            Systems above this line are not interacting during their entire life. They evolve as single stars.
        
        \item[-] \textbf{Case A/B boundary}:\\
            Boundary line between systems undergoing Case A and Case B mass transfer.
    \end{itemize}
    
    To get a better understanding of the implications of the phase diagram the most characteristic regions of Fig.\,\ref{fig:LMC_GRID_1.500} will be described in more detail.
    
    Binary systems with initial orbital periods beneath the ``Case~A/B'' boundary already have a mass-transfer phase during their time on the main-sequence. One can see that most of the systems with very tight orbits experience a contact phase. During this phase a large fraction of the donors envelope is transferred and accreted, leading to a mass ratio close to ${q=1}$ after the mass-transfer phase. Therefore, it is possible for most systems that either the secondary fills its Roche lobe and inverse mass transfer sets in, or mass is overflowing through the second Lagrangian point during the contact phase. Both scenarios are suspected to result in a merger. For systems with initially more extreme mass ratios ${q_\mathrm{\,i}\rightarrow0}$, the accretor is not luminous enough so that infalling material cannot be transferred to infinity by a radiation driven wind. These systems are also suspected to form a common envelope and merge. Only a certain fraction of the binary systems with wide enough orbits and large enough mass ratios can have a successful mass transfer phase where both stars can be modeled until carbon depletion. It is worth mentioning that the accretion efficiency is larger for case A systems and decreases with increasing initial orbital period whereas the mass transfer rate grows with increasing initial orbital period.
    
    Because of the effect of envelope inflation near the Eddington limit the most massive stars of our grid expand during their time on the main-sequence. This shifts the ``Case~A/B'' boundary to higher initial orbital periods for increasing initial masses. The ``Case~A/B'' boundary reaches its maximum at ${M_{1\mathrm{,\,i}}\approx63.1\,\msun}$, where almost all systems undergo Case~A mass-transfer.
    For higher initial masses we predict the ``Case~A/B'' boundary to go down to lower initial orbital periods again, as the winds become efficient enough to remove large parts of the envelope, allowing the donors to remain more compact during their main-sequence evolution (cf. Fig.\,\ref{fig:HRD_single}). 
    
    Binary systems that have a mass-transfer phase while they evolve off the main-sequence are located between the ``Case~A/B'' boundary and the ``not-interacting'' boundary. These systems are too wide to have a contact phase and only binary models with initial mass ratios close to ${q_\mathrm{\,i}\simeq1}$ are expected to have an inverse mass transfer phase. Models with extreme mass ratios ${q_\mathrm{\,i}\rightarrow0}$ are still unable to drive a wind that transfers the infalling material to infinity as they reach the upper limit on mass transfer. For the systems with the widest orbits ($P_\mathrm{\,i}\gtrsim \SI{300}{d}$), the maximum value for mass transfer is triggered.  These systems are expected to have donors that have evolved into a red supergiant phase and have large convective envelopes which drastically increase the mass-transfer. The stability of mass transfer of these systems is questionable and in this work it is assumed that these systems form a common envelope. Only a certain fraction of binary systems is suspected to be able to evolve both stars until core carbon depletion. 

    Binary systems with an initial orbital period larger than the ``not-interacting'' boundary ($P_\mathrm{\,i}\gtrsim \SI{3000}{d}$) evolve as single stars. As the stars are tidally synchronized their initial rotation is rather small and they can be considered as quasi non-rotating single stars.
    
    In total less than $<5\%$ of our systems encounter ``convergence errors'' of different kinds. By inspecting Figs.\,\ref{fig:LMC_GRID_1.500}, \ref{fig:LMC_GRID_1.800} and \ref{fig:LMC_GRID_1.900} one can see that the fraction of systems encountering these ``convergence errors'' increases with increasing initial donor mass.  At the highest masses ($M_\mathrm{ini,\,1} = 89$) the fraction goes up to $30\%$.
    
    At higher initial donor masses some of our primary models have numerical issues, typically during mass-transfer. For instance, when the primary already evolved to a RSG (${\log(P/\mathrm{d}) \gtrsim 2.5}$ it has a big convective envelope which rapidly expands, leading to too small time steps and thus to convergence errors. Other models in this region of the phase diagram end their evolution under the ``max MT limit''. We believe that most of the systems, which have numerical issues during mass-transfer, would have triggered one of the criteria associated with a following common envelope evolution (e.g. ``convergence errors'' or ``Post MS + Inv. MT'') anyways, meaning that these systems are not expected to contribute to our synthetic WR population even if they would not have encountered the convergence errors.
    
    With increasing initial donor mass also the masses of our secondaries increase. The secondary, which is evolved as a single star after the primary has depleted helium in its core, evolves in most cases into a RSG star. During this late evolutionary stage it can be in some cases that these models encounter numerical instabilities and trigger the  ``convergence errors'' flag. Implying that for these systems the primary who predominantly contributes to the WR population is fully modeled and included in our synthetic population. Consequently, the true fraction of systems that encounter numerical issues and are neglected in our final synthetic WR population is thought to be below $<1\%$.

\clearpage
\onecolumn

\section{Tables with the Observed Data of WN and WC Stars in the LMC}\label{app:tablesWR}
  
 {\small
 \begin{longtable}{rccccccccc} 
 	\caption{ Parameters for WN stars in the LMC. Parameters of stars which are excluded from our analysis (see Sect.\,\ref{sec:WR_pop_LMC}) are colored in gray.}\label{tab:WN_LMC} \\
 	        \hline \hline \rule{0cm}{2.2ex}%
 		    BAT99  & spectral subtype$\,^{(a)}$ &$\log(L)\,^{(a)}$  & $X_\mathrm{H}\,^{(a)}$	&Binary?$\,^{(a)}$	&P$\,^{(b)}$ &$q\,^{(b)}$&H-free&H-poor&H-rich\\
 		    \rule{0cm}{2.2ex}%
 		           &                            &$[\lsun\,]$        &                           &                   &$[\si{d}]$&$[M_2/M_\mathrm{WR}]$&&&\\
 			\hline \vspace{0.001cm}
\endfirsthead

            \multicolumn{7}{c}{{{\normalsize \tablename\ \thetable{} continued.}}} \vspace{0.3cm}\\
            \hline \hline \rule{0cm}{2.2ex}%
 		    BAT99  & spectral subtype$\,^{(a)}$ &$\log(L)\,^{(a)}$  & $X_\mathrm{H}\,^{(a)}$	&Binary?$\,^{(a)}$	&P$\,^{(b)}$ &$q\,^{(b)}$&H-free&H-poor&H-rich\\
 		    \rule{0cm}{2.2ex}%
 		           &                            &$[\lsun\,]$        &                           &                   &$[\si{d}]$&$[M_2/M_\mathrm{WR}]$&&&\\
 			\hline \vspace{0.001cm}
 			\endhead
     			
     		\hline
     	    \multicolumn{7}{l}{{{
         	\begin{minipage}{0.65\linewidth}
                \ignorespaces
     		    \rule{0cm}{4ex}%
                $^{(a)}$ Values adopted from ~\citet{hai1:14}. $^{(b)}$ Values taken from \citet{She1:19}.\\ $^{(c)}$ Showing X-ray emission (see \citet{hai1:14}).
            \end{minipage}}}}\\
            \endfoot
                
     		\hline
     	    \multicolumn{7}{l}{{{
         	\begin{minipage}{0.65\linewidth}
                \ignorespaces
     		    \rule{0cm}{4ex}%
                $^{(a)}$ Values adopted from ~\citet{hai1:14}. $^{(b)}$ Values taken from \citet{She1:19}. \\ $^{(c)}$ Showing X-ray emission (see \citet{hai1:14}).
            \end{minipage}}}}\\
            \endlastfoot
                  1&	WN3b       &	 5.30&	 0.0&	-   &	-   &	-   &  x  &  -  &  -  \\
                  2&	WN2b(h)    &	 5.37&	 0.0&	-   &	-   &	-   &  x  &  -  &  -  \\
                  3&	WN4b       &	 5.51&	 0.0&	-   &	-   &	-   &  x  &  -  &  -  \\
                  5&	WN2b       & 	 5.45&	 0.0&	-   &	-   &	-   &  x  &  -  &  -  \\
                \color{gray}  6&	\color{gray}O3f*+O     &	 \color{gray}6.45&	 \color{gray}0.2&	\color{gray}x   &	\color{gray}2.00&	\color{gray}-   &  \color{gray}-  &  \color{gray}-  &  \color{gray}-  \\
                  7&	WN4b       & 	 5.84&	 0.0&	-   &	-   &	-   &  x  &  -  &  -  \\
                 \color{gray}12&	\color{gray}O2If*/WN5  & 	 \color{gray}5.80&	 \color{gray}0.5&	\color{gray}x   &	\color{gray}3.24&	\color{gray}-   &  \color{gray}-  &  \color{gray}-  &  \color{gray}-  \\
                 13&	WN10       & 	 5.56&	 0.4&	-   &	-   &	-   &  -  &  x  &  -  \\
                 14&	WN4o(+OB)  & 	 5.86&	 0.0&	?   &	-   &	-   &  x  &  -  &  -  \\
                 15&	WN4b       & 	 5.57&	 0.0&	-   &	-   &	-   &  x  &  -  &  -  \\
                 16&	WN7h       & 	 5.80&	 0.3&	-   &	-   &	-   &  -  &  x  &  -  \\
                 17&	WN4o       & 	 5.69&	 0.0&	-   &	-   &	-   &  x  &  -  &  -  \\
                 18&	WN3(h)     & 	 5.63&	 0.2&	-   &	-   &	-   &  -  &  x  &  -  \\
                 19&	WN4b+O5:   & 	 6.14&	 0.0&	x$^c$ &17.99&1.79   &  x  &  -  &  -  \\
                 21&	WN4o(+OB)  & 	 6.30&	 0.0&	?	&	- 	&	-   &  x  &  -  &  -  \\
                 22&	WN9h       & 	 5.75&	 0.4&	- 	&	- 	&	-   &  -  &  x  &  -  \\
                 23&	WN3(h)     & 	 5.55&	 0.0&	- 	&	- 	&	-   &  x  &  -  &  -  \\
                 24&	WN4b       & 	 5.54&	 0.0&	- 	&	- 	&	-   &  x  &  -  &  -  \\
                 25&	WN4ha      & 	 5.55&	 0.2&	- 	&	- 	&	-   &  -  &  x  &  -  \\
                 26&	WN4b       & 	 5.62&	 0.0&	- 	&	- 	&	-   &  x  &  -  &  -  \\
                 \color{gray}27&	\color{gray}WN5b(+B1Ia)&	 \color{gray}7.30&	 \color{gray}0.2&	\color{gray}?	&	\color{gray}- 	&	\color{gray}-   &  \color{gray}-  &  \color{gray}-  &  \color{gray}-  \\
                 29&	WN4b+OB    & 	 5.50&	 0.0&	x	&	2.20&	-   &  x  &  -  &  -  \\
                 30&	WN6h       & 	 5.65&	 0.3&	- 	&	- 	&	-   &  -  &  x  &  -  \\
                 31&	WN4b       & 	 5.33&	 0.0&	?	&	- 	&	-   &  x  &  -  &  -  \\
                 32&	WN6(h)     & 	 5.94&	 0.2&	x	&	1.90&0.98   &  -  &  x  &  -  \\
                 \color{gray}33&	\color{gray}Ofpe/WN9?  &	 \color{gray}6.50&	 \color{gray}0.2&	\color{gray}- 	&	\color{gray}- 	&	\color{gray}-   &  \color{gray}-  &  \color{gray}-  &  \color{gray}-  \\
                 35&	WN3(h)     & 	 5.60&	 0.1&	- 	&	- 	&	-   &  -  &  x  &  -  \\
                 36&	WN4b/WCE+OB& 	 5.71&	 0.0&	?	&	- 	&	-   &  x  &  -  &  -  \\
                 37&	WN3o       & 	 5.65&	 0.0&	- 	&	- 	&	-   &  x  &  -  &  -  \\
                 40&	WN4(h)a    & 	 5.62&	 0.2&	?$^c$ &	- 	&	-   &  -  &  x  &  -  \\
                 41&	WN4b       & 	 5.60&	 0.0&	- 	&	- 	&	-   &  x  &  -  &  -  \\
                 \color{gray}42&	\color{gray}WN5b(h)(+B3I)&	 \color{gray}8.00&	 \color{gray}0.4&	\color{gray}?$^c$ &	\color{gray}- 	&	\color{gray}-   &  \color{gray}-  &  \color{gray}-  &  \color{gray}-  \\
                 43&	WN4o+OB    & 	 5.85&	 0.0&	x	&	2.82&	-   &  x  &  -  &  -  \\
                 44&	WN8ha      & 	 5.66&	 0.4&	- 	&	- 	&	-   &  -  &  x  &  -  \\
                 46&	WN4o       & 	 5.44&	 0.0&	- 	&	- 	&	-   &  x  &  -  &  -  \\
                 47&	WN3b       & 	 5.59&	 0.0&	?$^c$ &	- 	&	-   &  x  &  -  &  -  \\
                 48&	WN4b       & 	 5.40&	 0.0&	- 	&   -   &   -   &  x  &  -  &  -  \\
                 49&	WN4:b+O8V  & 	 6.34&	 0.6&	x	&  31.69&2.00   &  -  &  -  &  x  \\
                 50&	WN5h       & 	 5.65&	 0.4&	- 	&	- 	&	-   &  -  &  x  &  -  \\
                 51&	WN3b       & 	 5.30&	 0.0&	- 	&	- 	&	-   &  x  &  -  &  -  \\
                 54&	WN8ha      & 	 5.75&	 0.2&	- 	&	- 	&	-   &  -  &  x  &  -  \\
                 55&	WN11h      & 	 5.77&	 0.4&	- 	&	- 	&	-   &  -  &  x  &  -  \\
                 56&	WN4b       & 	 5.56&	 0.0&	- 	&	- 	&	-   &  x  &  -  &  -  \\
                 57&	WN4b       & 	 5.40&	 0.0&	- 	&	- 	&	-   &  x  &  -  &  -  \\
                 58&	WN7h       & 	 5.64&	 0.3&	- 	&	- 	&	-   &  -  &  x  &  -  \\
                 \color{gray}59&	\color{gray}WN4b+O8:   & 	 \color{gray}6.45&	 \color{gray}0.0&	\color{gray}?	&	\color{gray}2.20&	\color{gray}-   &  \color{gray}-  &  \color{gray}-  &  \color{gray}-  \\
                 60&	WN4(h)a    & 	 5.78&	 0.2&	- 	&	- 	&	-   &  -  &  x  &  -  \\
                 62&	WN3(h)     & 	 5.41&	 0.1&	- 	&	- 	&	-   &  -  &  x  &  -  \\
                 63&	WN4ha:     & 	 5.58&	 0.4&	- 	&	- 	&	-   &  -  &  x  &  -  \\
                 64&	WN4o+O9:   & 	 6.05&	 0.0&	x	&  37.59&	-   &  x  &  -  &  -  \\
                 65&	WN4o       & 	 5.75&	 0.0&	- 	&	- 	&	-   &  x  &  -  &  -  \\
                 66&	WN3(h)     & 	 5.78&	 0.2&	- 	&	- 	&	-   &  -  &  x  &  -  \\
                 67&	WN5ha      & 	 5.96&	 0.3&	?$^c$ &	- 	&	-   &  -  &  x  &  -  \\
                 \color{gray}68&	\color{gray}O3.5If*/WN7&	 \color{gray}6.00&	 \color{gray}0.6&	\color{gray}- 	&	\color{gray}- 	&	\color{gray}-   &  \color{gray}-  &  \color{gray}-  &  \color{gray}-  \\
                 71&	WN4+O8:    & 	 5.98&	 0.0&	x	&	5.21&	-   &  x  &  -  &  -  \\
                 72&	WN4h+O3:   & 	 5.80&	 0.4&	?	&	- 	&	-   &  -  &  x  &  -  \\
                 73&	WN5ha      & 	 5.72&	 0.4&	- 	&	- 	&	-   &  -  &  x  &  -  \\
                 74&	WN3(h)a    & 	 5.69&	 0.2&	- 	&	- 	&	-   &  -  &  x  &  -  \\
                 75&	WN4o       & 	 5.56&	 0.0&	- 	&	- 	&	-   &  x  &  -  &  -  \\
                 76&	WN9ha      & 	 5.66&	 0.2&	-   &	- 	&	-   &  -  &  x  &  -  \\
                 \color{gray}77&	\color{gray}WN7ha      & 	 \color{gray}6.79&	 \color{gray}0.7&	\color{gray}x$^c$ &	\color{gray}3.00&\color{gray}1.66   &  \color{gray}-  &  \color{gray}-  &  \color{gray}-  \\
                 78&	WN6(+O8V)  &	 5.70&	 0.2&	?$^c$ &	- 	&	-   &  -  &  x  &  -  \\
                 79&	WN7ha+OB   & 	 6.17&	 0.2&	?$^c$ &	- 	&	-   &  -  &  x  &  -  \\
                 \color{gray}80&	\color{gray}WN5h:a     & 	 \color{gray}6.40&	 \color{gray}0.2&	\color{gray}?$^c$ &	\color{gray}- 	&	\color{gray}-   &  \color{gray}-  &  \color{gray}-  &  \color{gray}-  \\
                 81&	WN5h       & 	 5.48&	 0.4&	- 	&	- 	&	-   &  -  &  x  &  -  \\
                 82&	WN3b       & 	 5.53&	 0.0&	?$^c$ &	- 	&	-   &  x  &  -  &  -  \\
                 86&	WN3(h)     & 	 5.33&	 0.0&	- 	&	- 	&	-   &  x  &  -  &  -  \\
                 88&	WN4b/WCE   & 	 5.80&	 0.0&	- 	&	- 	&	-   &  x  &  -  &  -  \\
                 89&	WN7h       & 	 5.78&	 0.2&	- 	&	- 	&	-   &  -  &  x  &  -  \\
                 91&	WN6(h)     & 	 5.42&	 0.2&	- 	&	- 	&	-   &  -  &  x  &  -  \\
                 \color{gray}92&	\color{gray}WN3:b(+O)+B1Ia&	 \color{gray}6.95&	 \color{gray}0.2&	\color{gray}x$^c$ &	\color{gray}4.31&	\color{gray}-   &  \color{gray}-  &  \color{gray}-  &  \color{gray}-  \\
                 \color{gray}93&	\color{gray}O3If*      &	 \color{gray}5.90&	 \color{gray}0.6&	\color{gray}?$^c$ &	\color{gray}- 	&	\color{gray}-   &  \color{gray}-  &  \color{gray}-  &  \color{gray}-  \\
                 94&	WN4b       & 	 5.80&	 0.0&	- 	&	- 	&	-   &  x  &  -  &  -  \\
                 95&	WN7h+OB    & 	 6.00&	 0.2&	x	&	2.11&2.20   &  -  &  x  &  -  \\
                 96&	WN8        & 	 6.35&	 0.2&	- 	&	- 	&	-   &  -  &  x  &  -  \\
                 \color{gray}97&	\color{gray}O3.5If*/WN7&	 \color{gray}6.30&	 \color{gray}0.6&	\color{gray}- 	&	\color{gray}- 	&	\color{gray}-   &  \color{gray}-  &  \color{gray}-  &  \color{gray}-  \\
                 \color{gray}98&	\color{gray}WN6        & 	 \color{gray}6.70&	 \color{gray}0.6&	\color{gray}- 	&	\color{gray}- 	&	\color{gray}-   &  \color{gray}-  &  \color{gray}-  &  \color{gray}-  \\
                 \color{gray}99&	\color{gray}O2.5If*/WN6&	 \color{gray}5.90&	 \color{gray}0.2&	\color{gray}x$^c$ &	\color{gray}92.6&	\color{gray}-   &  \color{gray}-  &  \color{gray}-  &  \color{gray}-  \\
                100&	WN7        & 	 6.15&	 0.2&	?$^c$ &	- 	&	-   &  -  &  x  &  -  \\
                \color{gray}102&	\color{gray}WN6        &	 \color{gray}6.80&	 \color{gray}0.4&	\color{gray}?$^c$ &	\color{gray}- 	&	\color{gray}-   &  \color{gray}-  &  \color{gray}-  &  \color{gray}-  \\
                103&	WN5(h)+O   & 	 6.25&	 0.4&	x$^c$ &	2.75&1.94   &  -  &  x  &  -  \\
                \color{gray}104&	\color{gray}O2If*/WN5  &	 \color{gray}6.06&	 \color{gray}0.4&	\color{gray}- 	&	\color{gray}- 	&	\color{gray}-   &  \color{gray}-  &  \color{gray}-  &  \color{gray}-  \\
                \color{gray}105&	\color{gray}O2If*      &	 \color{gray}6.40&	 \color{gray}0.6&	\color{gray}?$^c$ &	\color{gray}- 	&	\color{gray}-   &  \color{gray}-  &  \color{gray}-  &  \color{gray}-  \\
                \color{gray}106&	\color{gray}WN5h       & 	 \color{gray}6.51&	 \color{gray}0.4&	\color{gray}- 	&	\color{gray}- 	&	\color{gray}-   &  \color{gray}-  &  \color{gray}-  &  \color{gray}-  \\
                \color{gray}107&	\color{gray}O6.5Iafc+O6Iaf&  \color{gray}6.31&	 \color{gray}0.4&	\color{gray}x$^c$&\color{gray}153.89&\color{gray}0.81   &  \color{gray}-  &  \color{gray}-  &  \color{gray}-  \\
                \color{gray}108&	\color{gray}WN5h       & 	 \color{gray}6.87&	 \color{gray}0.4&	\color{gray}- 	&	\color{gray}- 	&	\color{gray}-   &  \color{gray}-  &  \color{gray}-  &  \color{gray}-  \\
                \color{gray}109&	\color{gray}WN5h       & 	 \color{gray}6.69&	 \color{gray}0.4&	\color{gray}- 	&	\color{gray}- 	&	\color{gray}-   &  \color{gray}-  &  \color{gray}-  &  \color{gray}-  \\
                \color{gray}110&	\color{gray}O2If*      &	 \color{gray}6.22&	 \color{gray}0.7&	\color{gray}- 	&	\color{gray}- 	&	\color{gray}-   &  \color{gray}-  &  \color{gray}-  &  \color{gray}-  \\
                111&	WN9ha      & 	 6.25&	 0.7&	?$^c$ &	- 	&	-   &  -  &  -  &  x  \\
                \color{gray}112&	\color{gray}WN5h       & 	 \color{gray}6.48&	 \color{gray}0.2&	\color{gray}?$^c$ &	\color{gray}8.2 &	\color{gray}-   &  \color{gray}-  &  \color{gray}-  &  \color{gray}-  \\
                \color{gray}113&	\color{gray}O2If*/WN5  &	\color{gray} 6.09&	\color{gray} 0.2&	\color{gray}x$^c$ &	\color{gray}4.70&\color{gray}0.32   & \color{gray} -  &  \color{gray}-  & \color{gray} -  \\
                \color{gray}114&	\color{gray}O2If*/WN5  &	 \color{gray}6.44&	 \color{gray}0.4&	\color{gray}?$^c$ &	\color{gray}- 	&	\color{gray}-   &  \color{gray}-  &  \color{gray}-  &  \color{gray}-  \\
                \color{gray}116&	\color{gray}WN5h:a     & 	 \color{gray}7.05&	 \color{gray}0.4&	\color{gray}?$^c$&\color{gray}154.55&\color{gray}0.92   &  \color{gray}-  &  \color{gray}-  &  \color{gray}-  \\
                \color{gray}117&	\color{gray}WN5ha      & 	 \color{gray}6.40&	 \color{gray}0.4&	\color{gray}- 	&	\color{gray}- 	&	\color{gray}-   &  \color{gray}-  &  \color{gray}-  &  \color{gray}-  \\
                \color{gray}118&	\color{gray}WN6h       & 	 \color{gray}6.66&	 \color{gray}0.2&	\color{gray}x$^c$ &	\color{gray}- 	&	\color{gray}-   &  \color{gray}-  &  \color{gray}-  &  \color{gray}-  \\
                \color{gray}119&	\color{gray}WN6h+?     & 	 \color{gray}6.57&	 \color{gray}0.2&	\color{gray}x$^c$&\color{gray}158.76&\color{gray}1.01   &  \color{gray}-  &  \color{gray}-  &  \color{gray}-  \\
                120&	WN9h       & 	 5.58&	 0.3&	- 	&	- 	&	-   &  -  &  x  &  -  \\
                122&	WN5h       & 	 6.23&	 0.2&	- 	&	- 	&	-   &  -  &  x  &  -  \\
                124&	WN4        & 	 5.45&	 0.0&	- 	&	- 	&	-   &  x  &  -  &  -  \\
                \color{gray}126&	\color{gray}WN4b+O8:   & 	 \color{gray}6.44&	 \color{gray}0.0&	\color{gray}?$^c$ &	\color{gray}25.5&	\color{gray}-   &  \color{gray}-  &  \color{gray}-  &  \color{gray}-  \\
                128&	WN3b       & 	 5.44&	 0.0&	- 	&	- 	&	-   &  x  &  -  &  -  \\
                129&	WN3(h)a+O5V&	 6.20&	 0.2&	x	&	2.77&1.64   &  -  &  x  &  -  \\
                130&	WN11h      & 	 5.68&	 0.4&	- 	&	- 	&	-   &  -  &  x  &  -  \\
                131&	WN4b       & 	 5.67&	 0.0&	- 	&	- 	&	-   &  x  &  -  &  -  \\
                132&	WN4b(h)    & 	 5.58&	 0.0&	- 	&	- 	&	-   &  x  &  -  &  -  \\
                133&	WN11h      & 	 5.69&	 0.4&	- 	&	- 	&	-   &  -  &  x  &  -  \\
                134&	WN4b       & 	 5.51&	 0.0&	- 	&	- 	&	-   &  x  &  -  &  -  \\
 \end{longtable} }

    \begin{table*} [htbp]
    \captionsetup{justification=centering}
 	\caption{\label{tab:WC_LMC} Parameters for WC stars in the LMC. Parameters of stars which are excluded from our analysis (see Sect.\,\ref{sec:WR_pop_LMC}) are colored in gray.}
 	{\small 
 	\begin{center}
 	    \begin{tabular}{rcccccc}\hline \hline \rule{0cm}{2.2ex}%
 		    BAT99  & spectral subtype$\,^{(a)}$ &$\text{M}_\mathrm{V}\,^{(a)}$	&$\log(L)$ & $\log(L_\mathrm{lit})$	&Binary?$\,^{(a,b)}$	&P$\,^{(c)}$\\
 		    \rule{0cm}{2.2ex}%
 		           &                            & $[\si{mag}]$                  &$[\lsun\,]$&$[\lsun\,]$&       &$[\si{d}\,]$\\
 			\hline \rule{0cm}{2.2ex}%
                9   & WC4               & -4.4	&5.49	&5.48$\,^{(d)}$	&-	&-\\
                8   & WC4               & -4.2	&5.41 	&5.42$\,^{(e)}$	&-	&-\\
                10  & WC4(+O9.5II:)     & -5.1  &5.77	&-	            &?	&-\\
                11  & WC4               & -5.5	&5.93	&5.70$\,^{(e)}$	&-	&-\\
                20  & WC4+O             & -4.6	&5.57	&-	            &-	&-\\
                28  & WC6+O5-6V-III(+O) & -6.5	&6.33	&-              &x	&14.926\\
                34  & WC4+OB            & -5.8	&6.05	&-	            &?	&-\\
                \color{gray}38  & \color{gray}WC4(+O8I:)        & \color{gray}-7.2	&\color{gray}6.61	&\color{gray}-	            &\color{gray}x	&\color{gray}3.0328\\
                39  & WC4+O6V-III(+O)   & -6.1	&6.17	&-	            &x	&1.9169\\
                52  & WC4               & -4.5	&5.53	&5.65$\,^{(e)}$	&-	&-\\
                53  & WC4(+OB)          & -5.4	&5.89	&5.35$\,^{(f)}$	&?	&-\\
                61  & WC4               & -5.0	&5.73	&5.68$\,^{(e)}$	&-	&-\\
                70  & WC4(+OB)          & -5.3	&5.85	&-	            &?	&-\\
                \color{gray}85  & \color{gray}WC4(+OB)          & \color{gray}-7.7	&\color{gray}6.81	&\color{gray}-	            &\color{gray}?	&\color{gray}-\\
                84  & WC4(+OB)          & -5.4	&5.89	&-	            &?	&-\\
                87  & WC4+OB            & -4.8	&5.65	&-	            &?	&-\\
                90  & WC4               & -4.6	&5.57	&5.44$\,^{(e)}$	&-	&-\\
                115 & WC4(+OB)          & -5.7	&6.01	&-	            &?	&-\\
                \color{gray}101 & \color{gray}WC4(+WN6+O)       & \color{gray}-7.5	&\color{gray}6.73	&\color{gray}-	            &\color{gray}?	&\color{gray}-\\
                121 & WC4               & -4.7	&5.61	&-	            &-	&-\\
                125 & WC4(+OB)          & -5.6	&5.97	&-	            &?	&-\\
                127 & WC4(+O)           & -6.2	&6.21	&-	            &?	&-\\

 		\hline
 		\end{tabular}
 	    \rule{0cm}{4ex}%
     	\begin{minipage}{0.65\linewidth}
            \ignorespaces
            $^{(a)}$ Values adopted from~\citet{bar1:01}. $^{(c)}$ We mark a binary with an ``x'' if it is known as an SB1 or SB2 binary, if it shows indications of binarity (see table 6 in~\citet{bar1:01}) it is marked with an ``?''. $^{(c)}$ Values taken from \citet{bar2:01}. $^{(d)}$Calculated by \citet{hil1:21}.$^{(e)}$Calculated by \citet{cro1:02}. $^{(f)}$ Calculated by \citet{ram1:18}. %
        \end{minipage}
    \end{center}
    }
 \end{table*} 
 
\clearpage

\end{appendix}

\end{document}